\documentclass[prd ,aps,12 pt,nofootinbib,groupedaddress, preprintnumbers,notitlepage]{revtex4-2}
\pdfoutput=1
\usepackage{graphicx}
\usepackage{amssymb}
\usepackage{mathtools}
\usepackage{epstopdf}
\usepackage{color}
\usepackage{bbold}
\usepackage{float}
\usepackage[title]{appendix}
\usepackage[colorlinks, citecolor=blue, linkcolor=black, urlcolor=purple]{hyperref}
\usepackage{amsfonts}
\usepackage{amsmath}
\usepackage{setspace}
\usepackage{epsfig}
\usepackage[usenames, dvipsnames]{xcolor}
\usepackage{tabularx}
\usepackage{empheq}
\usepackage{mathrsfs}
\usepackage{xcolor}
\usepackage{ulem}
\usepackage{braket}
\usepackage{tikz}
\usepackage{units}

\newcommand{\be}{\begin{eqnarray}}
\newcommand{\ee}{\end{eqnarray}}
\newcommand{\bea}{\begin{eqnarray}}
\newcommand{\eea}{\end{eqnarray}}
\newcommand{\nn}{\nonumber}

\newcommand{\beq}{\begin{equation}}
\newcommand{\eeq}{\end{equation}}

\newcommand{\MS}{{\rm \overline{MS}}}

\definecolor{bluDT}{cmyk}{1,0.5,0,0.3}

\definecolor{darkblue}{rgb}{0.2,0.2,0.9}
\definecolor{colorRTD}{rgb}{.2,.2,.7}

\definecolor{colorHD}{rgb}{.2,0.9,.0.9}

\usepackage{soul}

\begin{document}


\title{\Large The Two Scales of New Physics in Higgs Couplings \\ 
}

\author{Raffaele Tito D'Agnolo}
\author{Florian Nortier} 
\author{Gabriele Rigo} 
\author{Pablo Sesma}
\affiliation{Université Paris-Saclay, CNRS, CEA, Institut de Physique Théorique, 91191, Gif-sur-Yvette, France}


\begin{abstract}
\mbox{} \\
\indent Higgs coupling deviations from Standard Model predictions contain information about two scales of Nature: that of new physics responsible for the deviation, and the scale where new bosons must appear. The two can coincide, but they do not have to. The scale of new bosons can be calculated by going beyond an effective field theory description of the coupling deviation. We compute model-independent upper bounds on the scale of new bosons for deviations in Higgs to $WW$ and $ZZ$ couplings, finding that any measured deviation at present or future colliders requires the existence of new bosons within experimental reach. This has potentially interesting implications for naturalness. 
\end{abstract}

\maketitle
\onecolumngrid
\newpage
\tableofcontents
\newpage

\section{Introduction} 
The LHC broke our collective heart by discovering one of the most interesting particles in history\footnote{What appears to be the only fundamental scalar discovered so far.} and nothing else. The prevailing response in the particle community is to abandon model-building in favor of model-independent methods of interrogating the data. The wealth of information still hidden in LHC data should be extracted using Effective Field Theory (EFT) techniques and the Standard Model (SM) fields. In this work we strive to find a path in between the model-building feast of the past and the EFT austerity of the present. We go beyond a SM EFT description of the data, but we are still able to make a model-independent statement about Higgs couplings. This could not have been done by considering only operators built out of SM fields.

A deviation in Higgs couplings, compared to the SM prediction, implies the existence of new bosons below a calculable energy scale. We compute this scale for $hWW$ and $hZZ$ couplings. In general, new bosons appear at an energy greater or equal than the new particles responsible for the deviation. However, in the case of $hWW$ and $hZZ$, we find that any deviation observed at HL-LHC can only be generated by new bosons. Additionally, even deviations as small as those that can be probed at the most precise future lepton colliders require either new bosons roughly below 100 TeV, or new fermions with masses $M\simeq 100$~GeV and weak interactions with the SM, that can be discovered at HL-LHC. 

We obtain these results by writing theories that contain only new fermions and generate the coupling deviation. We find that these theories are unstable under renormalization group equation (RGE) running and have to be modified above a given scale. Only new bosons can stabilize the running by avoiding a deep AdS minimum in the Higgs potential that gives rise to an unacceptably large rate of SM vacuum decay. We imagine having already discovered the new fermions responsible for the coupling deviation and we then calculate the finite range of validity of these theories. This logic was already outlined in~\cite{Arkani-Hamed:2012dcq, Blum:2015rpa}, but never applied to $hWW$ and $hZZ$ coupling deviations that are the most sensitive to the presence of new bosons, but also the most laborious to calculate. Compared to~\cite{Arkani-Hamed:2012dcq, Blum:2015rpa} we introduce a second novelty. Instead of considering only a fixed set of low energy fermionic theories in small SM representations, we study the dependence of our results on the new fermions' representations and identify those that give the most conservative upper bound on the scale of new bosons, thus obtaining a model-independent result. Other works that discuss, in different contexts, the instability of the Higgs potential due to new fermions include~\cite{Gogoladze:2008ak, Reece:2012gi, Chen:2012faa, Joglekar:2012vc, Kearney:2012zi, Batell:2012ca, Altmannshofer:2013zba, Fairbairn:2013xaa, Xiao:2014kba, Ellis:2014dza, Angelescu:2016mhl, Goswami:2018jar, Gopalakrishna:2018uxn, Borah:2020nsz, Bandyopadhyay:2020djh, Arsenault:2022xty, Hiller:2022rla}.

The $hWW$ and $hZZ$ couplings are special because new fermions can modify them only at loop-level, but they exist at tree-level in the SM. As discussed in Section~\ref{sec:basic}, this implies the tightest upper bounds on the scale of new bosons if compared to other couplings exhibiting the same relative deviation. Additionally, $hWW$ and $hZZ$ are among the three Higgs couplings that will be most precisely measured at future colliders in terms of relative coupling deviation. This makes observing a deviation in these couplings particularly interesting, also in view of the possible connection between new bosons and naturalness.

The rest of the paper is organized as follows: in Section~\ref{sec:basic} we outline the conceptual steps that lead to a bound on the scale of new bosons, in Section~\ref{sec:reps} we introduce the fermionic theories that can induce a $hWW$ or $hZZ$ coupling deviation, in Section~\ref{sec:coupling} we calculate the coupling deviation and its dependence on the parameters of the fermionic theory, recalling the ingredients necessary to renormalize the electroweak sector of the SM at one loop. In Section~\ref{sec:repsII} we use these results to restrict the fermionic theories that we need to consider to obtain an upper bound on the scale of new bosons. The upper bound is then computed in Section~\ref{sec:results}, where we comment on the detectability of the new bosons.

\section{Basic Idea}\label{sec:basic}
New particles that stabilize the Higgs mass via loop diagrams
\begin{center}
\includegraphics[width=0.6\textwidth]{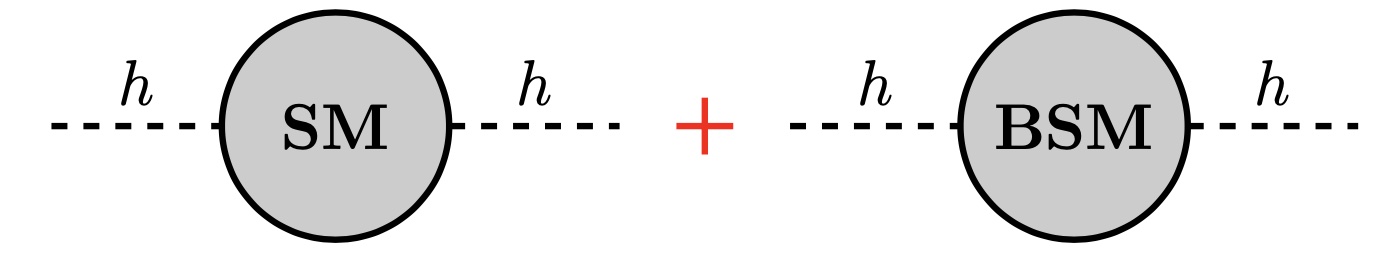}
\end{center}

can affect its production and decay rates through (almost) the same diagrams\footnote{The cartoon is adapted from~\cite{Arvanitaki:2011ck}.}

\begin{center}
\includegraphics[width=0.6\textwidth]{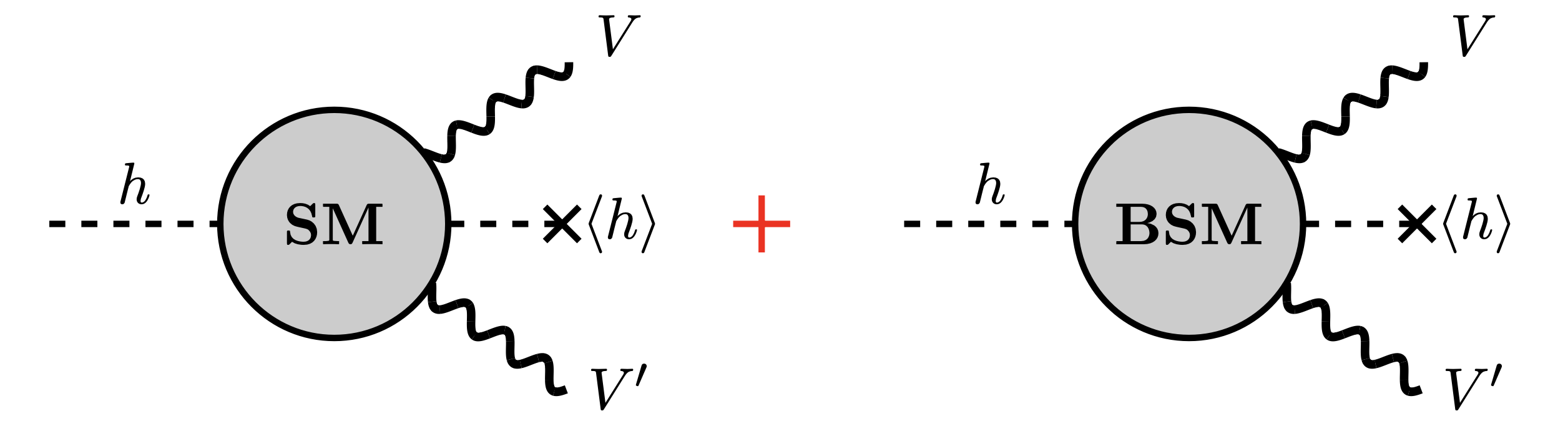}
\end{center}

As a consequence, theories that are natural in the traditional sense (i.e. those where a new weak-scale symmetry explains the observed value of $m_h^2$) predict deviations in Higgs couplings compared to the SM expectation. In practice it is extremely hard to turn this suggestive picture into quantitative and model-independent statements relevant for colliders, but in this work we discuss a rare case where this is possible. 

Measuring a deviation in Higgs couplings to $WW$ and/or $ZZ$ sets a calculable upper bound on the scale where new bosonic particles must appear in Nature. The bigger the deviation the smaller the upper bound. This is not an explicit statement about naturalness, but it has important implications for it. New bosons can be a definitive sign of unnaturalness if they have spin zero and come without symmetries (or anthropic roles) protecting them. However, they could equally well be the first sign of the long awaited symmetry that explains the value of $m_h^2$. The arguments in this work do not allow to distinguish between these two possibilities or to eliminate the third option (i.e. new bosons of spin $\geq 1$ having nothing to do with $m_h^2$), but allow to compute the scale of the new bosons. Experiment will then give us the answer about naturalness.

The scale of new bosons can be computed by considering low energy theories that contain only new fermions in addition to the SM and showing that they have a finite range of validity. This work is a ``proof by contradiction". We focus on the phenomenology of new fermions to prove that all theories with a $hWW$ or $hZZ$ coupling deviation must contain new bosons below an energy scale that can be calculated. 

Note that this is quite different compared to the traditional EFT intuition that associates a new physics scale to a Higgs coupling deviation. Consider for example a $d=6$ operator that contributes to a Higgs coupling to SM particles as $\delta g_h \simeq c(v^2/\Lambda^2)$. We imagine that we have already probed the scale $\Lambda \simeq v\sqrt{c/\delta g_h}$ and found only new fermions. Then we show that these theories are valid only up to a scale $\Lambda_B > \Lambda$ where new bosons must appear. In practice we find that $hWW$ or $hZZ$ coupling deviations within reach of HL-LHC imply $\Lambda_B \simeq \Lambda$. Even deviations as small as one part in a thousand (that can be probed at future lepton colliders) set tight upper bounds on $\Lambda_B$. We give more precise results in Section~\ref{sec:results}, where we also show that in most cases the new bosons are within reach of future hadron colliders or in some cases even the LHC.

\begin{figure}[!t]
\centering
\includegraphics[width=0.49\textwidth]{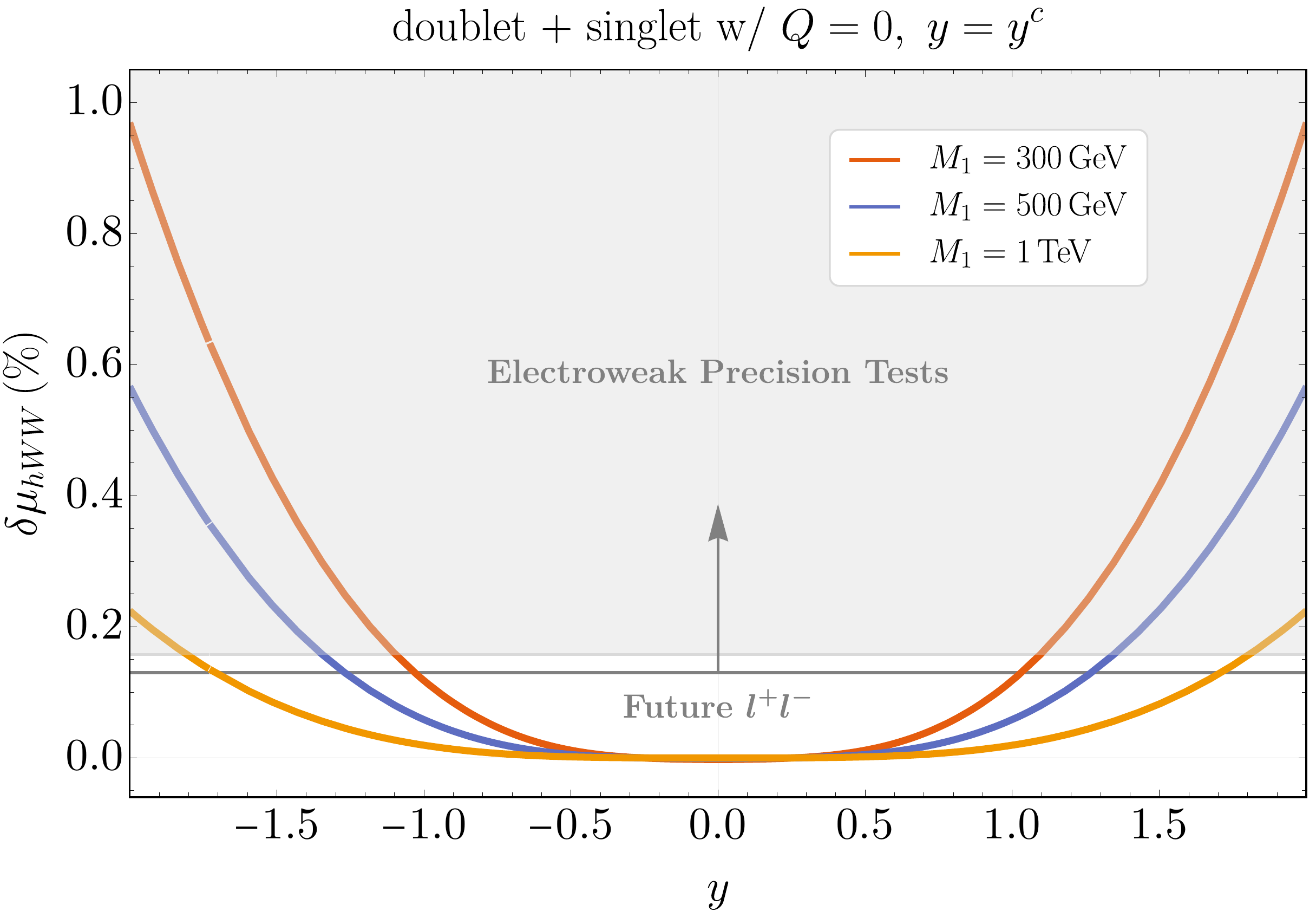}
\includegraphics[width=0.49\textwidth]{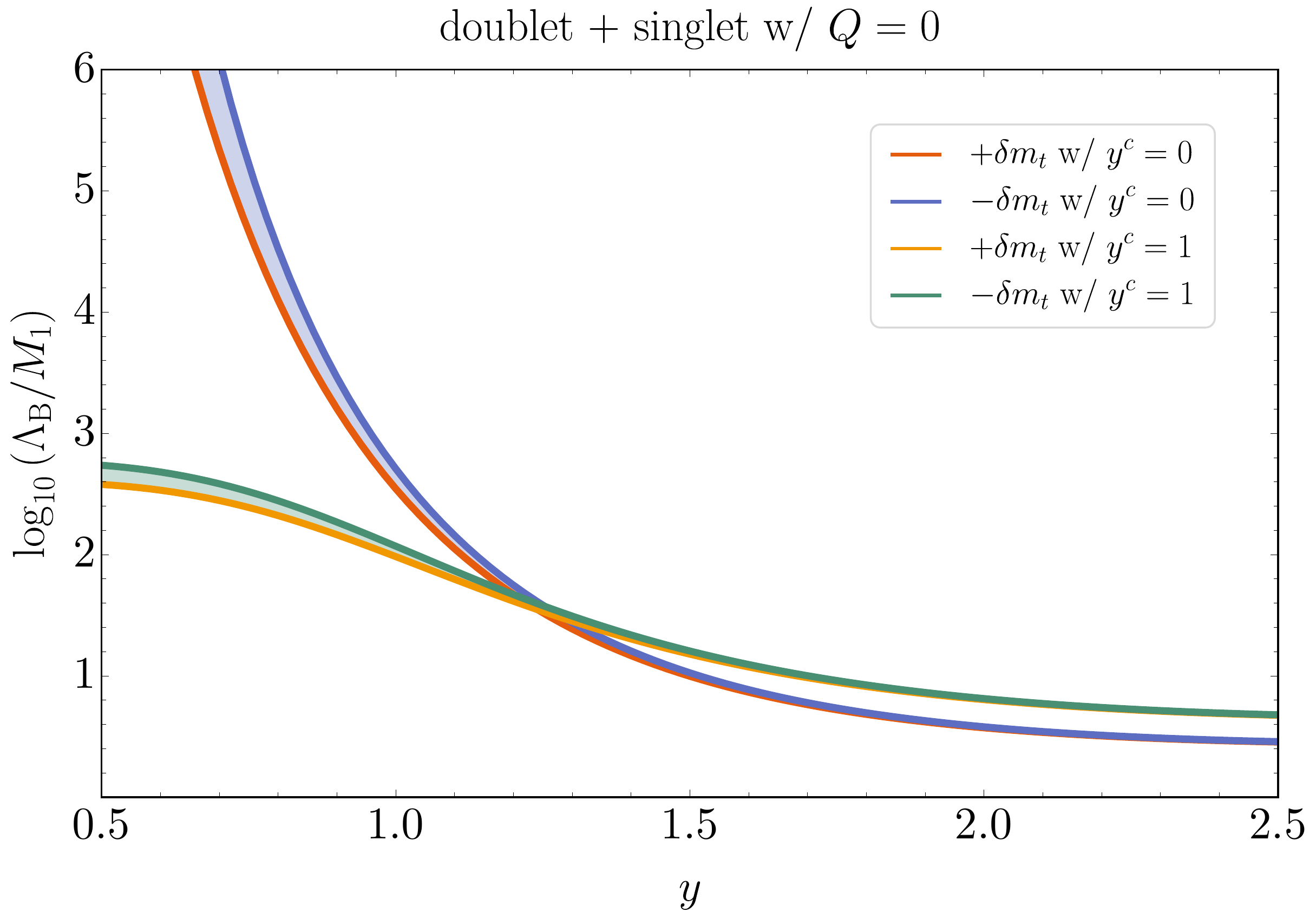}
\caption{Left panel: Relative $hWW$ coupling deviation for a configuration of parameters of the doublet+singlet model (introduced in Section~\ref{sec:4f}) that maximizes it. The future collider sensitivity is discussed in Section~\ref{sec:results}. Right panel: Instability scale $\Lambda_B$ of the doublet+singlet theory, as a function of one of its two Yukawa couplings (Eq.~\eqref{eq:L4}). The scale $\Lambda_B$ is normalized to the lightest new fermion mass $M_1$. The two pairs of curves correspond to $y^c=0$ (red and blue) and $y^c=1$ (green and yellow). Curves with the same $y^c$ differ by the choice of top Yukawa. We vary the top mass by twice its width to show that in these theories the instability scale is not very sensitive to the top Yukawa.}
\label{fig:basic_idea}
\end{figure}

Our calculation of $\Lambda_B$ proceeds as follows: the only renormalizable coupling between new fermions and the SM that can influence SM Higgs couplings is a Yukawa interaction. Higgs couplings to $WW$ and $ZZ$ exist at tree-level in the SM, but the leading contribution from new fermions is at one (weak) loop. Therefore we need a large Yukawa coupling to generate a visible deviation, larger than any coupling in the SM. These large Yukawa couplings $y$ dominate the RGEs of the model and can lead to two forms of instabilities. They can change the sign of the Higgs quartic coupling 
\be
\frac{d\lambda}{d\log \mu} \sim - \frac{y^4}{16\pi^2}\, , \label{eq:y4}
\ee
and/or hit a Landau pole\footnote{Here and in the following we use ``Landau pole" very loosely. As far as we know, in our theories there is no lattice proof that $y$ hits an actual Landau pole. We instead identify $\Lambda_B$ with the scale where $y$ becomes non-perturbative. We comment on what we mean by non-perturbative and on the validity of our perturbative calculations in Section~\ref{sec:results}.}
\be
\frac{dy}{d\log \mu} \sim  \frac{y^3}{16\pi^2}\, . \label{eq:y3}
\ee
Therefore in these theories we can associate a cutoff to any given Higgs coupling deviation. The precise definition of the cutoff $\Lambda_B$ can be found in Section~\ref{sec:LB}. We often refer to this scale, and sometimes plot it, in previous Sections. The reader who prefers to see a definition first can skip ahead to Section~\ref{sec:LB} that is self-contained.

\begin{figure}[!t]
\begin{center}
\includegraphics[width=0.6\textwidth]{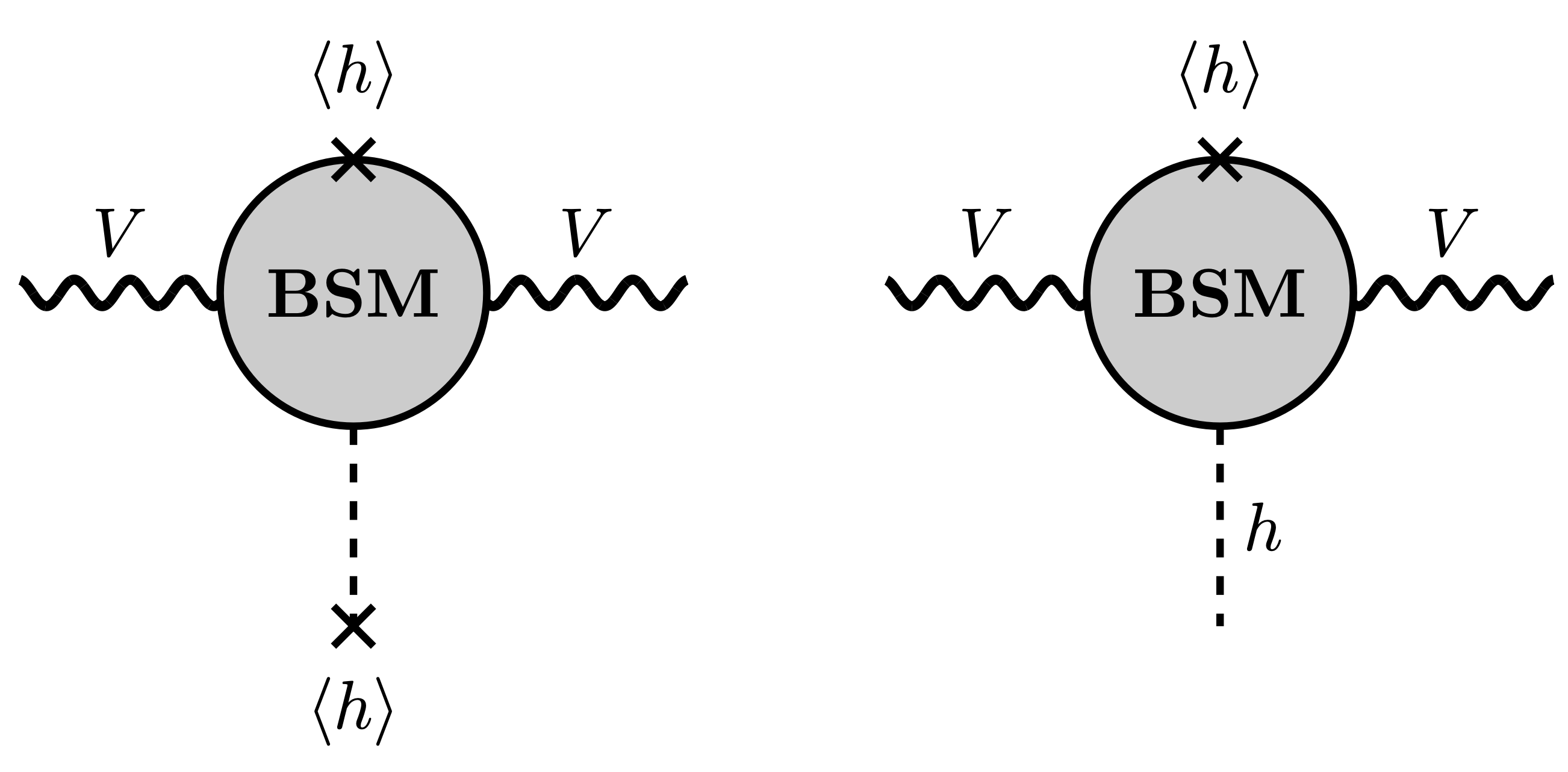}
\end{center}
\caption{Schematic view of diagrams that generate a $hVV$ coupling deviation (right) and at the same time affect electroweak precision measurements (left).}\label{fig:ST}
\end{figure}

To make contact with most studies of Higgs couplings at present and future colliders~\cite{Cepeda:2019klc, ILCInternationalDevelopmentTeam:2022izu, Robson:2018zje, Bernardi:2022hny, CEPCPhysicsStudyGroup:2022uwl, Forslund:2022xjq, deBlas:2022aow, deBlas:2022ofj}, we define an effective coupling from the Higgs partial width\footnote{When comparing decay widths we choose the following decays $Z^*\to e^+e^-$ and $W^*\to e\bar \nu_e$, neglecting the masses of the fermions whenever they give relative corrections smaller than $\delta \mu$.}
\be
g_{hVV}^2 \equiv (g_{hVV}^{\rm SM})^2\frac{\Gamma(h\to VV^{(*)})}{\Gamma^{\rm SM}(h\to VV^{(*)})}\, ,
\ee
and a relative coupling deviation
\be
\delta \mu_{hVV} \equiv \frac{g_{hVV}-g_{hVV}^{\rm SM}}{g_{hVV}^{\rm SM}}=\sqrt{\frac{\Gamma(h\to VV^{(*)})}{\Gamma^{\rm SM}(h\to VV^{(*)})}} -1\, . \label{eq:relg}
\ee
Since $|\delta \mu_{hVV}| \sim y^n$ with $n>0$, the bigger $|\delta \mu_{hVV}|$, the faster the running of $y$ and $\lambda$, giving a smaller scale at which our theory description must be modified. In Fig.~\ref{fig:basic_idea} we summarize this discussion showing the size of the coupling deviation and of $\Lambda_B$ for a given Yukawa coupling and a representative model discussed in Section~\ref{sec:4f}. 

Fig.~\ref{fig:basic_idea} is just an illustration of the more detailed results in Section~\ref{sec:results}, but it is sufficient to appreciate the main point of the paper. To measure any coupling deviation, even at the most precise $e^+e^-$ colliders that are currently being proposed, we need large Yukawa couplings and relatively light fermions $M_1 \lesssim 500$~GeV (left panel) corresponding to an instability scale a factor of 10 to a few above $M_1$ (right panel). Fig.~\ref{fig:basic_idea} contains also a quantitative illustration of the RGE domination of the new Yukawas: changing the top Yukawa by twice its error does not appreciably affect the scale of instability. To obtain all our results we compute the model RGEs using {\tt SARAH}~\cite{Staub:2008uz,Staub:2015kfa,Staub:2013tta,Staub:2012pb,Staub:2010jh,Staub:2009bi}. 

Even if we consider models with large Yukawas and light fermions, it is not easy to generate a large coupling deviation consistent with electroweak precision tests (EWPTs). The loop diagrams responsible for the deviation correct the $WW$ and $ZZ$ two-point functions, as shown schematically in Fig.~\ref{fig:ST}. The parameters that most conveniently describe these corrections are known at the permille level from LEP~\cite{ALEPH:2005ab} and in generic models they already set a bound on $hWW$ and $hZZ$ comparable to the sensitivity of future colliders, as shown in Fig.~\ref{fig:basic_idea}. Therefore, the impact of EWPTs on the maximally allowed $hWW$ and $hZZ$ deviations might be enough in itself to conclude that any observable deviation is generated by new bosons. For instance, we could have a gauge singlet scalar mixing with the Higgs after electroweak symmetry breaking that affects Higgs couplings at a measurable level, without compromising other electroweak precision observables.

However, even if we always show the EWPTs constraints on our models, we prefer to compute explicitly a conservative upper bound on $\Lambda_B$ from the more model-independent RGE argument outlined above. The bounds from EWPTs can always be partially evaded with some amount of model-building\footnote{However this comes at a price, it requires adding new Yukawas of the same size as those responsible for the Higgs coupling deviation~\cite{Arkani-Hamed:2012dcq}, lowering $\Lambda_B$ even further.} and the value of $\Lambda_B$ from RGEs is low enough (for any observable deviation) to be interesting in itself and single out $hWW$ and $hZZ$ couplings as prime direct probes of new bosons and prime indirect probes of naturalness.

Since we are proving our statement by contradiction we want to find the theories with the largest possible $\Lambda_B$ for any fixed $\delta \mu$. In the next three Sections we show that only a handful of new fermions representations need to be considered to achieve this goal. 

\section{Fermionic Low Energy Theories}\label{sec:reps}
We would like to extend the SM including new fermions that satisfy the following requirements: 1) They induce observable modifications in $hZZ$ and $hWW$ couplings, 2) They are consistent with all existing experimental constraints, 3) They are part of a consistent low energy theory (i.e. they do not introduce gauge anomalies). A comprehensive survey of representations that satisfy more general requirements was conducted in~\cite{Bizot:2015zaa}. The only difference with respect to our needs is that in~\cite{Bizot:2015zaa} deviations in any Higgs coupling to the SM were considered interesting. The new fermions relevant to our purposes are, therefore, a subset of those discussed in~\cite{Bizot:2015zaa}. We can exclude the most minimal extensions considered in~\cite{Bizot:2015zaa}, comprising only one or two new fermions, because they induce unobservably small deviations in $hZZ$ and $hWW$, as discussed in Appendix~\ref{app:reps}. The case of three chiral fermions was phenomenologically viable when~\cite{Bizot:2015zaa} was published, but it is currently excluded by measurements of the $hgg$ and $h\gamma\gamma$ couplings. We are left with SM extensions with three or more new fermions. 

\subsection{Three New Fermions}\label{sec:3f}
The minimal extension of the SM that is relevant to us contains three new fermions, one vector-like pair and a Majorana fermion. As shown in~\cite{Bizot:2015zaa} we can have two distinct possibilities. If we adopt the notation $(a, b)_Y$, with $a$ the dimension of the $SU(3)_c$ representation, $b$ that of the $SU(2)_L$ representation and $Y$ the hypercharge, the first possibility is
\be
L=(r,2n+1\pm1)_{-1/2}\, , \quad N=(r, 2n+1)_0\, , \quad r=\bar r\, ,
\ee
plus $L^c$, the vector-like partner of $L$. Note that we need the color representation to be self-conjugate ($r=\bar r$), as explained in~\cite{Bizot:2015zaa}. We can then write the interaction Lagrangian
\be
\mathcal{L}_3 \supset -y L H N -y^c L^c H^\dagger N - M_L L L^c - \frac{M_N}{2}N^2 +{\rm h.c.}\, . \label{eq:three}
\ee
Alternatively, we can consider a $SU(2)_L$ representation of even dimension for\footnote{Note that if $2n = 2+4 k$, $k\in \mathbb{N}$, we need $r$ to be even to avoid a $SU(2)_L$ anomaly~\cite{Bizot:2015zaa}.} $N$
\be
L=(r,2n\pm1)_{-1/2}\, , \quad N=(r, 2n)_0\, , \quad r=\bar r\, ,
\ee
but in this case we do not have a Majorana mass term and the Lagrangian reads
\be
\mathcal{L}_3^\prime \supset -y L H N -y^c L^c H^\dagger N - M_L L L^c +{\rm h.c.}\, . \label{eq:threeprime}
\ee
These are the minimal extensions of the SM consistent with our requirements. We have listed them for completeness, but in the following we mostly consider theories with two pairs of vector-like fermions. The reason is that we are not looking for a minimal extension of the SM, but for the extension that has the largest possible cutoff $\Lambda_B$ for a given Higgs coupling deviation. Including another fermion allows us to add another layer of generality to the low energy theory: the hypercharges are not fixed by the three conditions stated at the beginning of this Section and we have the extra freedom to dial $Y$ to increase the $hZZ$ coupling deviation. Going beyond four fermions does not introduce any other qualitative difference, but we discuss how our results scale if we include multiple copies of the four fermions presented in the next Section.

\subsection{The Main Character: A Four Fermion Extension of the SM}\label{sec:4f}
The simplest extension of the SM with four new fermions that can give a potentially large $hWW$ and/or $hZZ$ coupling deviation, without being obviously excluded by current measurements and without introducing gauge anomalies is given by
\be
L=(r, n)_{Y}\, , \quad N^c=(\overline r, n-1)_{-Y-1/2} \label{eq:fG}
\ee
and their vector-like partners
\be
L^c=(\overline r, n)_{-Y}\, , \quad N=(r, n-1)_{Y+1/2}\, . \label{eq:fG2}
\ee
Their renormalizable Lagrangian (leaving implied kinetic terms and gauge interactions) is\footnote{The case $r=\bar r, Y=0$ deserves special attention. If we keep the same particle content as above with $N$ distinct from $N^c$ the theory in Eq.~\eqref{eq:L4} is consistent and preserves a $U(1)$ ``lepton number" with charges $Q_{L, N}=1$, $Q_{L^c, N^c}=-1$. However we can break the $U(1)$ symmetry and include additional Yukawa couplings
\be
\mathcal{L}_2 \supset - M_L L L^c- M_N N N^c - y L H N^c - y^c L^c H^\dagger N  - y^\prime L H N- y^{c\prime} L^c H^\dagger N^c +{\rm h.c.}\, .
\ee
For $N, N^c$ in $SU(2)_L$ representations of odd dimension we can add also Majorana masses. We do not consider these cases in detail because they do not change our conclusions in Section~\ref{sec:results}.}
\be
\mathcal{L}_4= - y L H N^c - y^c L^c H^\dagger N - M_L L L^c - M_N N N^c +{\rm h.c.}\, . \label{eq:L4}
\ee

In the following we often discuss a subset of these models for which it is useful to have a name. We call {\it doublet+singlet} the model with $r=1, n=2, Y=-1/2$ and {\it doublet+triplet} the model with $r=1, n=3, Y=0$. In the doublet+singlet model we call $\ell, \ell^c$ the two doublets that have the same quantum numbers as SM lepton doublets, and $n,n^c$ the two singlets that have the same quantum numbers as right-handed neutrinos. The particles in the doublet+triplet model have the same quantum numbers as a pair of vector-like winos plus two Higgsinos.

When we think about the limit of heavy new fermions, we have to take $M_{L,N}$ large. Taking the Yukawa couplings large, while leaving $M_{L,N}$ fixed is already excluded by $hgg$ and $h\gamma\gamma$ measurements, as mentioned at the beginning of Section~\ref{sec:reps} when discussing chiral fermions.  

Note also that the four parameters $y, y^c, M_L, M_N$ cannot all be taken to be real and positive without loss of generality. There is one physical phase in this model. However, we want to avoid stringent constraints on CP violation that increase $M_{L, N}$ and reduce our cutoff $\Lambda_B$. Therefore, we always take the only physical phase to be 0 or $\pi$, allowing at most for a relative sign between the two Yukawa couplings.

We are going to study these theories in their perturbative limit. The Yukawas that we consider can be larger than one, but at the mass scale of the new fermions they are perturbative. This gives the most conservative upper bound on the scale of new physics that we want to compute. A simple argument is enough to see this: imagine a composite Higgs theory that in the UV is described by new fermions that condense at low energy. We call $f$ the scale of the $\sigma$-model that describes the pions of the confining sector, including the Higgs. A low energy observer measures Higgs coupling deviations at $\mathcal{O}(v^2/f^2)$ from a series of irrelevant operators~\cite{Giudice:2007fh}. The scale where new particles (and new bosons) appear is $m_* = g_* f$, for a strongly coupled theory $m_* \sim 4 \pi f$. This should be compared with our weakly coupled theories where the Higgs coupling deviations arise at one-loop at $\mathcal{O}(y^4 v^2/16 \pi^2 M^2)$ with $M$ a vector-like mass and $y$ a coupling. We can compare with an effective $f_{\rm eff} \simeq 4\pi M/y^2$ and conclude that these theories can be extrapolated to much higher energies than $4\pi f_{\rm eff}$ and this is why we focus on them to get a conservative upper bound on the scale of new physics. When $y \lesssim \mathcal{O}(1)$ we will indeed find that our perturbative theories have much larger cutoffs than $4\pi f_{\rm eff}$. In some special cases, with large coupling deviations and large $y$, the running is sufficiently rapid to give cutoffs lower than $4\pi f_{\rm eff}$. These latter cases must be taken with a grain of salt. First of all, our two-loop approximation is not adequate to capture the precise value of the cutoff in these limiting cases. Secondly, one could get a more conservative upper bound in a strongly coupled theory, so in practice any cutoff below $4\pi f_{\rm eff}$ is not the largest one that is possible to achieve.

To conclude this Section, note that in most of these models (the only exception being the doublet+singlet model with a neutral singlet) we generate also a deviation in the $h\gamma\gamma$ coupling. This is a much bigger relative effect than the deviation in $hWW$ and $hZZ$ because we are comparing a one-loop new physics effect with a one-loop coupling in the SM. For completeness we compute the scale of new bosons for $h\gamma\gamma$ in Section~\ref{sec:final_results}, but this case was already discussed in~\cite{Arkani-Hamed:2012dcq, Blum:2015rpa, Joglekar:2012vc, Kearney:2012zi, Batell:2012ca}.

\begin{figure}[!t]
\begin{center}
\includegraphics[width=0.95\textwidth]{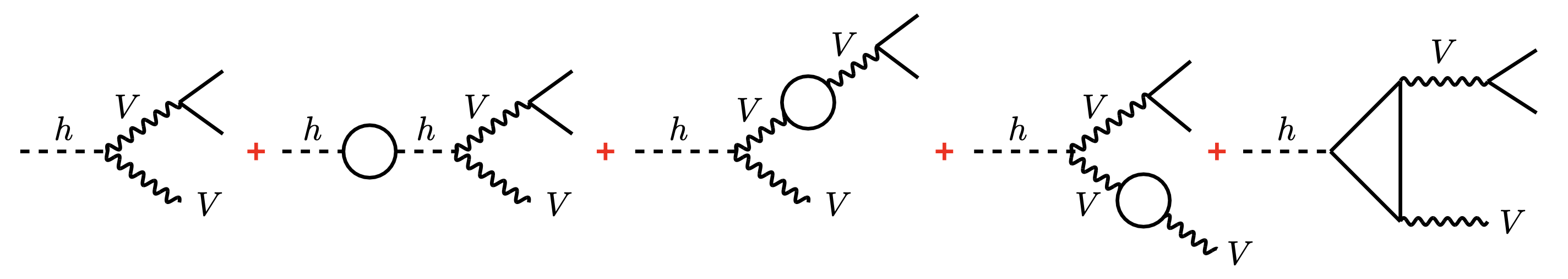}
\end{center}
\caption{Schematic view of the leading diagrams contributing to the $h\to V V^*\to V \psi_a \psi_a$ decay width.}\label{fig:diagrams}
\end{figure}

\section{Calculation of the Coupling Deviations}\label{sec:coupling}
The coupling deviation that we want to compute receives contributions from the diagrams that are listed schematically in Fig.~\ref{fig:diagrams}. We have included in the Figure only loop corrections from the new fermions, because SM loop corrections cancel in the ratio of Eq.~\eqref{eq:relg}. 

To implement the calculation consistently, in addition to the diagrams in Fig.~\ref{fig:diagrams}, we have to consider additional processes, involving the new fermions, that renormalize the electroweak sector of the SM at one loop. We start by considering the diagrams already contained in Fig.~\ref{fig:diagrams}. The bubble diagrams on the external legs have two effects: they modify the tree-level relation between the pole mass of SM bosons and the corresponding $\MS$ Lagrangian parameters, and they give a field strength renormalization that we need to include when computing $S$-matrix elements from Feynman diagrams. In the $ZZ$ case they also induce a mixing with the photon that we discuss after the $WW$ calculation. 

Following the notation in~\cite{Schwartz:2014sze}, we call $\hat m_Z, \hat e, ...$ the measured parameters and $m_Z, e, ...$ the $\MS$ Lagrangian parameters. We can compute the $\MS$ $V$-boson mass ($V=W, Z$) from the measured mass $\hat m_V$ as
\be
m_V^2 &=& \hat m_V^2\left(1 + \frac{\text{Re}[\Pi_{VV}(\hat m_V^2)]}{\hat m_V^2}\right)\, , \label{eq:physMZ}
\ee
where
\be
\mathcal{M}^{\mu\nu}(V(p)\to V(p)) &=& g^{\mu\nu}\Pi_{VV}(p^2)+...\, .
\ee
The field strength renormalization is given by
\be
Z_V = \frac{1}{1-\Pi^\prime_{VV}(\hat m_V^2)}\, ,
\ee
where the prime denotes derivation by $p^2$. Note that the sign in front of $\Pi^\prime$ is opposite for the Higgs boson, if we use the standard convention for scalar propagators~\cite{Schwartz:2014sze, Peskin:1995ev}. These equations are useful to set the notation, but we do not give further details because this is a standard calculation for EWPTs, that can be found, for instance, in Chapter 31 of~\cite{Schwartz:2014sze}.

The bubble diagram on the internal $V$-boson leg in Fig.~\ref{fig:diagrams} is slightly more tricky. Summing all 1PI diagrams we obtain the propagator
\be
\frac{-i g^{\mu\nu}}{p^2-m_V^2-\Pi_{VV}(p^2)}+\mathcal{O}(p^\mu p^\nu)\, .
\ee
We can omit terms proportional to $p^\mu p^\nu$ since the internal gauge boson is contracted with a light SM current.  
The function $\Pi_{VV}(p^2)$ contains $1/\epsilon$ and $\mu$-dependent terms that might be modified non-trivially by the integral over $p^2$. To make all the cancellations manifest we rearrange the terms in the propagator as
\be
\frac{-i g^{\mu\nu}}{p^2-m_V^2-\Pi_{VV}(p^2)}&=&\frac{-i Z_V g^{\mu\nu}}{p^2-\hat m_V^2-\Xi_{VV}(p^2)}\, , \label{eq:propagator}
\ee
where we have defined the new function $\Xi_{VV}(p^2)$ as
\be
\Xi_{VV}(p^2)&\equiv & \Pi_{VV}(p^2)-\text{Re}[\Pi_{VV}(\hat m_V^2)]-(p^2-\hat m_V^2)\Pi_{VV}^\prime(\hat m_V^2)\, .
\ee
It is easy to show that $\Xi(p^2)$ is UV-finite and $\mu$-independent. The divergence and $\mu$-dependent pieces in $Z_V$ are independent of $p^2$ and can be factored out of the integral.

The triangle diagram\footnote{For $h\to W W^*$ there is a single diagram, in the case of $h\to ZZ^*$ the Figure represents schematically three independent diagrams. Two with two $Z$-bosons that can be obtained by exchanging the momenta and Lorentz indexes of the two $Z$'s. The last one with an on-shell $Z$ and an off-shell photon.} in Fig.~\ref{fig:diagrams} gives a complicated function of external momenta, let us call this function $T$, then the amplitude for the decay $h \to W W^*$ reads
\be
&&\mathcal{M}(h\to W(p_1) W^*(q_{23})\to W(p_1) \psi_a(p_2) \psi_b(p_3))= \nn \\
&=&i\frac{g}{\sqrt{2}}\varepsilon_{\nu}^*(p_1)\left[\Bar{u}(p_3)\gamma_{\mu}P_Lv(p_2)\right]\frac{m_W(g+T(p_1, q_{23}))}{q_{23}^2-\hat{m}_W^2+\Xi(q_{23}^2)}\sqrt{Z_W^{3}Z_h}\, , \label{eq:AWW}
\ee
where we have included also the tree-level contribution from the first diagram in Fig.~\ref{fig:diagrams}. To get a physical answer we express $g$ and $m_W$ in terms of measured quantities using the one-loop renormalization conditions~\cite{Schwartz:2014sze,Bardin:1999ak}
\be
e^2&=&\hat e^2(\hat m_Z)\left[1-\frac{\Pi_{\gamma\gamma}(\hat m_Z^2)}{\hat m_Z^2}\right]\, , \nn \\
m_Z^2 &=& \hat m_Z^2\left(1-\frac{\text{Re}[\Pi_{ZZ}(\hat m_Z^2)]}{\hat m_Z^2}\right)\, , \nn \\
s^2_W &=& \hat s^2_W\left[1+\frac{\hat c^2_W}{\hat c^2_W-\hat s^2_W}\left(\frac{\text{Re}[\Pi_{ZZ}(\hat m_Z^2)]}{\hat m_Z^2}-\frac{\Pi_{\gamma\gamma}(\hat m_Z^2)}{\hat m_Z^2}-\frac{\text{Re}[\Pi_{WW}(0)]}{\hat m_W^2}\right)\right]\, . \label{eq:ren}
\ee
These conditions incorporate also the relevant effects of the new fermions that are not illustrated by Fig.~\ref{fig:diagrams}.
The measured parameters that we use are~\cite{ParticleDataGroup:2022pth, Schwartz:2014sze}
\be
\hat \alpha(0)&=&(137.035 999 074 \pm 0.000 000 044)^{-1} \rightarrow \hat e^2(\hat m_Z) = 4\pi \hat \alpha(m_Z) = 4\pi (127.944 \pm 0.014)^{-1}\, , \nn \\
\hat m_Z&=&(91.1876\pm0.0021)\; {\rm GeV}\, , \nn \\
\hat G_F&=&(1.1663787\pm0.0000006)\times 10^{-5}\;{\rm GeV}^{-2}\, . \label{eq:mes}
\ee
Here and in the following we call $c_W$ and $s_W$ the cosine and sine of Weinberg's angle and we neglect the SM running between $m_Z$ and $m_h$. Note that $g$ and $m_W$ are not the same in the SM and in our vector-like fermion theory because the two-point functions $\Pi_{XX}$ differ in the two cases. Combining the three previous equations we obtain
\be
\mathcal{M}(h\to W(p_1) W^*(q_{23})\to W(p_1) \psi_a(p_2) \psi_b(p_3))= \mathcal{\hat M}\left(1+\frac{\Pi_g}{2}\right)\, , \label{eq:AWWf}
\ee
where we found convenient to define
\be
\Pi_g&\equiv&\frac{3 \left(\Pi'_{WW}(\hat{m}_W^2)\right)}{2}-\frac{\left(\Pi'_{hh}(\hat{m}_h^2)\right)}{2}-\frac{\Pi_{\gamma\gamma}(\hat{m}_Z^2)}{\hat{m}_Z^2}-\frac{{\rm Re}\left[\Pi_{ZZ}(\hat{m}_Z^2)\right]}{2\hat{m}_Z^2}-\frac{\Pi_R}{\hat c_W^2-\hat s_W^2}\left(\frac{\hat s_W^2}{2}+\hat c_W^2 \right)\, , \nn \\ \label{eq:Pig}
\ee
with
\begin{equation}
\Pi_R=-\frac{\Pi_{\gamma\gamma}(\hat{m}_Z^2)}{\hat{m}_Z^2}+\frac{\Pi_{ZZ}(\hat{m}_Z^2)}{\hat{m}_Z^2}-\frac{\Pi_{WW}(0)}{\hat{m}_W^2},
\end{equation}
and $\mathcal{\hat M}$ is the matrix element in the second line of Eq.~\eqref{eq:AWW} where all $\MS$ Lagrangian parameters are replaced by measured quantities, $Z_W=Z_h=1$ and we drop terms $\sim T\times \Pi_g$ that are of second order in the loop expansion.

Using Eq.s~\eqref{eq:AWWf} and~\eqref{eq:Pig} we can compute $\delta \mu_{hWW}$ by summing over all SM fermions in the final states (except the top quark). Taking $y^c=y$, $M_L=M_N=M$ and $M\gg yv$ we have, for the doublet+singlet model defined in the previous Section ($Y=-1/2$), 
\be
\delta \mu_{hWW}^{\rm DS}&=&\frac{y^2 v^2}{80\pi^2 M^2}\left[y^2\frac{\left(1+c_W^2\right)}{c_W^2-s_W^2}-\frac{m_h^2}{v^2}-\frac{g^2}{4}\frac{s^2_W}{c^2_W-s^2_W}\right]+\frac{1}{2}\sum_{i=1}^3 c_i^W\frac{m_h^3}{m_W M^2}\frac{P_W^{(i)}(x)}{R_W(x)}\, , \label{eq:WWDS} \\ 
c_1^W&=&\frac{13 y^2}{240 \pi ^2},\quad c_2^W=-c_3^W=-\frac{7 y^2}{240 \pi ^2}\, , \quad x\equiv \frac{m_W^2}{m_h^2}\, , \nn
\ee
where all parameters are the measured ones and we neglected terms without new Yukawas. We left the ``hats'' implied to improve readability. We have already accounted for one-loop renormalization conditions and one can express the above parameters in terms of the measured quantities in Eq.~\eqref{eq:mes} using the SM tree-level relations (i.e. $e=g s_W$, $m_W=c_W m_Z$). 
The functions $P_W^{(i)}(x)$ can be found in Appendix~\ref{app:fs}, they correspond to the three higher-dimensional operators: $O_1=h \partial_\nu W_\mu \partial^\nu W^\mu$, $O_2=h \partial_\nu W_\mu \partial^\mu W^\nu$, and $O_3=h (\square W_\mu)W^\mu$. Note that these operators exist only after EW symmetry breaking and are suppressed by an extra $m_W/M$ compared to their naive scaling dimension.
All other operators are either redundant or proportional to $\partial_\mu W^\mu$ which is zero for the on-shell $W$ and when contracted with a massless SM current. Corrections proportional to the SM fermions' masses are smaller than the smallest values of $\delta \mu$ that we consider in this work.
  
\begin{figure}[!t]
\begin{center}
\includegraphics[width=0.45\textwidth]{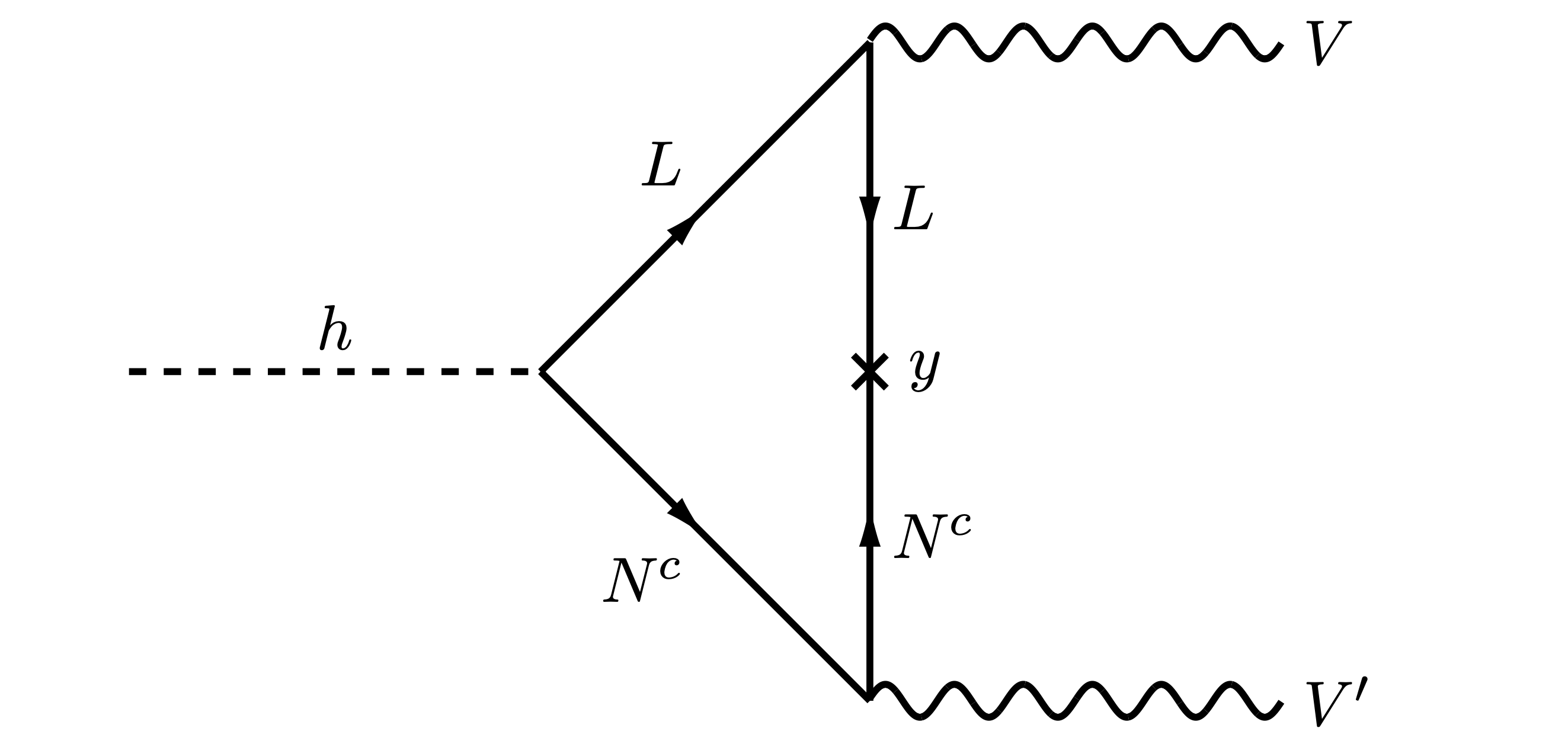}
\includegraphics[width=0.45\textwidth]{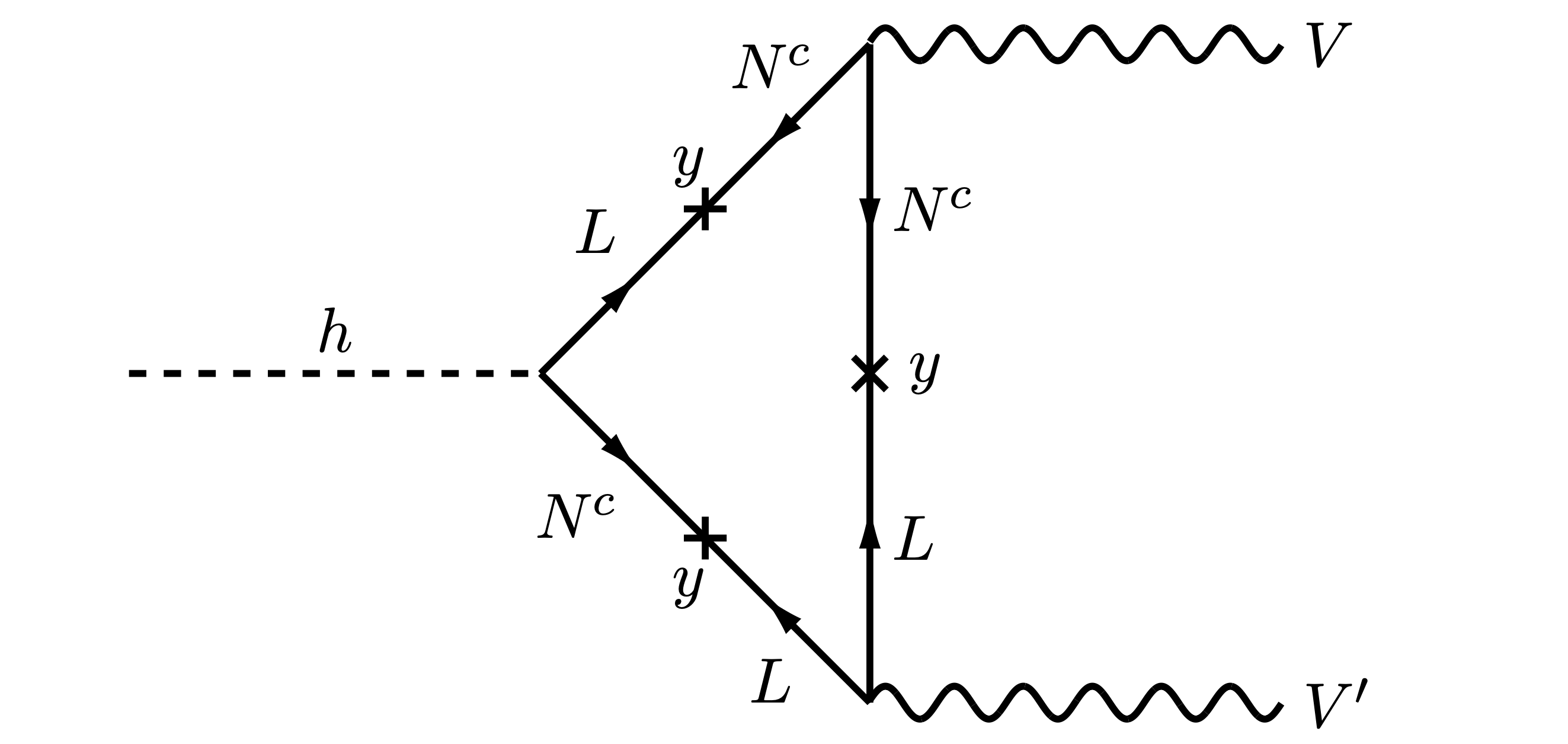}
\end{center}
\caption{Left panel: Naive leading diagram for the $hVV$ width from vector-like leptons. Right panel: Leading diagram for the $hVV$ width from vector-like leptons in the limit of large Yukawa couplings ($y \gtrsim g_{\rm SM}$).}\label{fig:twoY}
\end{figure}

The parametric form of the result can be understood from the two diagrams in Fig.~\ref{fig:twoY}. From the left panel we see that at least two insertions of $y$ are needed to close the loop. Additionally, we have two powers of a SM gauge coupling from the two gauge boson vertices. We might naively conclude that the coupling deviation from the triangle diagrams scales as $\delta g_{hWW}\sim g^2 y^2 v^3/16\pi^2 M^2$. If we restore units to $\hbar$ we see immediately that this is wrong because $\delta g_{hWW}$ has dimensions of $\hbar^{-1} \times v$. If we take into account the $\hbar$ from the loop, the coupling deviation from these diagrams must scale as $(g^2 y^2 v^3/16\pi^2 M^2)\times g_*^2$ where $g_*$ has the dimensions of a gauge coupling ($\hbar^{-1/2}$). The largest coupling in the theory is $y$, so we need to evaluate the diagram in the right panel of Fig.~\ref{fig:twoY} which gives parametrically the leading term in Eq.~\eqref{eq:WWDS}\footnote{In practice we always work with mass eigenstates and resum all $y$ insertions. We show Fig.~\ref{fig:twoY} just to illustrate the parametrics.}.

For the general $SU(2)_L$ representations in Eq.~\eqref{eq:fG} and~\eqref{eq:fG2} and $Y=-1/2$, for $y^c=(-1)^{n-1}y$ (the choice that makes the mass matrices symmetric) we have
\be
\delta \mu_{hWW}&=& \frac{n y^2 v^2 }{240 \pi^2 M^2}\left[y^2\frac{(2n-1)(c_W^2+\frac{s_W^2}{2})}{(n-1)(c^2_W-s^2_W)}-\frac{3 m_h^2}{2 v^2}-g^2\frac{(23-10 n)s_W^2}{8(c^2_W-s^2_W)}\right] \nn \\
&+& \frac{1}{2}\sum_{i=1}^3 c_i^W\frac{m_h^3}{m_W M^2}\frac{P_W^{(i)}(x)}{R_W(x)}\, , \quad x\equiv \frac{m_W^2}{m_h^2}\, , \nn \\
c_1^W&=&n\frac{12 y^2}{640 \pi^2}+\left(\frac{1}{2}+\frac{n}{5}-(n-1)^2-\frac{(n-1)^3}{2}\right)\frac{y^2}{36\pi^2},\nn \\ 
c_2^W&=& -n\frac{y^2}{480 \pi^2}+\left(-\frac{1}{2}-\frac{n}{20}+(n-1)^2+\frac{(n-1)^3}{2}\right)\frac{y^2}{36\pi^2},\nn \\ 
c_3^W&=&n\frac{17y^2}{960\pi^2}+n\frac{y^2}{320 \pi^2}\, . \label{eq:generalWW} 
\ee
As before we took $M\gg yv$, dropped terms that do not contain powers of $y$ and left the ``hats'' implied, but all parameters are the measured ones. The growth of $\delta \mu_{hWW}$ as $n^3$ at large $n$ is expected on general grounds from the Higgs low energy theorems~\cite{Ellis:1975ap, Shifman:1979eb, Kniehl:1995tn, Gillioz:2012se} that relate the coupling deviation to the $W$ two-point function. The two-point function scales as the Dynkin index of the representation that for $SU(2)$ at large $n$ goes as $T(n)\sim n^3$.

The case of $hZZ$ is similar, but not identical. There are three additional diagrams: 1) the bubble diagram on the internal $Z$-leg can mix the $Z$ with a photon, 2) the triangle diagram in Fig.~\ref{fig:diagrams} can have a photon in the internal gauge boson line, 3) there are two triangle diagrams with two $Z$'s that differ by the exchange of momenta and Lorentz indexes of the two gauge bosons. Including these diagrams, we have 
\be
\mathcal{M}(h\to Z^* Z \to Z \bar \psi \psi)=\mathcal{\hat M}_Z\left(1+\frac{\Pi_g^Z}{2}\right)\, ,
\ee
where
\be
\Pi_g^Z&\equiv& \frac{3 \left(\Pi'_{ZZ}(\hat{m}_Z^2)\right)}{2}-\frac{\left(\Pi'_{hh}(\hat{m}_h^2)\right)}{2}-\frac{\Pi_{\gamma\gamma}(\hat{m}_Z^2)}{\hat{m}_Z^2}-\frac{{\rm Re}\left[\Pi_{ZZ}(\hat{m}_Z^2)\right]}{2\hat{m}_Z^2} \nn \\
 &+&\frac{a_2 \hat s_W^2 +2 a_4 \hat s_W^4}{2(a_0+a_2 \hat s_W^2 +a_4 \hat s_W^4)}\left[\frac{\hat c_W^2}{\hat c_W^2-\hat s_W^2}\Pi_R-\frac{\hat c_W}{\hat s_W}\left(\Pi'_{Z\gamma}(0)\right)\right]\, .
\ee
$a_{1,2,3}$ are $\mathcal{O}(1)$ numbers that depend on the SM fermion in the final state. We can obtain them from the SM $Z$-boson current,
\be
(a_0+a_2 \hat s_W^2+a_4 \hat s_W^4) \equiv (T_3- \hat s_W^2 Q)^2\, .
\ee
The explicit result for the doublet+singlet model for $Z^*\to e^+ e^-$, taking $y^c=y$, $M_L=M_N=M$ and $M\gg v$, is 
\be
\delta \mu_{hZZ}^{\rm DS}&=&\frac{y^4v^2}{40\pi^2M^2}\frac{1-2c_W^4+4(-1+2c_W^4)s_W^2+8(1-2c_W^2)s_W^4}{c_{2W}(1-4s_W^2+8s_W^4)} \nn \\
&+& \frac{y^2 v^2}{480\pi^2M^2}\left[-6\frac{m_h^2}{v^2}+g^2\frac{s_{W}^2(4s_W^2-1)}{c_{2W}(1-4s_W^2+8s_W^4)} \right]+\frac{1}{2}\sum_{i=1}^3 c_i^Z\frac{m_h^2}{M^2}\frac{P_Z^{(i)}(z)}{R_Z(z)}\, , \nn \\
c_1^Z&=&\frac{13 y^2}{240 \pi^2}\, , \quad c_2^Z=-c_3^Z=-\frac{7 y^2}{240 \pi^2}\, , \quad z\equiv \frac{m_Z^2}{m_h^2}. 
\ee
All parameters are the measured ones and we neglected terms without new Yukawas. We left the ``hats'' implied to improve readability and neglected terms that do not contain powers of the new Yukawas.  The functions $P_Z^{(i)}(x)$ can be found in Appendix~\ref{app:fs}. The result for general $SU(2)_L$ representations and $Y=-1/2$ is 
{\allowdisplaybreaks
\be
\delta \mu_{hZZ}&=&\frac{n(2n-1)}{n-1}\frac{y^4v^2}{240\pi^2M^2}\frac{(1-2c^4_W)(1-4s^2_W)+8s^4_W(1-2c^2_W)}{c_{2W}(1-4s^2_W+8s^4_W)}\nn \\
&+&\frac{ny^2v^2}{960M^2\pi^2}\left[-6\frac{m_h^2}{v^2}+\frac{g^2}{3}(23-10n)\frac{s^2_W(-1+4s^2_W)}{c_{2W}(1-4s^2_W+8s^4_W)} \right]+\frac{1}{2}\sum_{i=1}^{4}c_i^Z\frac{m_h^2}{M^2}\frac{P_Z^{(i)}(z)}{R_Z(z)}\, , \nn \\
c_1^Z&=&n\frac{y^2}{1440\pi^2 c_W^2}\left(-1+60s_W^2-20s_W^4+20(n-1)c_W^2+20(n-1)^2c_W^4\right), \nn \\
c_2^Z&=&-n\frac{y^2}{1440\pi^2 c_W^2}\left(-19+60s_W^2-20s_W^4+20(n-1)c_W^2+20(n-1)^2c_W^4\right), \nn \\
c_3^Z&=&n\frac{7y^2}{480\pi^2 c_W^2}, \nn \\
c_4^Z &=& n(n-2)\frac{y^2}{144\pi^2}\frac{t_W^2\left(2s_W^2-\frac{1}{2}\right)}{2s_W^4-s_W^2+\frac{1}{4}}\left(3-2s_W^2+2(n-1)c_W^2 \right)\, .
\ee
}
The parametric form of the results for $hZZ$ has the same explanation as that of $hWW$, given above. The expansions for large $M$ are useful for the arguments in Section~\ref{sec:repsII}, but we often have to compute $\delta \mu$ also when $M \simeq y v$, for example in the doublet+singlet model introduced in the previous Section.
When the vector-like masses are comparable to $y v$, we compute numerically the loop function $T$ and $\Pi_g$ in Eq.~\eqref{eq:AWW} (and the equivalent functions for $hZZ$), using {\tt Package-X}~\cite{Patel:2015tea, Patel:2016fam} and then we integrate them numerically over phase space using {\tt Mathematica}. 

\section{Fermion Representations}\label{sec:repsII}
Before making quantitative predictions on the scale of new bosons, we want to understand what color, $SU(2)_L$ and hypercharge representations we need to consider to obtain the most conservative upper bound on this scale. We can considerably reduce the number of representations that we need to study, compared to Eq.s~\eqref{eq:fG} and \eqref{eq:fG2}, by following a few scaling arguments presented in the next Sections. We also discuss the impact of adding multiple copies of the four fermions introduced in Section~\ref{sec:reps}.

\subsection{Color and Fermion Multiplicity}
We can begin by showing that we do not need to consider colored particles. In Section~\ref{sec:coupling} we have discussed the parametric form of the expected coupling deviation. It is straightforward to include the scaling with the $SU(3)_c$ representation dimension $r$, 
\be
\delta g_{hVV} \sim r\frac{\alpha_V}{4\pi }\frac{y^4 v^2}{M^2} \lesssim r\frac{\alpha_V}{4\pi }\frac{y^4 v^2}{M^2_{\rm exp}}\, , \label{eq:dgnaif}
\ee
where $M_{\rm exp}$ is the experimental bound on the mass of the new states and $\alpha_V$ is an electroweak gauge coupling that depends on the choice of final state. We can use this scaling because LHC bounds require $M_{\rm exp} \gg yv$ for colored states. We have assumed small $SU(2)_L$ representations and hypercharge. For simplicity we have also taken $y^c=y$ and $M_L=M_N=M$, more general choices do not affect the conclusions of this Section. Eq.~\eqref{eq:dgnaif} is valid both for the four fermions extensions in Section~\ref{sec:4f} and the three fermions in Section~\ref{sec:3f}.

As discussed in Section~\ref{sec:basic}, above a certain cutoff we do not have perturbative control of the theory or the Higgs potential becomes strongly unstable. We want to find the value of $r$ that gives the largest cutoff, given a fixed $\delta g_{hVV}/g_{hVV}^{\rm SM}$. Focusing on the representation that gives the largest cutoff allows us to set an upper bound on the scale at which new physics {\it must} appear for any fixed value of $\delta g_{hVV}/g_{hVV}^{\rm SM}$.

We can easily conclude that we do not need to consider $r > 8$, since larger representations generate a Landau pole in the QCD coupling $g_s$ a factor of a few above the mass of the new particles~\cite{Blum:2015rpa}, independently of the value of $\delta g_{hVV}/g_{hVV}^{\rm SM}$. 

We can do better and show that we need to consider only color singlets ($r=1$). For small $SU(3)_c$ representations, $r\leq 8$, the leading effects that make us lose perturbative control of the theory or destabilize the Higgs potential are due to the new couplings $y, y^c$. If we hit a Landau pole before any Higgs instability, from the RGEs we can conclude that a fixed value for the couplings $y^{(c)}_r$ defined as
\be
y_r^{(c)} \equiv y^{(c)}\sqrt{r}
\ee
gives the same scale of the Landau pole for any $r$. $y_r$ and $y_r^c$ play a similar role as the 't Hooft coupling in gauge theories at large-$N$~\cite{tHooft:1973alw}. Therefore what is most relevant for us is the scaling of $\delta g_{hVV}$ at fixed $y_r^{(c)}$ and not at fixed values of the Yukawas $y^{(c)}$. This is given by
\be
\delta g_{hVV} \sim r\frac{\alpha_V}{4\pi }\frac{y^4 v^2}{M^2} = \frac{\alpha_V}{4\pi }\frac{y^4_r v^2}{r M^2}\lesssim \frac{\alpha_V}{4\pi }\frac{y^4_r v^2}{r M^2_{\rm exp}}\, .  \label{eq:gscaling}
\ee
Colored states give a smaller $\delta g_{hVV}$ at fixed $y_r^{(c)}$ compared to their colorless counterparts. This effect is further enhanced by the dependence of $M_{\rm exp}$ on $r$. Collider bounds on colored particles masses $M_{\rm exp}$ are stronger due to their larger production cross sections. So it is sufficient to consider vector-like leptons to get the most conservative upper bound on the scale of new bosons.  

If a Higgs instability arises before any Landau pole, the couplings that give a fixed scale of the instability for any $r$ are
\be
y_r^{(c)\prime} \equiv y^{(c)\prime}r^{1/4}\, ,
\ee
and the Higgs coupling deviation scales as
\be
\delta g_{hVV} \sim r\frac{\alpha_V}{4\pi }\frac{y^4 v^2}{M^2} = \frac{\alpha_V}{4\pi }\frac{y^{\prime 4}_r v^2}{M^2}\lesssim \frac{\alpha_V}{4\pi }\frac{y^{\prime 4}_r v^2}{M^2_{\rm exp}}\, .
\ee
Once again we can conclude that colored states give a smaller $\delta g_{hVV}$ at fixed $y_r^{(c)}$ compared to their colorless counterparts, since collider bounds on their masses $M_{\rm exp}$ are stronger due to their larger production cross section.

Ultimately in our analysis it is sufficient to consider only vector-like leptons
\be
L=(1, n)_{Y-1/2}\, , \quad N^c=(1, n-1)_{-Y}\, , \quad L^c=(1, n)_{-Y+1/2}\, , \quad N=(1, n-1)_{Y}\, . \label{eq:VLL}
\ee
If a Higgs coupling deviation is measured it will be instructive to repeat our analysis for $r\neq 1$ in order to go beyond our scaling arguments and obtain precise predictions for the scale of new bosons for any $r$. We leave this study to future work.

Models with new vector-like (or chiral) leptons have already been studied extensively in the literature in relation to neutrino masses, the muon anomalous magnetic moment, flavor models, and a variety of other subjects. A comprehensive review is beyond the scope of this work, but we refer to~\cite{DeRujula:1972pc, Fritzsch:1975sr, Fujikawa:1976qy, Cheng:1976ii, Slavnov:1976dp, Azimov:1976kc, Mikaelian:1976ut, Ma:1977da, McKay:1978wn, Goldberg:1981gc, delAguila:1982fs, Enqvist:1983xh, Cabibbo:1983bk, Ohba:1984uyo, Kalyniak:1984zz, Nakazato:1984dy, Choudhury:1984sh, Ellis:1985nu, Hagiwara:1985wt, Krucken:1985cx, Tanimoto:1985vj, Rizzo:1986wf, Rizzo:1987ye, Kovalchuk:1987aq, Stoker:1988kf, Pitkanen:1988ke, Geng:1988ar, Bhattacharyya:1989am, Singh:1990qd, Nagawat:1990ui, Nagawat:1990fa, Singh:1990ha, Ginzburg:1990kv, Hung:1992tz, Hayashi:1992cu, Bhattacharyya:1992np, Boudjema:1992em, Nardi:1992am, Nardi:1992nq, Djouadi:1993pe, Azuelos:1993qu, Pleitez:1994pu, Gonzalez-Garcia:1996oxz, doAmaralCoutinho:1996grs, Montvay:1996iu, Diaz-Cruz:1997mbr, Diener:1997nx, Thomas:1998wy, CiezaMontalvo:1998mi, Rosner:1999ub, Tafirout:2000gy, Allanach:2001sd, Rakshit:2001xs, Boos:2001vc, Eboli:2001hi, CiezaMontalvo:2001cy, CiezaMontalvo:2002gk, Bailin:2002fd, Singhal:2004wx, EspiritoSanto:2004tf, Diaz:2005th, Holdom:2006mr, Dorsner:2006fx, Bajc:2006ia, Bajc:2007zf, Abada:2007ux, delAguila:2008pw, Cynolter:2008ea, Abada:2008ea, Gogoladze:2008ak, Biggio:2008in, Strumia:2008cf, delAguila:2008hw, Kikukawa:2009mu, He:2009tf, Arhrib:2009xf, Arhrib:2009mz, Aguilar-Saavedra:2009fxa, Bandyopadhyay:2009xa, Patra:2009bu, Cacciapaglia:2009ky, Cacciapaglia:2009cv, Ma:2009kh, Ibanez:2009du, Picek:2009is, Gavela:2009cd, Yue:2010zzb, Bandyopadhyay:2010wp, Inan:2010af, Gross:2010ce, AristizabalSierra:2010mv, Kumericki:2011hf, Bandyopadhyay:2011aa, Eboli:2011ia, Ren:2011mh, Law:2011qe, Ari:2011is, Chen:2012wz, Voloshin:2012tv, Bae:2012ir, Law:2012mj, Chen:2012faa, Joglekar:2012vc, Almeida:2012bq, Arkani-Hamed:2012dcq, Kearney:2012zi, Ibrahim:2012ds, Endo:2012cc, Djouadi:2012ae, Dermisek:2012ke, Barr:2012ma, Majhi:2012bx, Arina:2012aj, Moreau:2012da, Feng:2013mea, Chang:2013eia, Schwaller:2013hqa, Dermisek:2013gta, Altmannshofer:2013zba, Fairbairn:2013xaa, Aguilar-Saavedra:2013twa, McDonald:2013hsa, Joglekar:2013zya, Kyae:2013hda, Ishiwata:2013gma, Banerjee:2013kna, Falkowski:2013jya, Batell:2013bka, Halverson:2014nwa, Xiao:2014kba, Dermisek:2014cia, Dermisek:2014qca, Biondini:2014dfa, Ellis:2014dza, Dorsner:2014wva, Ma:2014zda, Gopalakrishna:2015wwa, Voloshin:2015ona, Dermisek:2015oja, Bhattacharya:2015qpa, Cogollo:2015fpa, Okada:2015nca, Dermisek:2015hue, Majhi:2015bxa, Huo:2015exa, Ishiwata:2015cga, Fan:2015sza, Kumar:2015tna, Blum:2015rpa, Bizot:2015zaa, Doff:2015nru, Abdullah:2015zta, Chaudhuri:2015pna, Ruiz:2015zca, Angelescu:2015uiz, Chen:2016lsr, Tang:2016oba, Alvarado:2016par, Zeng:2016tmw, Baek:2016kud, Cai:2016ymq, Abdullah:2016avr, Herrero-Garcia:2016uab, Boucenna:2016qad, Dermisek:2016via, Hebbar:2016gab, vonderPahlen:2016cbw, Gopalakrishna:2016tku, Angelescu:2016mhl, Lee:2016wiy, Bahrami:2016has, Raby:2017igl, Megias:2017dzd, Poh:2017tfo, Bertuzzo:2017wam, Mann:2017wzh, Goswami:2017jqs, Ghosh:2017vhe, Dermisek:2017ihj, Bhattacherjee:2017cxh, Barrie:2017eyd, Borah:2017dqx, Ellis:2017nrp, Agostinho:2017biv, Dhargyal:2017cqo, Raby:2017igl, Dhargyal:2018bbc, Colucci:2018qml, Abada:2018nio, Abada:2018sfh, Carone:2018eka, Aboubrahim:2018hll, Xu:2018pnq, Dermisek:2018hxq, Dhargyal:2018ppo, Angelescu:2018dkk, Goswami:2018jar, Gopalakrishna:2018uxn, Song:2019aav, Zheng:2019kqu, Bhattiprolu:2019vdu, Kawamura:2019hxp, Bell:2019mbn, Arnan:2019uhr, Kowalska:2019qxm, Biswas:2019ygr, Suematsu:2019kst, Biggio:2019eeo, Fuentes-Martin:2020hvc, DeJesus:2020yqx, Gu:2020nic, Hieu:2020hti, Ma:2020qyz, Bissmann:2020lge, Borah:2020nsz, Frank:2020smf, Chala:2020odv, Crivellin:2020ebi, Ashanujjaman:2020tuv, Angelescu:2020yzf, Chun:2020uzw, Chakrabarty:2020jro, Bandyopadhyay:2020djh, Endo:2020tkb, Dermisek:2020cod, Rehman:2020ana, Matsedonskyi:2020mlz, Matsui:2021khj, Guedes:2021oqx, Jana:2020qzn, Das:2020gnt, Das:2020uer, Hernandez:2021tii, Kawamura:2021ygg, OsmanAcar:2021plv, Sen:2021fha, Morais:2021ead, Bharadwaj:2021tgp, Dermisek:2021ajd, Iguro:2021kdw, Yang:2021dtc, CarcamoHernandez:2021yev, Lee:2021gnw, Cherchiglia:2021syq, Karozas:2021dlh, Sahoo:2021vug, Ashanujjaman:2021zrh, Criado:2021qpd, Djouadi:2021wvb, Escribano:2021css, Bonilla:2021ize, Cacciapaglia:2021gff, Koren:2021ypn, Baspehlivan:2022qet, Chakraborty:2022uqo, Drewes:2022akb, Lee:2022nqz, Zhou:2022cql, Li:2022vsg, Carvunis:2022yur, Raju:2022zlv, Dermisek:2022xal, Brune:2022gnu, Kawamura:2022fhm, Abada:2022wvh, Hiller:2022rla, Vatsyayan:2022rth, Li:2022kkc, Ghosh:2022vpb, Ashanujjaman:2022cso, Arsenault:2022xty, Li:2022xty, Li:2022hzl, Hamaguchi:2022byw, Arora:2022uof, Coy:2023xtw, Arcadi:2023qgf, De:2023sqa, Ajjath:2023ugn, Shang:2023rfv, Bernreuther:2023uxh, Guo:2023jkz, Antusch:2023kli, Mishra:2023cjc, Mahapatra:2023zhi, Kulkarni:2023fyq, Das:2023tna} for a list of relevant studies. 

\begin{figure}[!t]
\begin{center}
\includegraphics[width=0.3\textwidth]{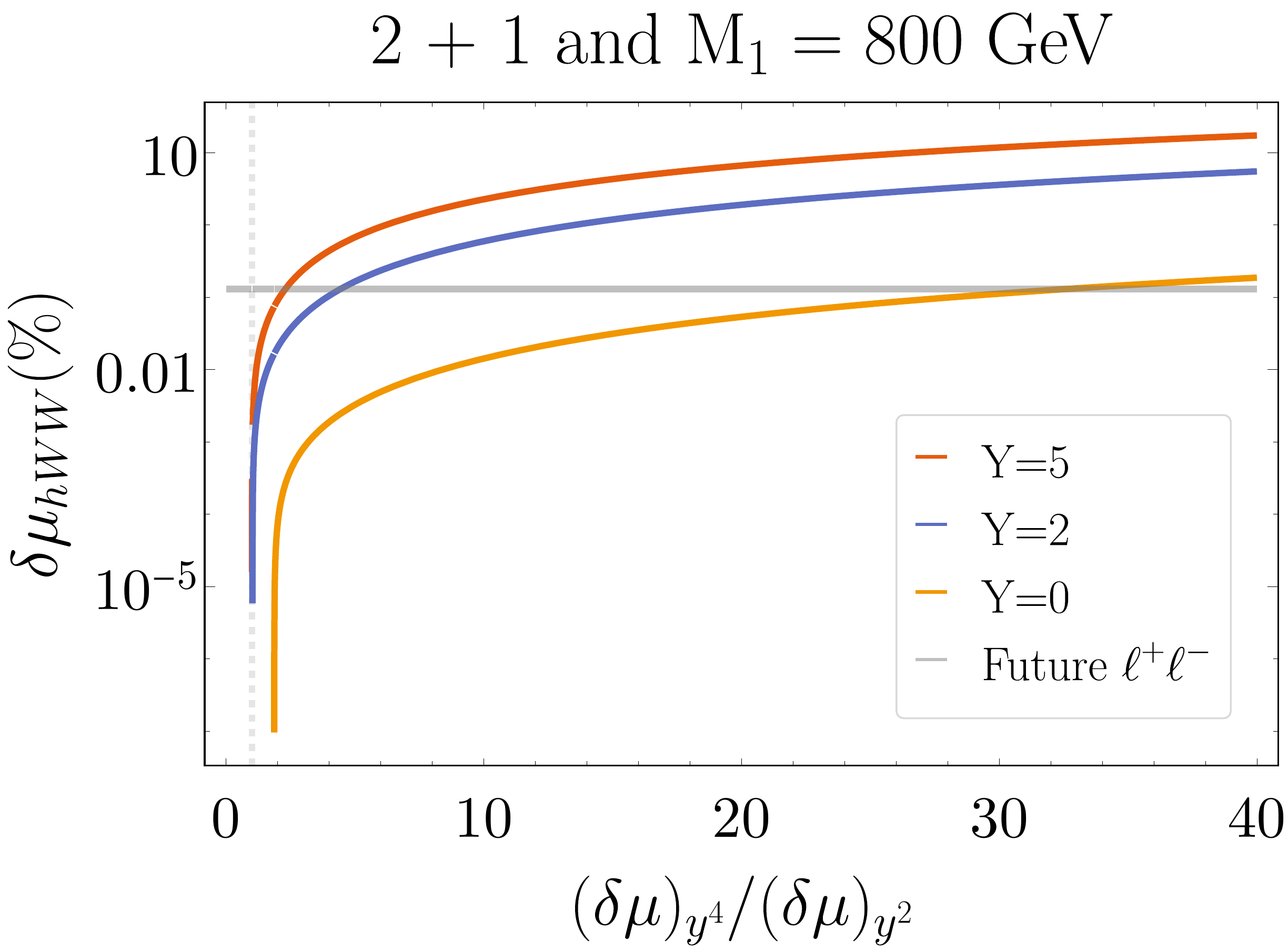}
\includegraphics[width=0.3\textwidth]{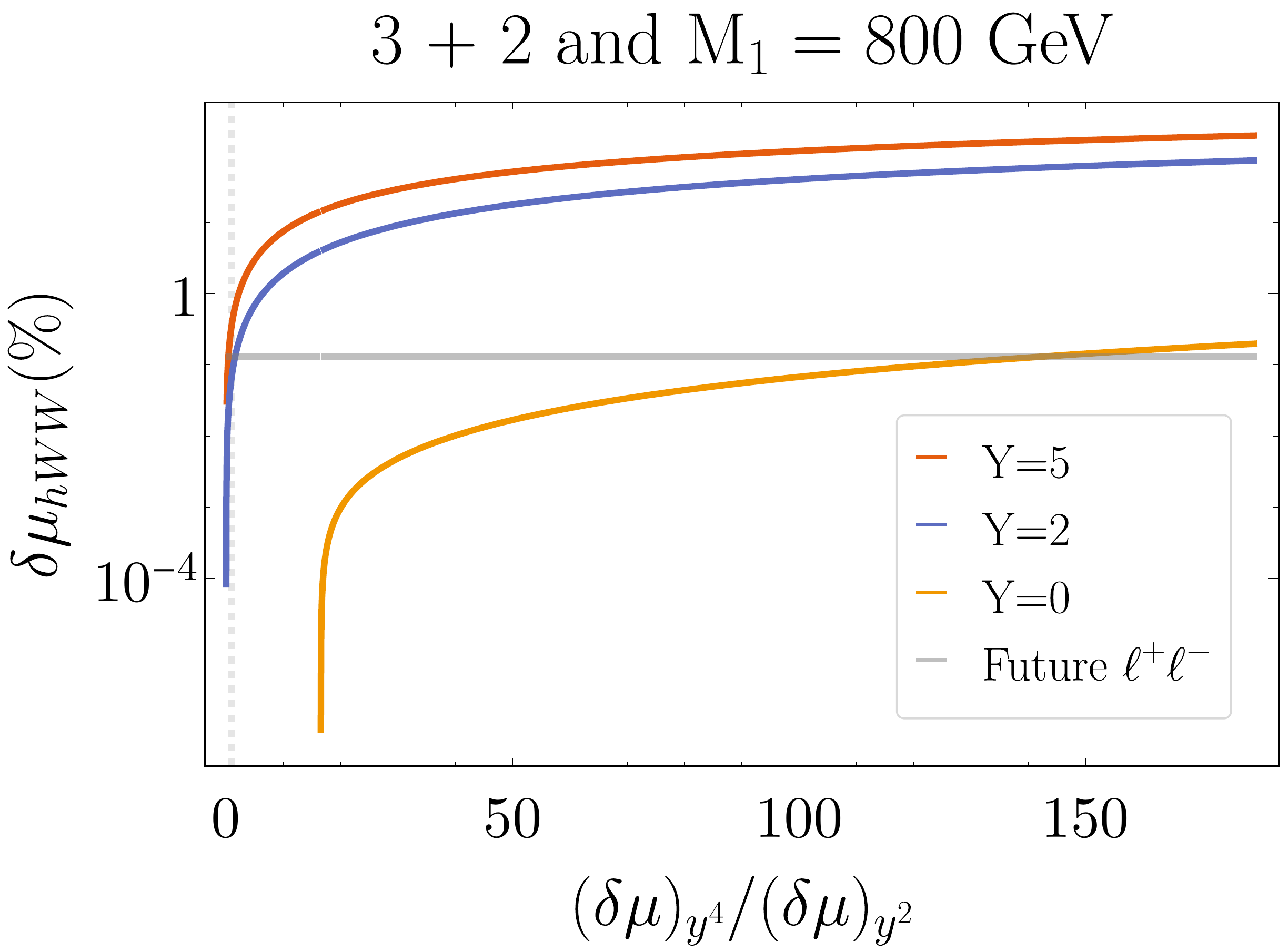}
\includegraphics[width=0.3\textwidth]{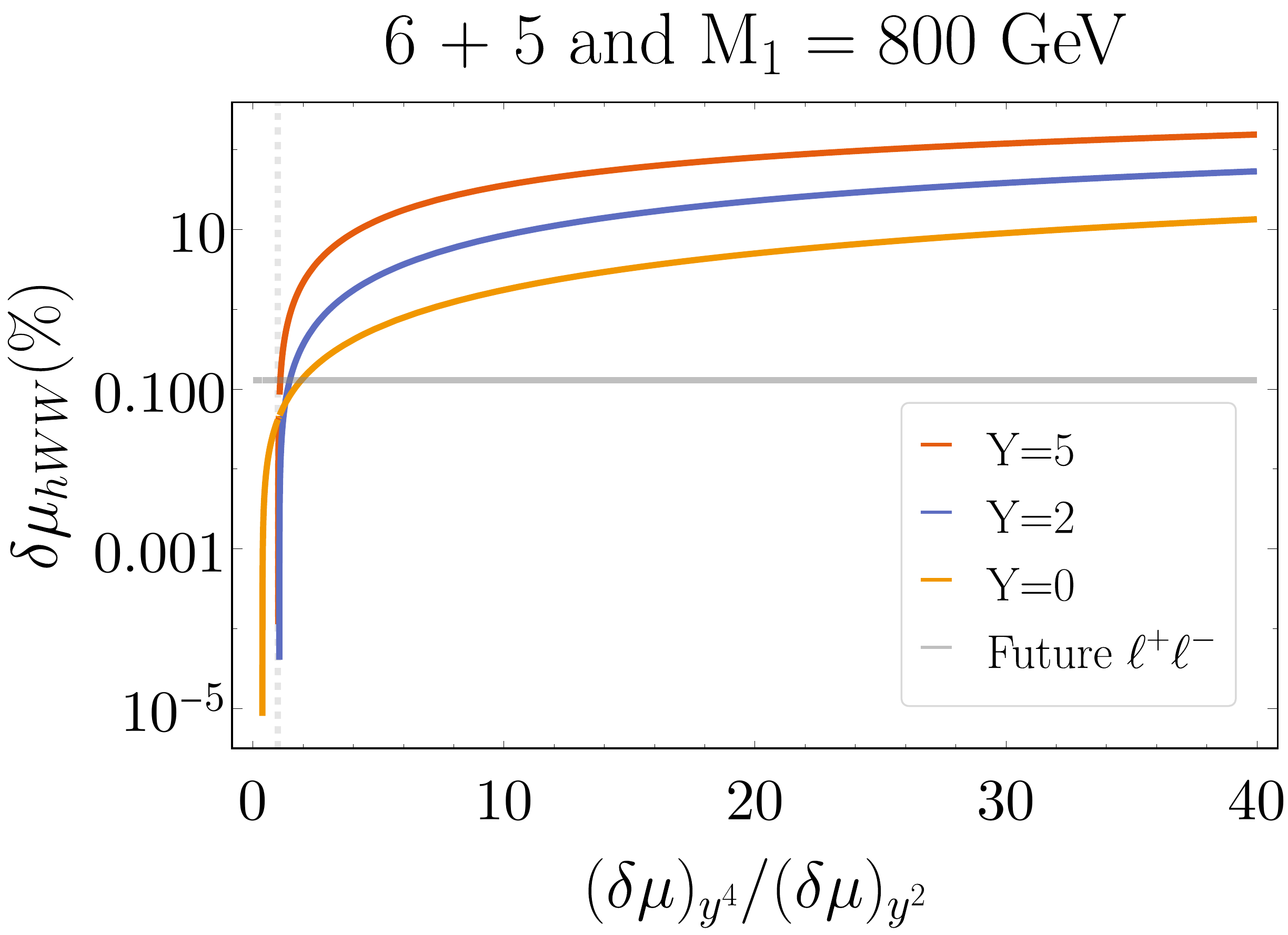}
\end{center}
\caption{Relative importance of the $y^4$ and $y^2 g_{\rm SM}^2$ terms in the $hWW$ coupling deviation. The gray dashed line signals where the two terms are equal. The $y^4$ terms dominate for coupling deviations large enough to be detected at future lepton colliders (solid gray line) or HL-LHC. The future collider sensitivity is discussed in Section~\ref{sec:results}.}\label{fig:y4y2WW}
\end{figure}

\begin{figure}[!t]
\begin{center}
\includegraphics[width=0.3\textwidth]{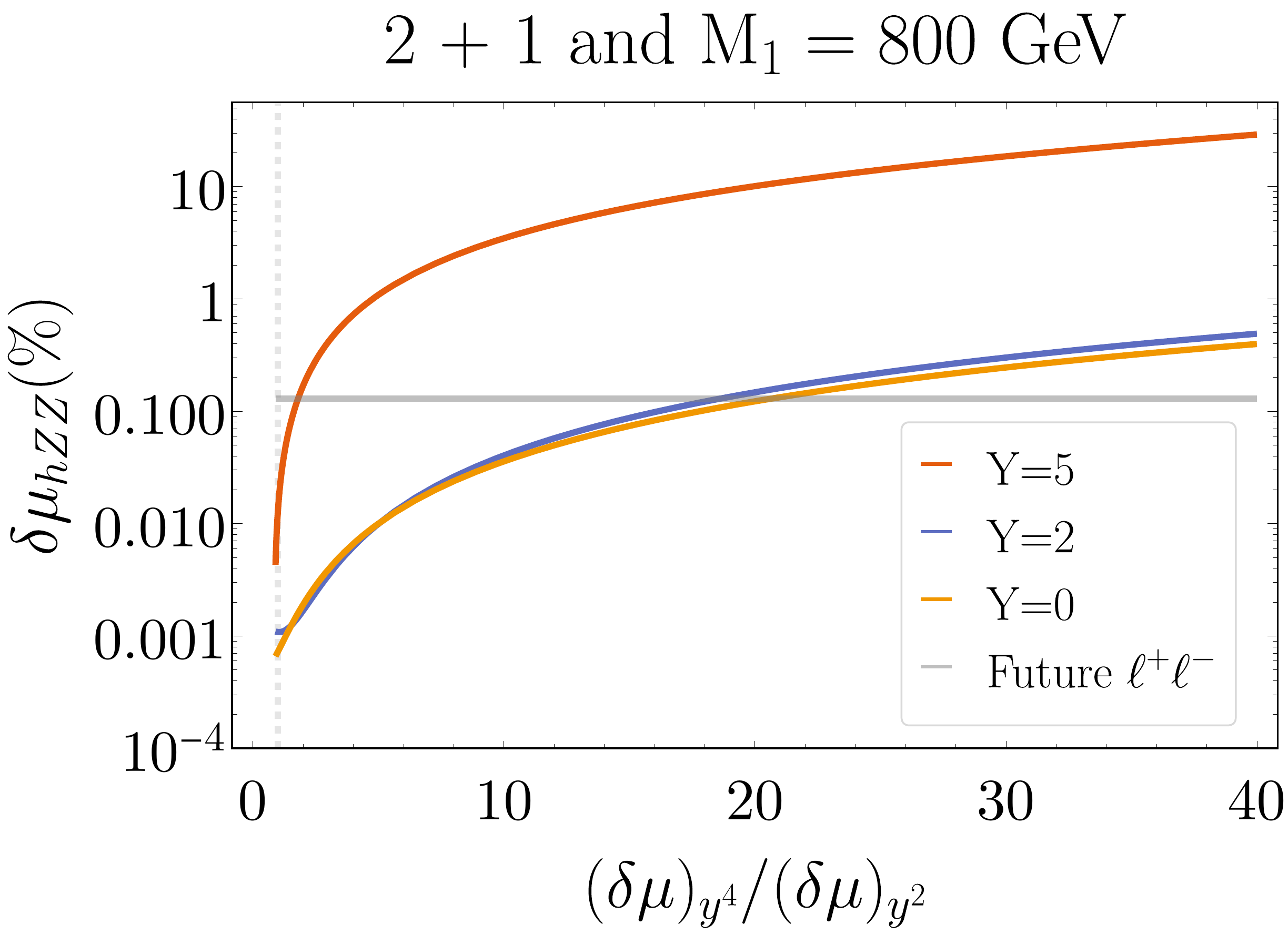}
\includegraphics[width=0.3\textwidth]{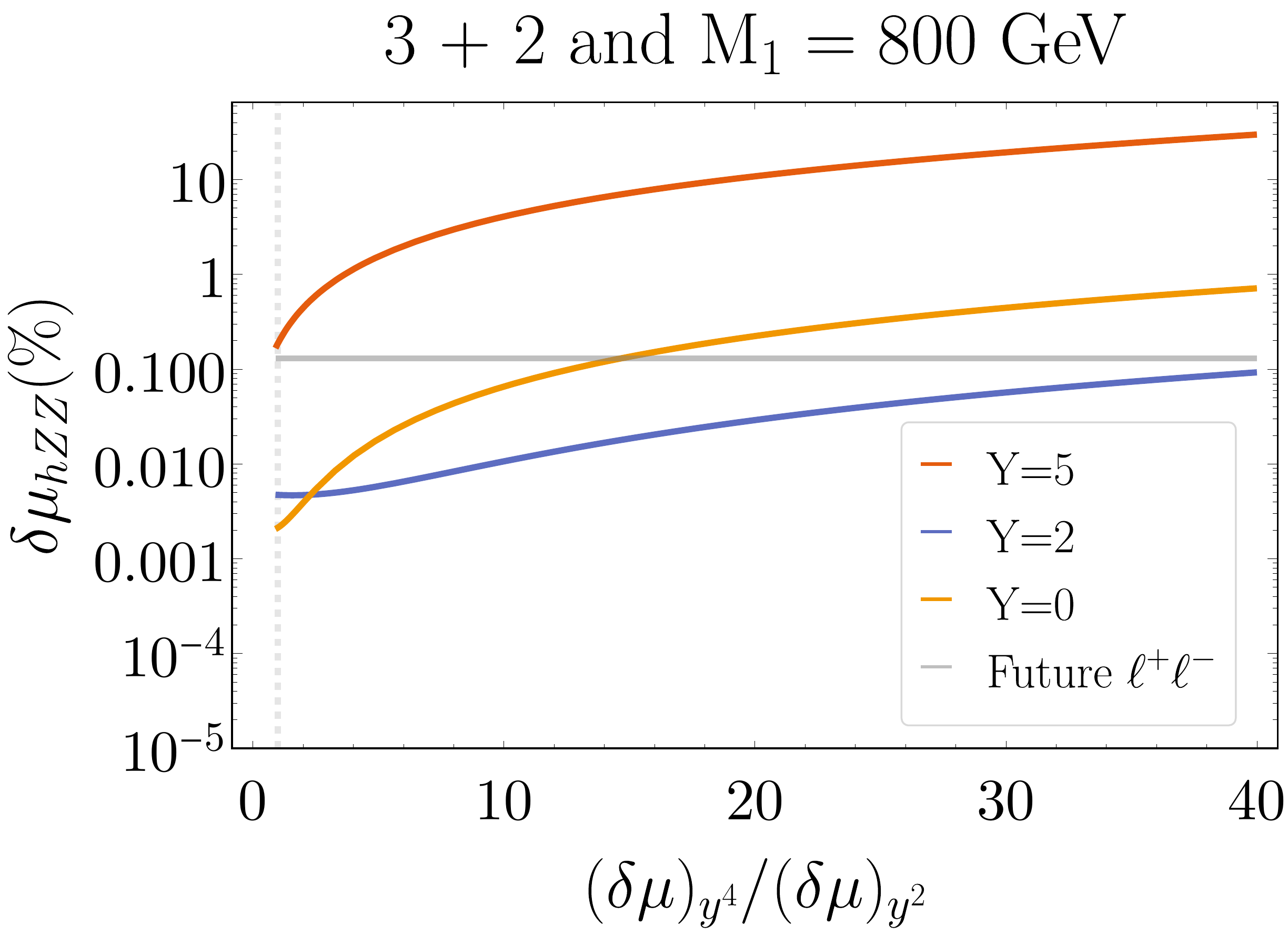}
\includegraphics[width=0.3\textwidth]{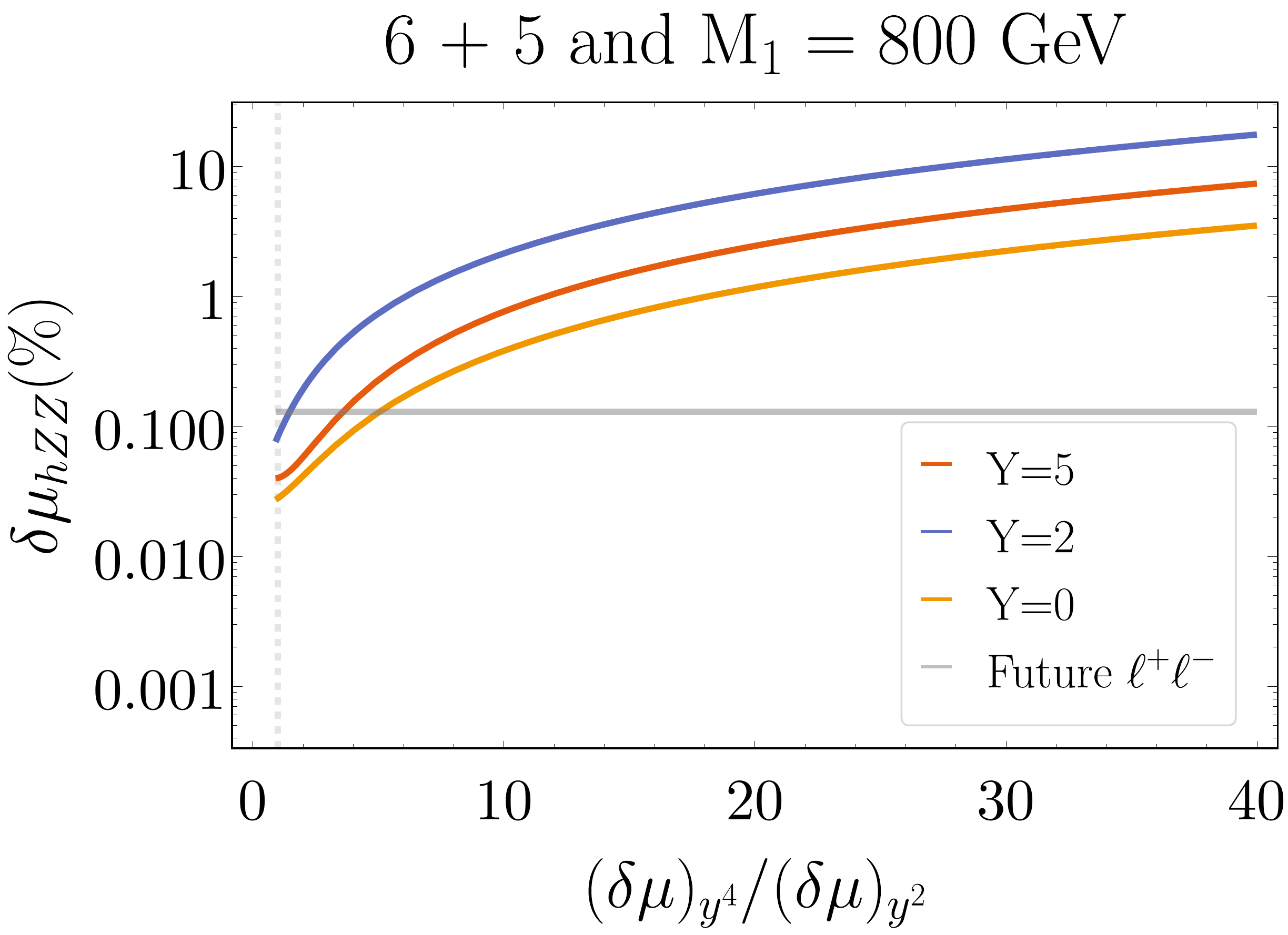}
\end{center}
\caption{Relative importance of the $y^4$ and $y^2 g_{\rm SM}^2$ terms in the $hZZ$ coupling deviation. The gray dashed line signals where the two terms are equal. The $y^4$ terms dominate for coupling deviations large enough to be detected at future lepton colliders (solid gray line) or HL-LHC. The future collider sensitivity is discussed in Section~\ref{sec:results}.}\label{fig:y4y2ZZ}
\end{figure}

The only possible loophole in our argument is that we focused on the part of the coupling deviation proportional to $y^4$, but there are also terms proportional to $y^2 g_{\rm SM}^2$. The $y^2 g_{\rm SM}^2$ terms proportional to $y y^c$ grow with the dimension of the $SU(2)_L$ representation as $n^3$, versus the more modest linear growth of the $y^4$ terms, so one might worry that states with large $r$ and $n$ invalidate our argument. However, we verified numerically that for $\delta \mu_{hVV} \gtrsim 0.13\%$ (the sensitivity of the most precise future lepton colliders) the $y^4$ term always dominates if $M_1$ is compatible with LHC bounds on colored particles. This is shown in Fig.s~\ref{fig:y4y2WW} and~\ref{fig:y4y2ZZ}. The mass of the lightest new states in the Figures is fixed at $M_1=800$ GeV to represents a conservative lower bound on masses of new colored states at the LHC. The upper bound on $SU(2)_L$ representations and hypercharges in the Figure is determined by low energy Landau poles, as detailed in Section~\ref{sec:SU2}.

Before turning to $SU(2)_L$ and hypercharge quantum numbers, it is worth pointing out that adding multiple copies of the fermions in Eq.~\eqref{eq:VLL} does not allow to raise the maximal UV cutoff. The scaling arguments that lead to this conclusion are exactly the same as those given for the color representation, replacing the dimension of the representation $r$ with the number of copies $\mathcal{N}$. In making this statement we are implicitly assuming that all $\mathcal{N}$ fermions are close in mass. However, if they are not, one generation dominates the coupling deviation and we are effectively in the $\mathcal{N}=1$ case (or a case with even lower cutoff if multiple Yukawas are large). 

One might envision adding $\mathcal{N}$ fermions that are close enough in mass to contribute comparably to $\delta g_{hVV}$, but still far enough not to show up as a $\mathcal{N}^2$ enhancement of the production cross-section. In this case we cannot strictly say that $M_{\rm exp}$ in the previous equation increases compared to the $\mathcal{N}=1$ case, as we did for colored fermions, but even if $M_{\rm exp}$ were to remain the same, the above arguments are still valid and show that the UV cutoff with $\mathcal{N}>1$ copies is not larger than that obtained for $\mathcal{N}=1$. Note that this is not true for Higgs couplings to massless gauge bosons, due to the different scaling of the coupling deviation with $y$. In those cases $r> 1$ and/or $\mathcal{N}>1$ can give a bigger cutoff for a fixed coupling deviation~\cite{Arkani-Hamed:2012dcq}.

\subsection{$SU(2)_L$ and Hypercharge}\label{sec:SU2}
We can obtain an upper bound on the largest $n$ and $Y$ that we need to consider from Landau poles in gauge couplings. If, for a given choice of fermions representations, we get a Landau pole right above the masses of the new fermions, we do not need to consider larger representations because the theory needs to be extended very close to where the new fermions appear, independently of $\delta g_{hVV}$.

\begin{figure}[!t]
\centering
\includegraphics[width=0.49\textwidth]{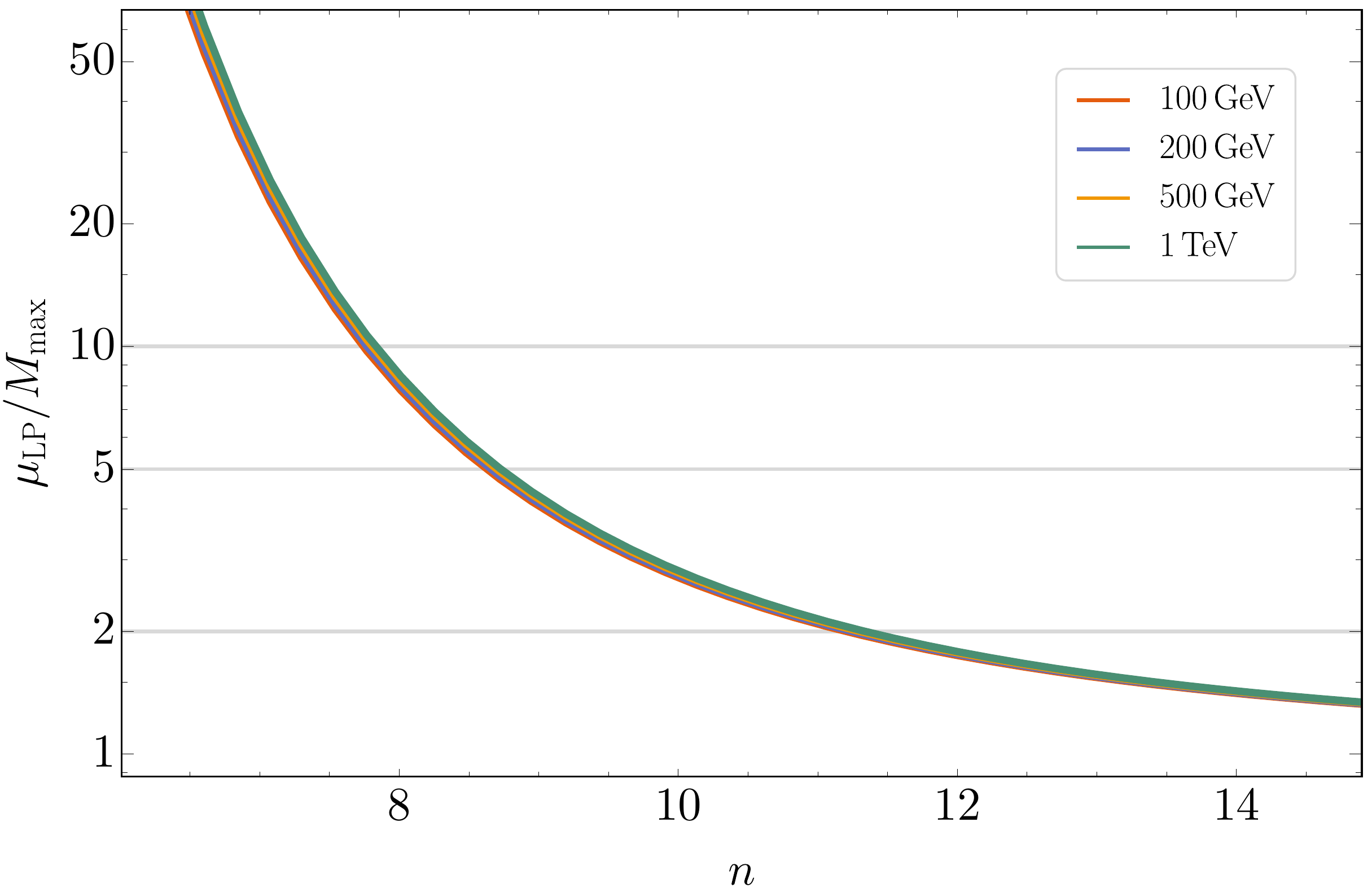}
\includegraphics[width=0.49\textwidth]{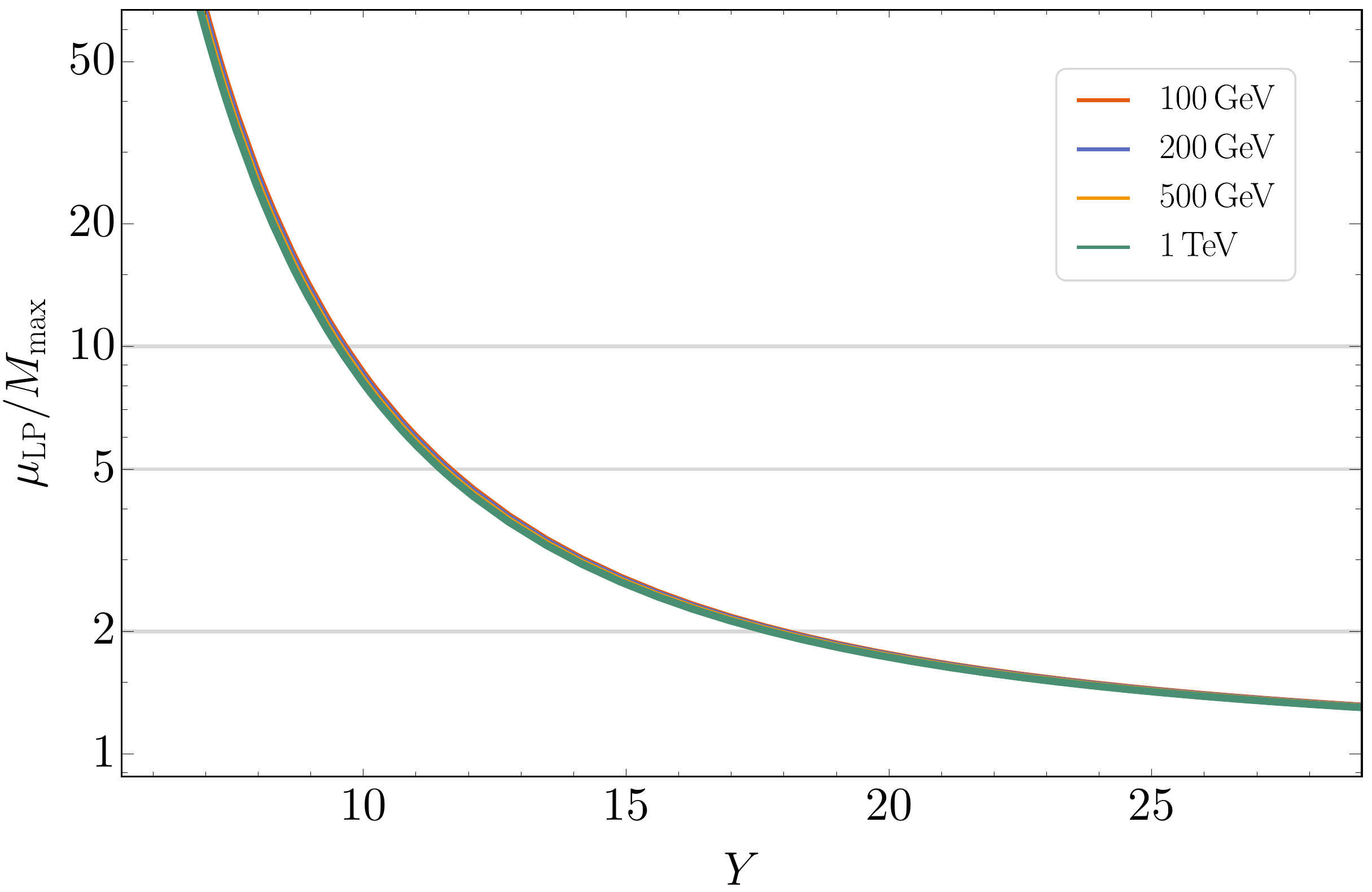}
\caption{Location of the $SU(2)_L$ Landau pole (left panel) and $U(1)_Y$ Landau pole (right panel), normalized to the heaviest mass of the vector-like fermions $M_{\rm max}$, as a function of the dimension of their representations.}
\label{fig:landau}
\end{figure}

In Fig.~\ref{fig:landau} we show the location of the Landau pole in $g$ and $g^\prime$ for $y=y^c=0$. For each plot we take the smallest value of the other parameter that we want to bound, so $n=2$ for the hypercharge Landau pole and $Y=0$ for the $SU(2)_L$ Landau pole. This is conservative, since larger values move the Landau pole to lower energies, as can be seen from the model RGEs. We have verified that taking the new Yukawas to be different from zero does not affect appreciably the location of the Landau pole in the gauge coupling, unless we take them so large that we hit an instability even earlier.

For $n=8$, i.e. when we add to the theory a vector-like pair of $8$'s and a vector-like pair of $7$'s, the $SU(2)_L$ Landau pole is a factor of 10 above the heaviest new lepton mass, $M_{\rm max}$, signalling that the theory is barely under control perturbatively. For this reason we consider only $n\leq 7$ in what follows. In Section~\ref{sec:results} we show explicitly $\Lambda_B$ only up to $n=4$, as larger representations give even lower values of $\Lambda_B$ at fixed $\delta \mu$. We could be more aggressive and consider only smaller representations, but we are trying to be conservative when setting an upper bound on $\Lambda_B$. 

Our results for the Landau pole are very close to the analytic estimate that one can perform at one loop
\be
 \mu_\textnormal{LP}=\exp\left[\frac{1}{\beta_\textnormal{SM}+\Delta \beta}\left(\frac{8\pi^2}{g^2\left(M_Z\right)}+\beta_\textnormal{SM}\log\left(M_Z\right)+\Delta \beta\log(M)\right)\right]\, ,
\ee
where $\beta_{\rm SM}$ is the one-loop $SU(2)_L$ $\beta$-function in the SM and 
\be
\Delta \beta=\frac{4}{3}\left[T(n)+T(n-1)\right]
\ee
is the contributions from the new leptons, where $T(n)$ is the Dynkin index of a $SU(2)_L$ representation of dimension $n$
\be
T(n)= \frac{n(n-1)(n+1)}{12}\, .
\ee
Beyond this discussion, there is no other simple scaling argument that can exclude large $SU(2)_L$ representations from our analysis. At large $n$ and large $M$ the coupling deviation scales as
\be
\delta g_{hVV} \sim \frac{\alpha_V}{4\pi} \left(A n^3 \frac{g^2_{\rm SM} y^2 v^2}{M^2} + B n \frac{y^4 v^2}{M^2} \right)\, , \label{eq:su2n}
\ee
with $A$ and $B$ both $\mathcal{O}(1)$ numbers.
The $n^3$ term in the previous expression is expected on general grounds, since the $VV$ two-point function scales as the Dynkin index $T(n) \sim n^3$ and the coupling deviation for small external momenta can be obtained from it through the Higgs low energy theorem~\cite{Ellis:1975ap, Shifman:1979eb, Kniehl:1995tn, Gillioz:2012se}, that reproduces exactly our results in the appropriate limit. To try to reduce the representations in our analysis we have to compare the scaling in Eq.~\eqref{eq:su2n} to the scaling of the relevant RGEs,
\be
16 \pi^2\frac{d y}{dt } \sim n y^3\, , \quad 16 \pi^2\frac{d \lambda}{dt } \sim - n y^4+ n\lambda y^2\, .
\ee

For example, in the case of a Higgs instability we can define
\be
y_n \equiv n^{1/4} y\, ,
\ee
that gives the same instability scale for any $n$ and write the coupling deviation as
\be
\delta g_{hVV} \sim \frac{\alpha_V}{4\pi} \left(A n^{5/2} \frac{g^2 y^2_n v^2}{M^2} + B \frac{y^4_n v^2}{M^2} \right)\, , \label{eq:su2n2}
\ee
so the numerator of the coupling deviation grows with $n$ (if $n$ is large enough that the first term dominates). However, the denominator also grows with $n$, due to the stronger collider (and indirect) constraints on $M$ for representations with large charges. States in large-$n$ representations have exotic charges and can decay to final states with effectively zero background. Additionally, the doublet+singlet model ($n=2$) is still very unconstrained by the LHC, with LEP setting the most stringent bounds (see Section~\ref{sec:bounds}) and Eq.~\eqref{eq:su2n2} does not appropriately describe the coupling deviation in this case (i.e. we cannot take $M\gg v$). So we cannot reach a conclusion on the $SU(2)_L$ representations with the largest $\Lambda_B$ from these simple arguments. The only case where we can say something definite is when the first problem with the theory is a $SU(2)_L$ Landau pole. In this case the important RGE is $dg/dt \sim n^3 g^2$ and we see by rescaling $g$ that smaller representations give a bigger coupling deviation for a fixed scale of the Landau pole.

In summary our simple scaling arguments are inconclusive and in Section~\ref{sec:results} we consider both the doublet+singlet model and higher-dimensional $SU(2)_L$ representations. In practice we find that smaller representation in general give a bigger $\Lambda_B$, so that we can restrict our analysis to $n=4$ and below. 

A similar reasoning holds for the hypercharge of the new fermions, taking now $T(Y)=Y^2$. To avoid low-energy Landau poles independent of $\delta g_{hVV}$, we consider only $Y\leq 5$ in what follows (see Fig.~\ref{fig:landau}). This is more ``aggressive" than our upper bound on $SU(2)_L$ representations, if we judge by the position of the gauge Landau pole in Fig.~\ref{fig:landau}. However, when computing $\Lambda_B$, we find that gauge Landau poles dominate over other instabilities already for $Y= 5$, as discussed in the next Section.

As for $SU(2)_L$, it is not possible to replicate the scaling arguments that led to consider only the lowest color representations, because there is no obvious rescaling of $y,y^c$ that gives a fixed scale of the instability for any value of $Y$. The only exception arises if $Y$ is so large that the first issue with the theory is the hypercharge Landau pole. In this case the one-loop RGE for the gauge coupling $d g_Y/dt \sim Y^2 g_Y^3$ suggests the rescaling $g_Y^\prime = g_Y/Y$. In this regime there is no gain in going to large $Y$ because $\delta g_{h ZZ} \sim g_Y^2 Y^2 (v^3/M^2) \sim (g_Y^\prime)^2 (v^3/M^2)$ and states with large $Y$ are easier to detect, effectively giving a smaller $\delta g_{h ZZ}$ for a fixed scale of the Landau pole. However at the weak scale we can have $y, y^c \gg g_Y$ and for values of $0 < Y \leq 5$ one can be in a regime where the instability of the Higgs quartic or the Landau poles in $y, y^c$ are not appreciably affected by increasing $Y$ (that enters their running at two-loops in association with the comparatively small hypercharge gauge coupling), while $\delta g_{h ZZ}$ increases. For this reason in the following we show results for the scale of new bosons also in models where $Y$ is large. 

\section{The Scale of New Bosons}\label{sec:results}

\subsection{Definition of $\Lambda_B$}\label{sec:LB}
We have seen in the previous Sections that the models with only new vector-like leptons give the most conservative upper bound on the scale of new bosons $\Lambda_B$. Before turning to our results for $\Lambda_B$ we comment briefly on how to compute it. If $\Lambda_B$ comes from a loss of perturbativity, we fix it conventionally to be the scale where the coupling hits $4\pi$. This is somewhat arbitrary given that we do not have complete control of the theory already at smaller values of the coupling. However, the running is fast when the Yukawas approach $4\pi$, and changing the upper bound on the coupling does not appreciably affect $\Lambda_B$. For example, it was shown in~\cite{Blum:2015rpa} that reducing the threshold to $\sqrt{4\pi}$ does not qualitatively affect the result for $\Lambda_B$ within the two-loop approximation for the RGEs that we employ also in this work. If a Higgs coupling deviation is measured it will be worth refining these results, but our choice $y \simeq 4\pi$ is good enough for our illustrative purposes. 

The case where $\Lambda_B$ is due to an instability in the Higgs potential is slightly more subtle. In this work we adopt a manifestly gauge invariant criterion to determine the scale where the Higgs potential becomes unstable. We compute at two loops the RGE evolution of the Higgs quartic $\lambda$ and require a stable theory to satisfy
$\lambda(\mu)^{-1} > -14.53 + 0.153\log[{\rm GeV}/\mu]$, for any scale $\mu$~\cite{Isidori:2001bm}. The physical meaning of this criterion has already been discussed quite extensively in the literature (see for instance~\cite{Isidori:2001bm, Degrassi:2012ry}). The scale $\Lambda_B$ is then given by $\lambda(\Lambda_B)^{-1} = -14.53 + 0.153\log[{\rm GeV}/\Lambda_B]$. 

Intuitively the stability bound is on $\lambda$ because the effective Higgs potential is well approximated by $V_{\rm eff}(h)\simeq (\lambda_{\rm eff}(h)/4) h^4$ when $h\gg v$. One can compute the bounce action for tunnelling from the SM vacuum ($h=v$) to a point $h^*$ in the region $h\gg v$, where $V_{\rm eff}(v)=V_{\rm eff}(h^*)$. The bounce action depends on $\lambda(\mu)$ and the inequality that we use corresponds to a tunnelling rate equal to the lifetime of the universe~\cite{Isidori:2001bm}. This criterion is manifestly gauge invariant given that $\lambda(\mu)$ is a measurable, gauge-independent quantity and so is the tunnelling rate. However, this choice does not capture all the corrections to the Higgs effective potential at this order, i.e. $\lambda_{\rm eff}(\mu) \neq \lambda(\mu)$. 

\begin{figure}[!t]
\centering
\includegraphics[width=0.5\textwidth]{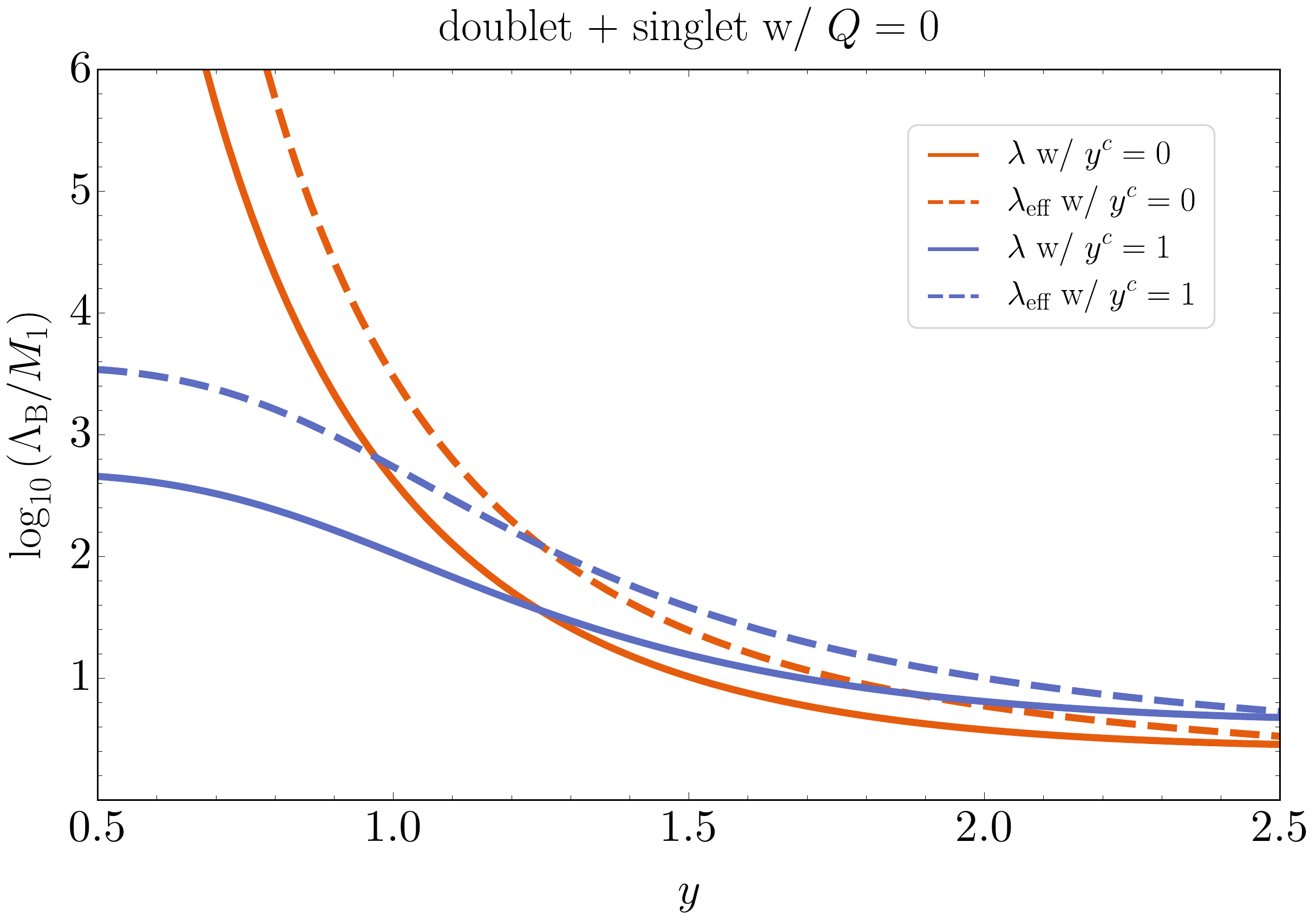}
\caption{Instability scale of the Higgs potential $\Lambda_B$ normalized to the lightest new fermion mass $M_1$ in the {\it doublet+singlet} theory introduced in Section~\ref{sec:4f}, as a function of one of its two Yukawa couplings (Eq.~\eqref{eq:L4}). The solid lines are obtained by running the two-loop RGEs of the Higgs quartic $\lambda$. The dashed lines are obtained by extracting the effective Higgs quartic $\lambda_{\rm eff}$ from the two-loop improved effective potential in Landau gauge.}
\label{fig:comp_lambda}
\end{figure}

In Fig.~\ref{fig:comp_lambda} we show that using $\lambda_{\rm eff}(\mu)$, as computed from the two-loop improved effective potential\footnote{To compute the effective potential we use the same methodology described in detail in the appendix of~\cite{Blum:2015rpa} and obtain the same results.} in Landau gauge, gives results for $\Lambda_B$ that can differ from those obtained from $\lambda(\mu)$ by up to a factor of ten for the models considered in this work. However $\lambda_{\rm eff}(\mu)$ is gauge-dependent. It is a well-known problem that $V_{\rm eff}$ is not gauge invariant~\cite{Jackiw:1974cv}. In principle it is possible to extract gauge-invariant quantities from it and the metastability bound that we are interested in should be one of them. However to date we do not know of a way to obtain a gauge-independent result at fixed order in perturbation theory~\cite{Andreassen:2014gha, Andreassen:2014eha}. The gauge dependence arises from electroweak corrections to $\lambda(\mu)$, it is numerically quite mild and often instability scales are quoted in Landau gauge~\cite{Degrassi:2012ry, Blum:2015rpa}. However, in this work we prefer the theoretically cleaner criterion of bounding $\lambda(\mu)$ and finding a gauge-independent bound. 

\subsection{Direct and Indirect Constraints on New Fermions}\label{sec:bounds}
The main goal of the paper is to set a (conservative) upper bound on the scale of new bosons, given an observed deviation in Higgs couplings to $WW$ or $ZZ$. The lighter the new fermions can be for a fixed Yukawa, the larger the upper bound on $\Lambda_B$. In this Section we compute current constraints, in the spirit of understanding how far we can go in this direction. Our goal is not to study in detail the constraints on the new fermions, we prefer to make conservative statements to understand how large $\Lambda_B$ can be. A more thorough study of the bounds on the new fermions will be appropriate if a deviation in $hWW$ or $hZZ$ is discovered.

The strongest candidate for the maximization of $\Lambda_B$ is the doublet+singlet model discussed in Section~\ref{sec:4f}, since the masses of its constituents are the hardest to constrain at the LHC. This theory is similar, but not identical, to a Higgsino-Bino system. They would have been identical if we had considered the model with a Majorana fermion $N=N^c$ and identified $M_{\widetilde B}=M_N$, $M_L=\mu$, $g^\prime v_u = y v$, $g^\prime v_d = y^c v$. However the intuition from SUSY searches that this system is poorly constrained at the LHC holds also in our case.

First of all, it is useful to notice that our coupling deviation is mainly determined by the vector-like mass of the doublet $M_L$. Taking $M_N=0$ does not appreciably increase $|\delta \mu_{hVV}|$ compared to $M_N=M_L$. To see this explicitly we integrate out the doublets when the singlets are light. As in Section~\ref{sec:4f} we call the doublets $\ell, \ell^c$ and the singlets $n, n^c$. At the weak scale we are left with
\be
\mathcal{L} &\supset& i \left(\frac{|y^c|^2}{|M_L|^2}(H n)^\dagger \bar \sigma^\mu D_\mu (H n)+\frac{|y|^2}{|M_L|^2}(H n^c)^\dagger \bar \sigma^\mu D_\mu (H n^c)\right)+\left(\frac{y y^c (H n)^\dagger H n^c}{M_L}+{\rm h.c.}\right)\nn \\
&+&\mathcal{O}(1/M_L^3)\, . 
\ee
From this Lagrangian we can estimate the coupling deviation coming purely from light singlets in the loop. We have
\be
\frac{\delta g_{hVV}}{g_{hVV}^{\rm SM}} \sim y y^c |y^c|^2 |y|^2 \frac{v^5}{M_L^5}\, ,
\ee
which is very subleading to the $\mathcal{O}(v^2/M_L^2)$ corrections that one obtains by integrating out the doublets at one loop. This simple exercise shows that the limit $M_N \ll M_L$ in Eq.~\eqref{eq:L4} is not relevant for us, if $M_L$ is constrained to be larger than the weak scale. 

This simplifies our analysis and allows us to get some intuition on the collider constraints on the model. Our theory contains two neutral particles with mass $M_{1,2}$ and one charged particle with mass $M_L$. To better understand the spectrum let us consider a few relevant limits where we can get some analytic intuition. We have discussed how a big hierarchy between $M_L$ and $M_N$ does not increase the coupling deviation. We can therefore start by considering $M_L=M_N$. In this limit we have
\be
M_2 > M_L > M_1 \quad {\rm if} \quad M_L > \frac{y y^c v}{\sqrt{2} (y+y^c)}\, ,\label{eq:ML1} 
\ee
and $M_1>M_L$ in the opposite case. Even if we take $M_N=0$ we have a parametrically similar conclusion
\be
M_2 > M_L > M_1 \quad {\rm if} \quad M_L > \frac{y y^c v}{\sqrt{2(y^2+y^{c2})}}\, . \label{eq:ML2}
\ee
Taking $M_{L,N}$ small at fixed $y,y^c$ maximizes the coupling deviation. However these results show that, as we make $M_{L, N}$ smaller, at some point we hit a configuration where our charged particle is the lightest of all\footnote{If we choose to include a relative sign and take for instance $y=-y^c$, the charged state of mass $M_L$ is always the lightest in the spectrum.}. The collider constraints on a stable $Q=1$ particle are quite stringent. CMS gets $M_{\rm exp} \gtrsim 400$~GeV from DY pair production with 2.5 fb$^{-1}$~\cite{CMS:2016kce}. The ATLAS bound with 36 fb$^{-1}$ on the charged component of a $SU(2)_L$ doublet is $M_{\rm exp}\gtrsim 840$~GeV~\cite{ATLAS:2019gqq}. 

\begin{figure}[!t]
\centering
\includegraphics[width=0.49\textwidth]{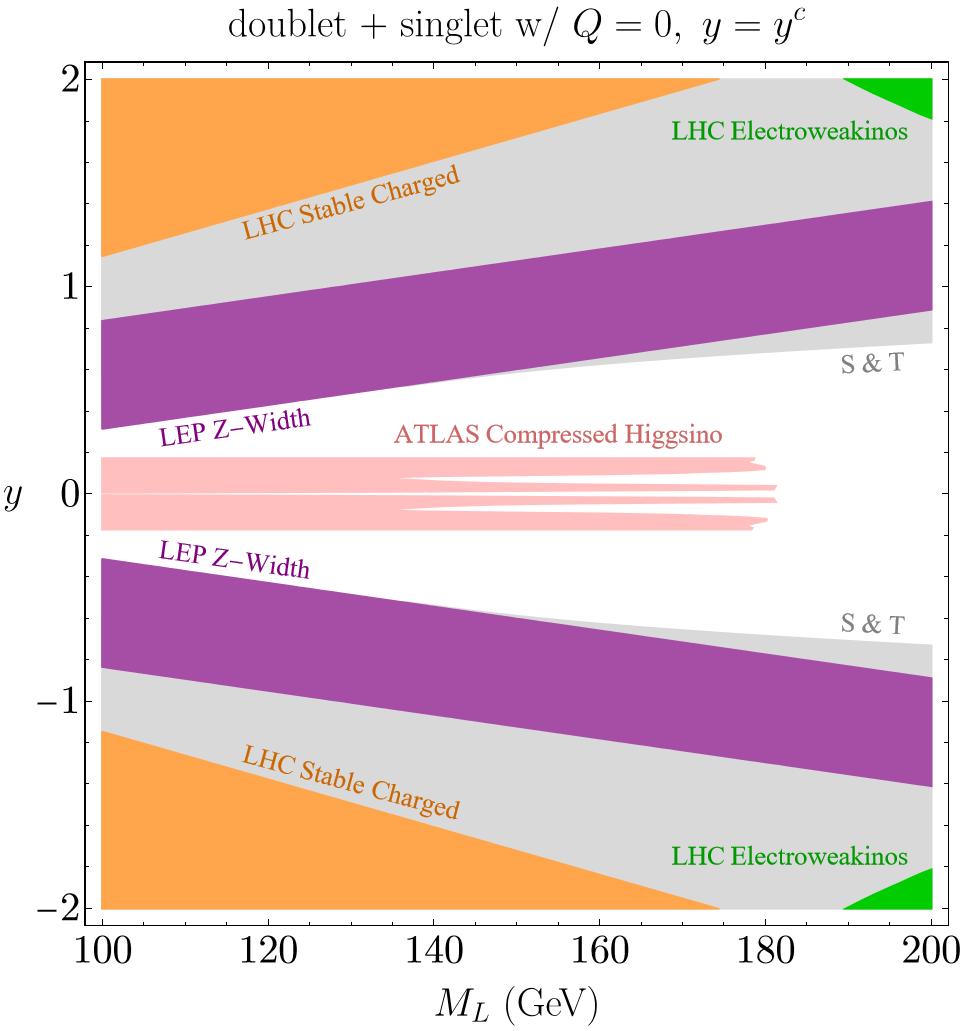}
\includegraphics[width=0.49\textwidth]{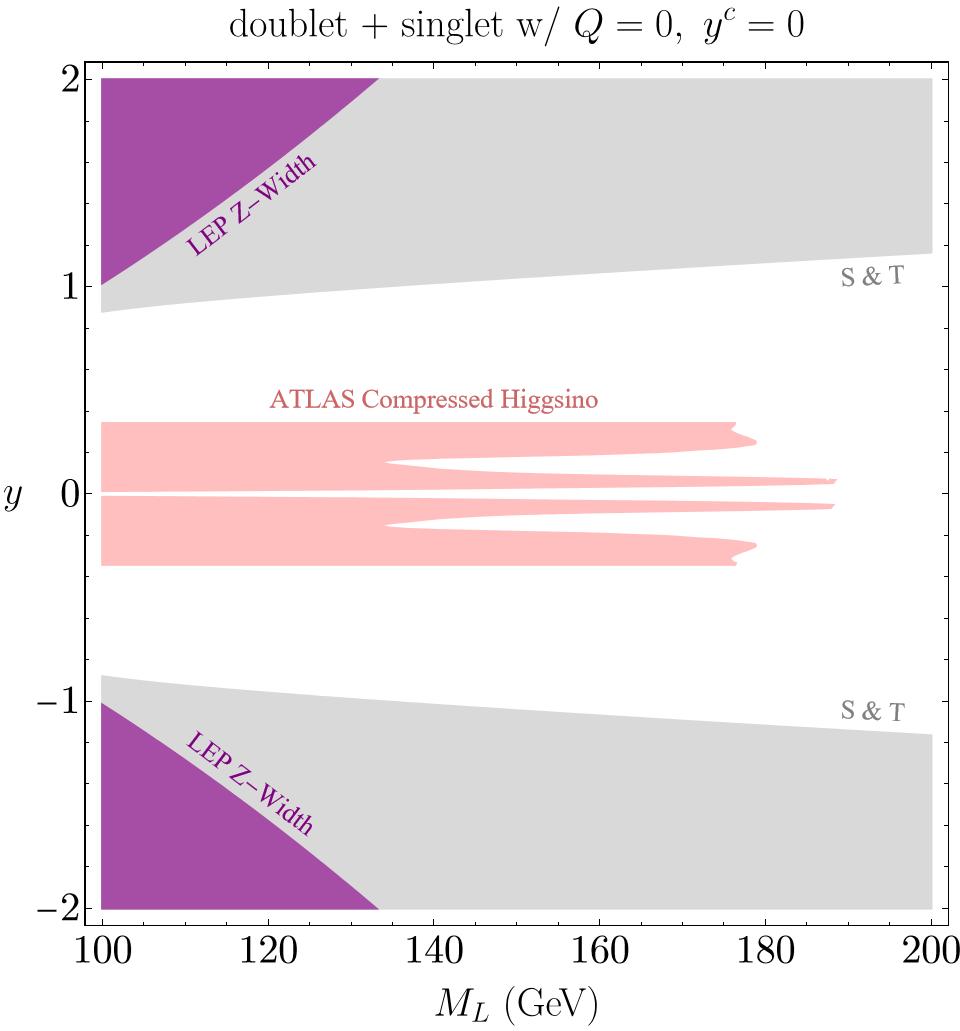}
\caption{Experimental constraints on the doublet+singlet model from direct searches at colliders and electroweak precision measurements for $y=y^c$ (left panel) and $y^c=0$ (right panel).}
\label{fig:collider}
\end{figure}

Searches for heavy stable charged particles at the LHC have very low background~\cite{CMS:2016kce, ATLAS:2019gqq, ATLAS:2022pib} and the same is true for long-lived particles decaying visibly to soft SM particles, so we might try to elude these bounds by adding a small mixing between $\ell, \ell^c$ and the SM lepton doublets, just large enough to make the lightest new state decay promptly into SM leptons, but small enough to avoid all indirect constraints from flavor. However, after more than ten years of LHC data analysis, the bounds are stringent also in this case: $M\gtrsim 790$~GeV~\cite{CMS:2019hsm} for a new lepton doublet decaying predominantly to third generation leptons\footnote{Since rare events at the LHC are those containing electrons or muons and the $\tau$ leptonic branching ratio is smaller than its hadronic one, recasting this analysis would give a stronger constraint on decays to the first two generations.}. The small gap at low masses between the LEP bound and this analysis is bridged by CMS and ATLAS searches for charginos that exclude our model for masses around $M_L=100$~GeV~\cite{ATLAS:2018eui, ATLAS:2019lng, ATLAS:2021moa, ATLAS:2021yqv, ATLAS:2020pgy, ATLAS:2021yqv, ATLAS:2019lff, ATLAS:2021yqv, ATLAS:2022zwa, ATLAS:2022hbt, CMS:2018szt, CMS:2020bfa, CMS:2021cox, CMS:2022sfi, CMS:2021edw}.

In summary if we want $M_L$ as small as possible in order to maximize the coupling deviation we have to take it at least $M_L \gtrsim \mathcal{O}(y v)$. This is not a particularly stringent constraint, as we could take either $y=0$ or $y^c=0$ and always have a lightest neutral state in the spectrum (as one can see from Eq.s~\eqref{eq:ML1} and~\eqref{eq:ML2}). However it is useful to keep it in mind when reading the plots, since the most important LHC bounds at low mass are just the condition $M_L > M_1$.

If we stick to the region of parameter space where the hierarchy between mass eigenstates is $M_2 > M_L > M_1$ the LHC does not strongly constrain the model, since we can have all three masses close enough to make the SM decay products hard to detect, without tuning our parameters. This, together with the relatively small cross sections of our doublet plus singlet model, makes current searches for vector-like leptons or electroweakinos insensitive to our new particles. In Fig.~\ref{fig:collider} we show a  summary of existing constraints. The purple bands show where $M_1 < m_Z/2$, at odds with the LEP measurement of the $Z$-width~\cite{ALEPH:2005ab}. The region in orange is where $M_L < M_1$ and we are excluded by stable charged particles searches at the LHC~\cite{CMS:2016kce, ATLAS:2019gqq, ATLAS:2022pib}. We see that this excluded region sets a bound on the largest coupling deviation that we can achieve, since parametrically it is similar to imposing $M_L \gtrsim y v$. The pink region is excluded by the ATLAS search for compressed Higgsinos~\cite{ATLAS:2021moa} and the green one by their more general searches for electroweakinos~\cite{ATLAS:2018eui, ATLAS:2019lng, ATLAS:2021moa, ATLAS:2021yqv, ATLAS:2020pgy, ATLAS:2021yqv, ATLAS:2019lff, ATLAS:2021yqv, ATLAS:2022zwa, ATLAS:2022hbt, CMS:2018szt, CMS:2020bfa, CMS:2021cox, CMS:2022sfi, CMS:2021edw}. 

The main qualitative message of Fig.~\ref{fig:collider} is that a light doublet+singlet system is still compatible with collider constraints. What really limits the maximal $|\delta \mu_{hVV}|$ are the bounds from electroweak precision measurements on oblique parameters. For $M_1 \geq 100$~GeV, these are well captured by the deviations in the Peskin-Takeuchi $S$ and $T$ parameters \cite{Peskin:1991sw} at $U=0$. To implement the constraint and check the size of $U$, we use the following standard definitions~\cite{Schwartz:2014sze},
\be
T&\equiv& \frac{1}{\alpha}\left(\frac{\Pi_{WW}^{\rm new}(0)}{m_W^2}-\frac{\Pi_{ZZ}^{\rm new}(0)}{m_Z^2}\right)\, , \nn \\
S&\equiv &\frac{4 c_W^2 s_W^2}{\alpha}\left(\frac{\Pi_{ZZ}^{\rm new}(m_Z^2)-\Pi_{ZZ}^{\rm new}(0)}{m_Z^2}-\frac{c_W^2-s_W^2}{c_W^2 s_W^2}\frac{\Pi_{\gamma Z}^{\rm new}(m_Z^2)}{m_Z^2}-\frac{\Pi_{\gamma \gamma}^{\rm new}(m_Z^2)}{m_Z^2}\right)\, , \nn \\
U&\equiv&\frac{4 c_W^2 s_W^2}{\alpha}\left(\frac{\Pi_{WW}^{\rm new}(m_W^2)-\Pi_{WW}^{\rm new}(0)}{m_W^2}-\frac{c_W}{s_W}\frac{\Pi_{\gamma Z}^{\rm new}(m_Z^2)}{m_Z^2}-\frac{\Pi_{\gamma \gamma}^{\rm new}(m_Z^2)}{m_Z^2}\right)-S\, ,
\ee
and compare the calculation with the measured $S-T$ ellipse ($95\%$ CL) from the {\tt Gfitter} collaboration~\cite{Haller:2018nnx}. The constraint is shown in gray in Fig.~\ref{fig:collider}. For smaller masses, $M_1 \leq 100$~GeV, the constraints from $U$ and other parameters relevant for light new physics ($V, W, X$ and $Y$~\cite{Maksymyk:1993zm, Barbieri:2004qk}) become important, but do not affect qualitatively the constraints that we show\footnote{For these small masses, we used the more conservative $S-T$ ellipse ($95\%$ CL) from the {\tt Gfitter} collaboration for $U \neq 0$ \cite{Baak:2012kk}.}. We leave to future work a more complete calculation of these bounds, in the hope that deviations in $hWW$ and $hZZ$ are discovered.

The situation for higher $SU(2)_L$ representations or larger $Y$ is different. In this case also LHC constraints have an important role to play. Searches for electroweakinos~\cite{ATLAS:2018eui, ATLAS:2019lng, ATLAS:2021moa, ATLAS:2021yqv, ATLAS:2020pgy, ATLAS:2021yqv, ATLAS:2019lff, ATLAS:2021yqv, ATLAS:2022zwa, ATLAS:2022hbt, CMS:2018szt, CMS:2020bfa, CMS:2021cox, CMS:2022sfi, CMS:2021edw} can exclude our doublet+triplet model for $M_L\lesssim 240$~GeV, for mass splittings as small as $8$~GeV (Fig.~16 in~\cite{ATLAS:2021moa}). Hypercharge assignments that do not allow for a lightest neutral state (i.e. $Y\neq \pm (n-1)/2$ with $n$ the dimension of the largest $SU(2)_L$ representation) lead to much stronger constraints, comparable or more stringent than those discussed for a stable particle of $Q=1$ from a $SU(2)_L$ doublet. For higher $SU(2)_L$ representations we take conservatively the exclusion on the doublet+triplet model as a benchmark for the lightest mass that we should show in the plots, but show the results also for larger values of $M_1$.

In practice, in the next Section we include the constraints from EWPTs as shaded areas in the plots for $\Lambda_B$, showing the result also for regions that are excluded in the simplest models. We implement the collider constraints by limiting the values of $M_1$ (the lightest mass of the new fermions) that we show in the plots for $\Lambda_B$, but we always compute the value of $\Lambda_B$ for multiple choices of $M_1$, also much above the most conservatives constraints. 

\subsection{Results}\label{sec:final_results}
The HL-LHC is going to measure $\delta\mu_{hVV}$ at the 1.5\% level in the $WW$, $ZZ$ and $\gamma\gamma$ channels~\cite{Cepeda:2019klc}. Future lepton colliders (ILC~\cite{ILCInternationalDevelopmentTeam:2022izu}, CLIC~\cite{Robson:2018zje}, FCC-$ee$~\cite{Bernardi:2022hny}, CEPC~\cite{CEPCPhysicsStudyGroup:2022uwl}, MuC~\cite{Forslund:2022xjq, deBlas:2022aow}), in particular FCC-$ee$ and the muon collider, have the potential to reach a precision of $\simeq 0.13\%$~\cite{deBlas:2022ofj} comparable to that of LEP on $Z$ couplings. In the plots we show lines corresponding to the $1\sigma$ sensitivity in the $\kappa$-framework for HL-LHC~\cite{Cepeda:2019klc} and in the $g^{\rm eff}$-framework of~\cite{deBlas:2022ofj} for future colliders. These lines are meant mostly to guide the eye, if a deviation is ever found we will conduct a more thorough study (for instance map our results onto higher-dimensional operators and include in the analysis their correlation matrix).

We find that \% level deviations measurable at HL-LHC cannot be generated by perturbative theories containing only new fermions. Future lepton colliders can probe deviations small enough to push $\Lambda_B$ almost to the GUT scale, but only for extremely light new fermions, with masses $\simeq 50$~GeV, within reach of HL-LHC. If the new fermions are heavier, we find that new bosons are kinematically within reach of future hadron colliders or at most around $100$~TeV if the new fermions are lighter than 150 GeV. Even if it is implicitly obvious, it is important to explicitly point out that new bosons must appear below $\Lambda_B$. For example $\Lambda_B \simeq 10\, M$, with $M$ a typical mass for the new fermions, means that new bosons must already exist at the same scale as the new fermions, otherwise they do not have a chance to cure the instabilities of the running of the fermionic theory. 

In the rest of this Section we show plots of $\Lambda_B$ as a function of $\delta \mu_{hVV}$ at fixed values of the lightest new fermions mass $M_1$. The values of $M_1$ that we plot reflect the discussion of experimental bounds in the previous Section. For example, we mentioned how ATLAS and CMS exclude models almost identical to our doublet+triplet scenario up to $M_1\simeq240$~GeV, even for mass splittings as small as $8$~GeV. In our $\Lambda_B$ plots for the doublet+triplet model we conservatively allow $M_1$ to go down to 200~GeV. We do this because a more thorough analysis of experimental constraints might unveil (tuned) regions of parameter space where the lower bounds on $M_1$ discussed in the previous Section are relaxed. We also show larger values of $M_1$ in line with the bounds.

In the title of each plot of this Section we specify the relation between $y$ and $y^c$ used to compute $\Lambda_B$. We choose their relative value to maximize the cutoff. In the vast majority of cases this corresponds to $y=\pm y^c$. At very light $M_1$ and large $\delta \mu$ also the $y^c=0$ case becomes relevant.

To conclude this Section we show the value of $\Lambda_B$ as a function of deviations in $h\gamma\gamma$. All the fermionic theories that we consider induce a deviation measurable at HL-LHC, with the only exception of the doublet+singlet theory with a neutral singlet. We show these results for completeness, since they might be the first experimental sign of the vector-like leptons that we discuss in this work.

\subsubsection{$WW$ and $ZZ$}

\begin{figure}[!t]
\centering
\includegraphics[width=0.49\textwidth]{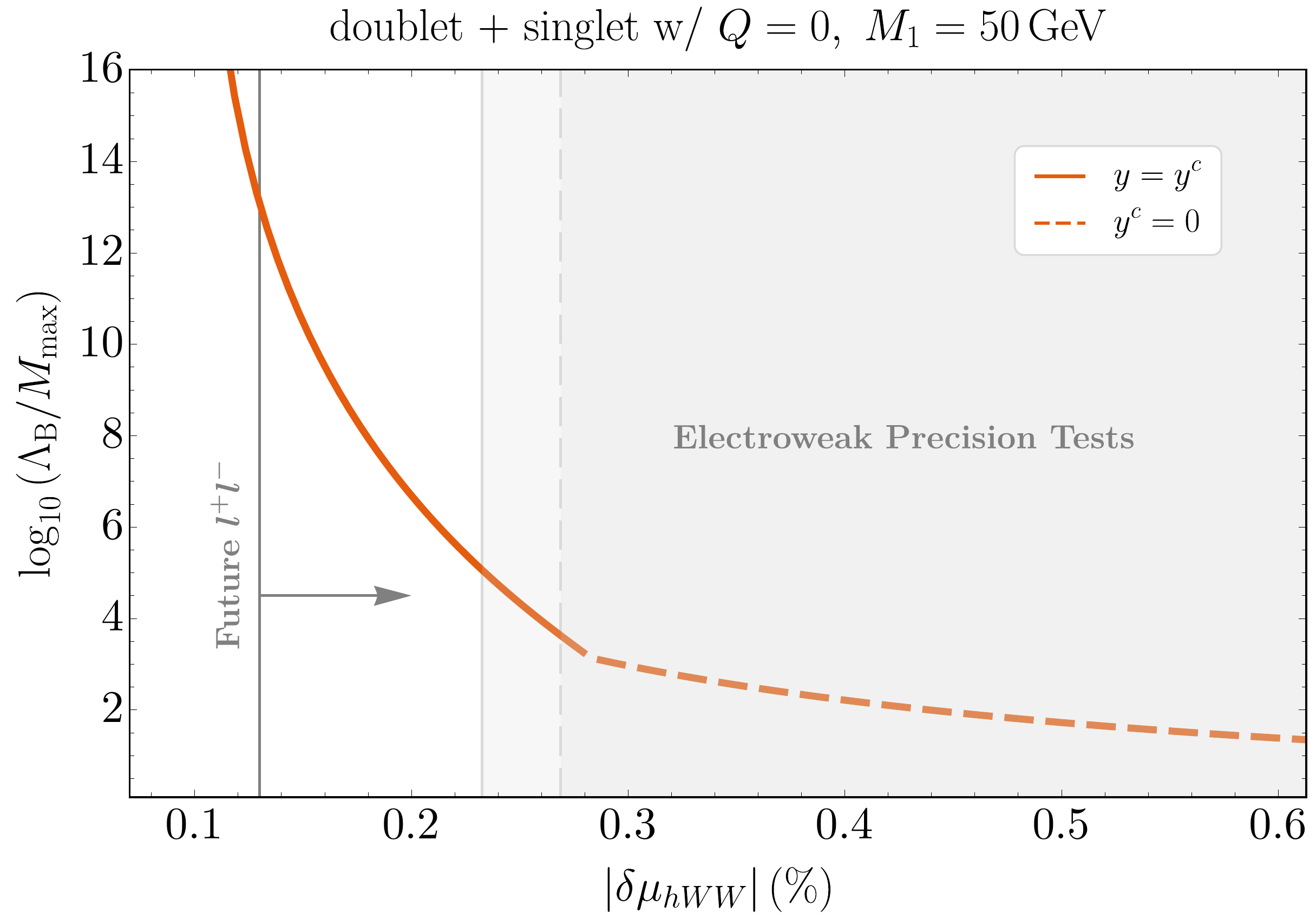}
\includegraphics[width=0.49\textwidth]{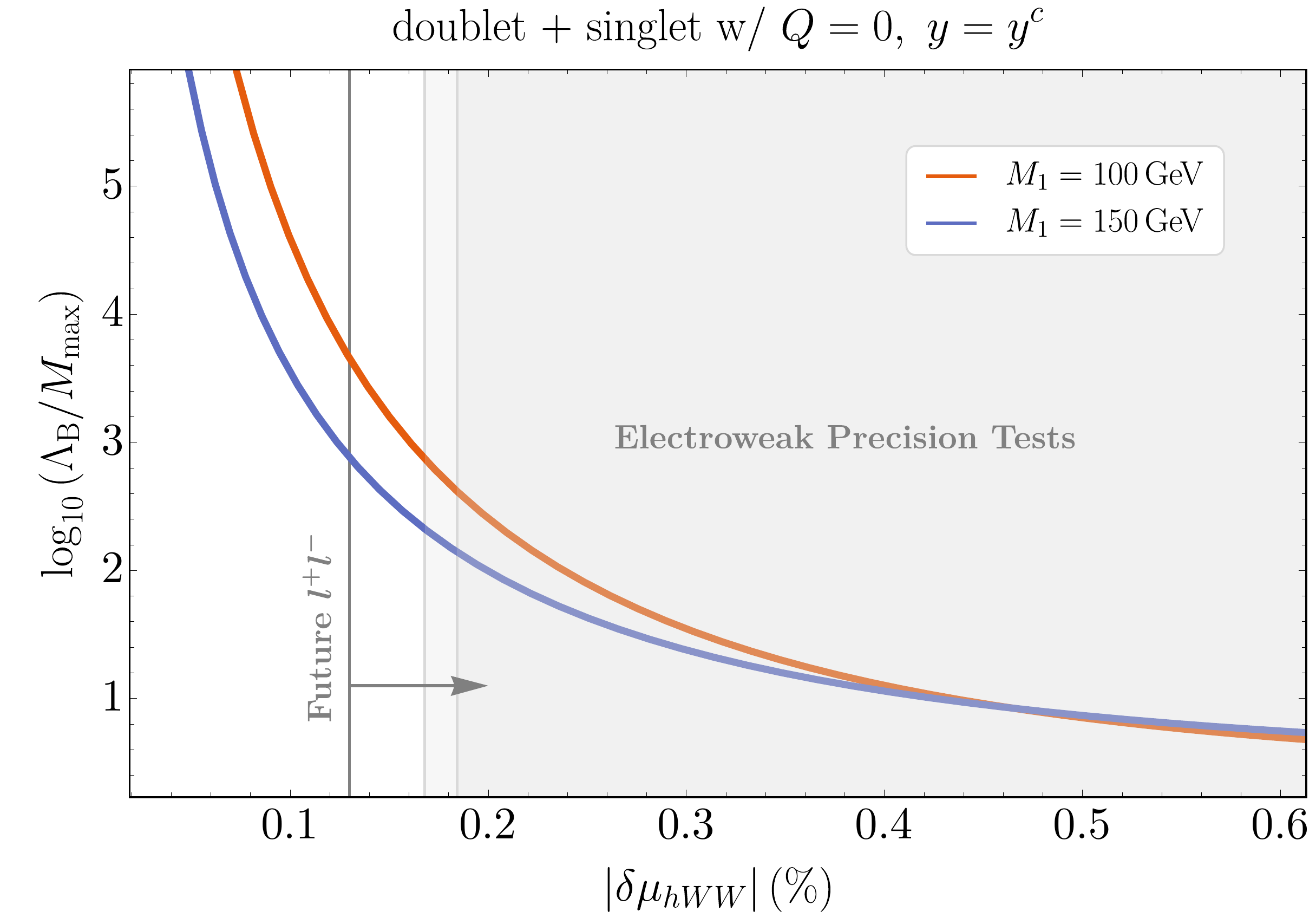}
\includegraphics[width=0.49\textwidth]{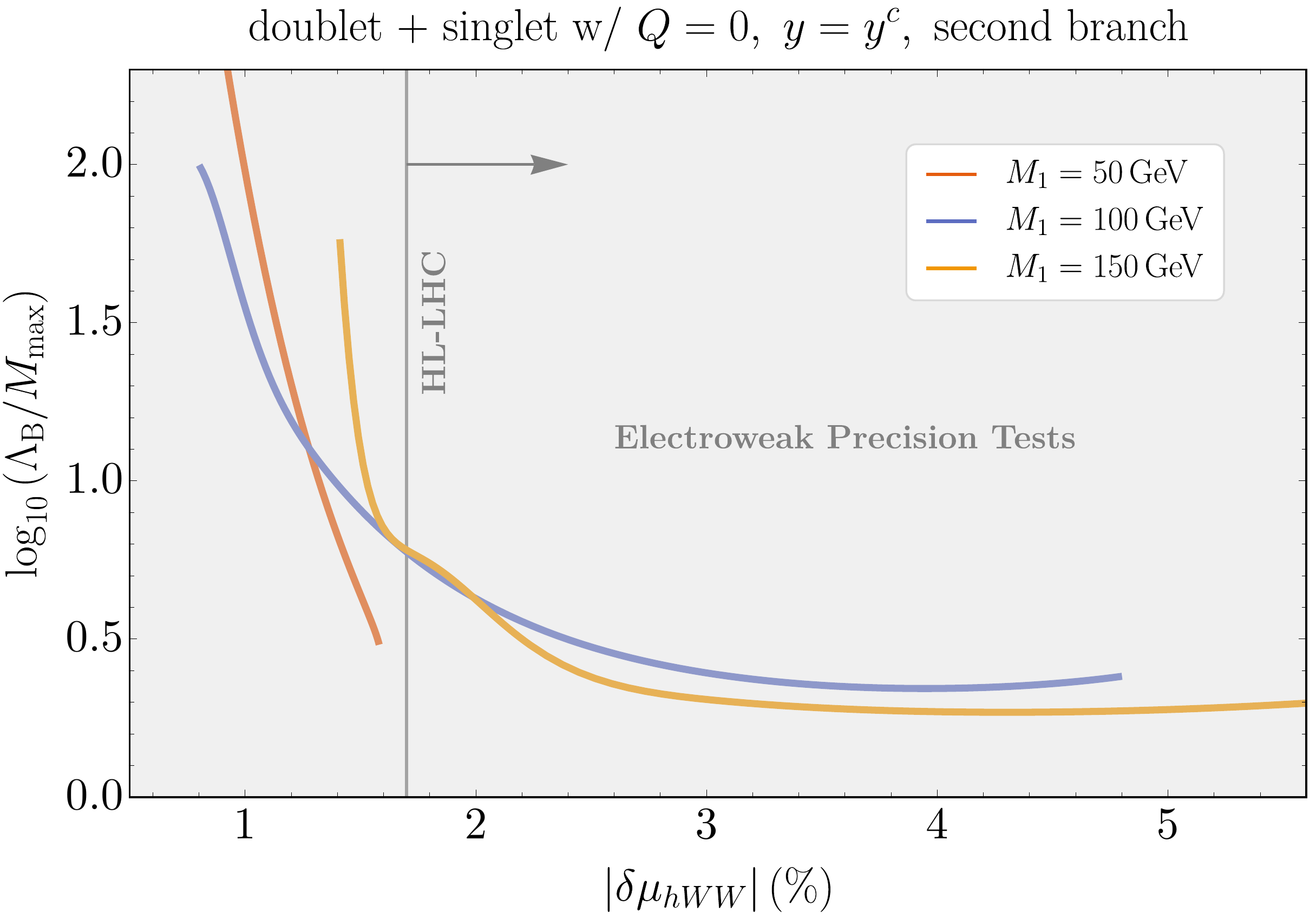}
\includegraphics[width=0.49\textwidth]{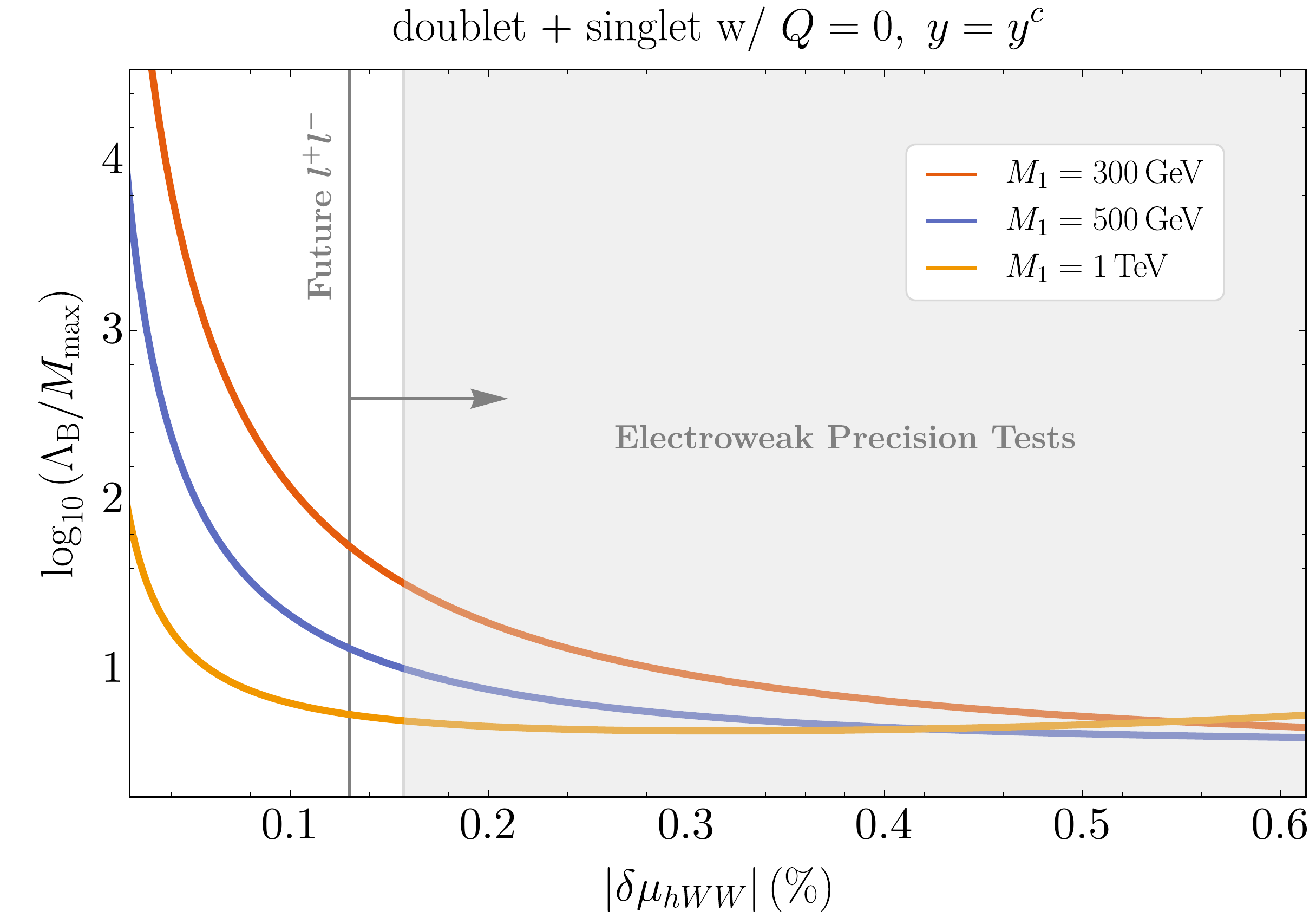}
\caption{Instability scale of the RGEs $\Lambda_B$ (i.e. upper bound on the scale of new bosons) as a function of relative $hWW$ coupling deviation in the doublet+singlet model defined in Section~\ref{sec:4f}. $M_{\rm max}$ on the $y$-axis is the largest of the new fermions masses, while $M_1$ is the smallest one. The smallest values of $M_1$ in the plots reflect a conservative estimate of collider constraints on the models. The gray shaded areas represent the constraint from EWPTs. In the bottom right panel the constraint is the same for all values of $M_1$. In the top right panel the EWPT constraint at lowest $\delta \mu$ is on the line with lowest $M_1$. The constraint gets monotonically weaker at higher masses.  The choice $y=y^c$ indicated in the title of each plot maximizes $\Lambda_B$, except for the first panel, where $y^c=0$ gives a larger $\Lambda_B$ for $\delta \mu_{hWW}\gtrsim 0.3\%$. The charge $Q$ in the title refers to the singlet. The bottom left panel is the only Figure in the paper where we picked the solution $M=M_1-y v/\sqrt{2}$. For larger values of $M_1$ or smaller values of $\delta \mu$ the only existing solution is $M=M_1+y v/\sqrt{2}$. We stop the $M_1=50$~GeV line at the value of $y$ beyond which $\delta \mu$ starts decreasing. For such small $M_1$ different terms in the coupling deviation are comparable and start to cancel.}
\label{fig:WWDS}
\end{figure}

\begin{figure}[!t]
\centering
\includegraphics[width=0.49\textwidth]{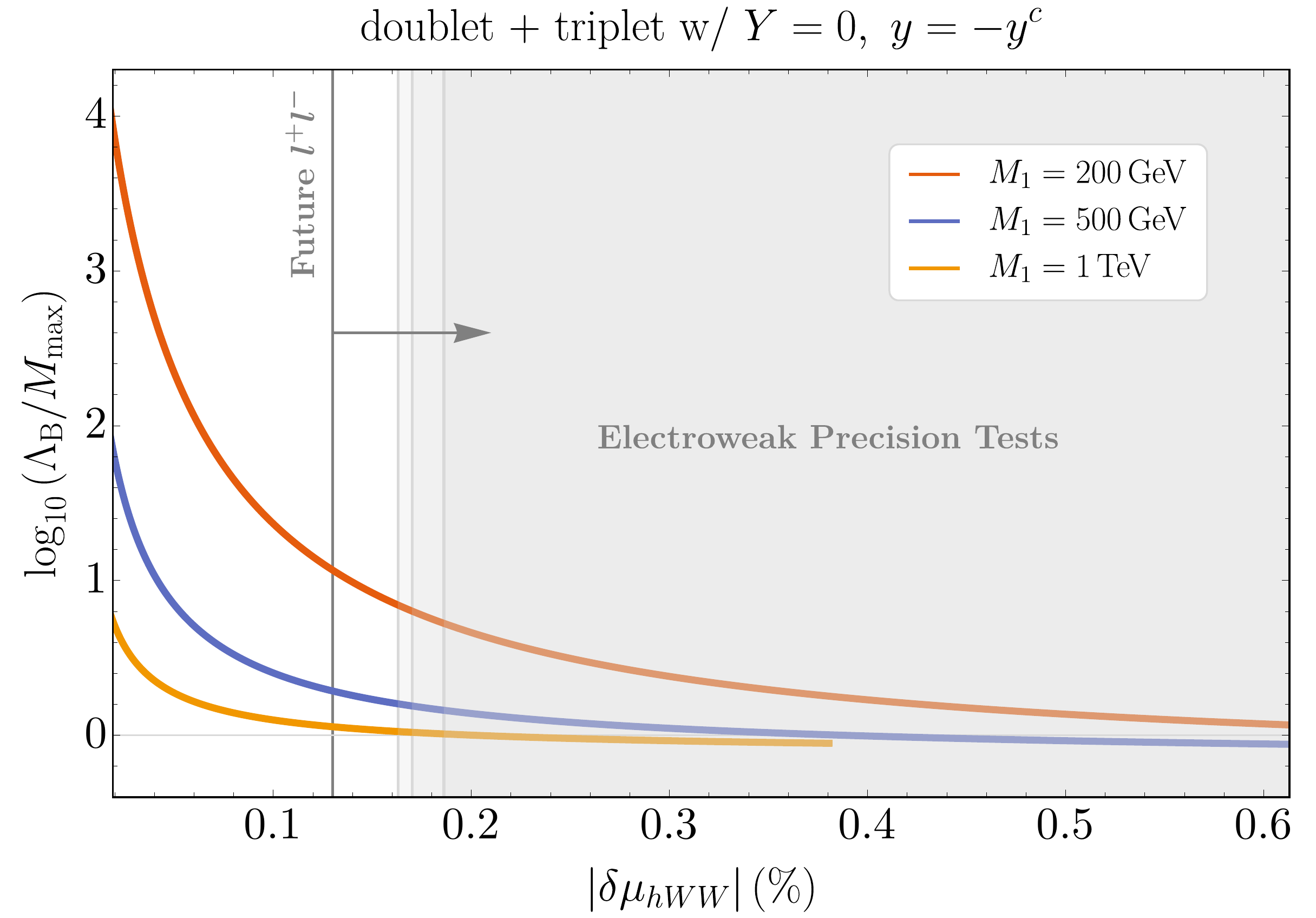}
\includegraphics[width=0.49\textwidth]{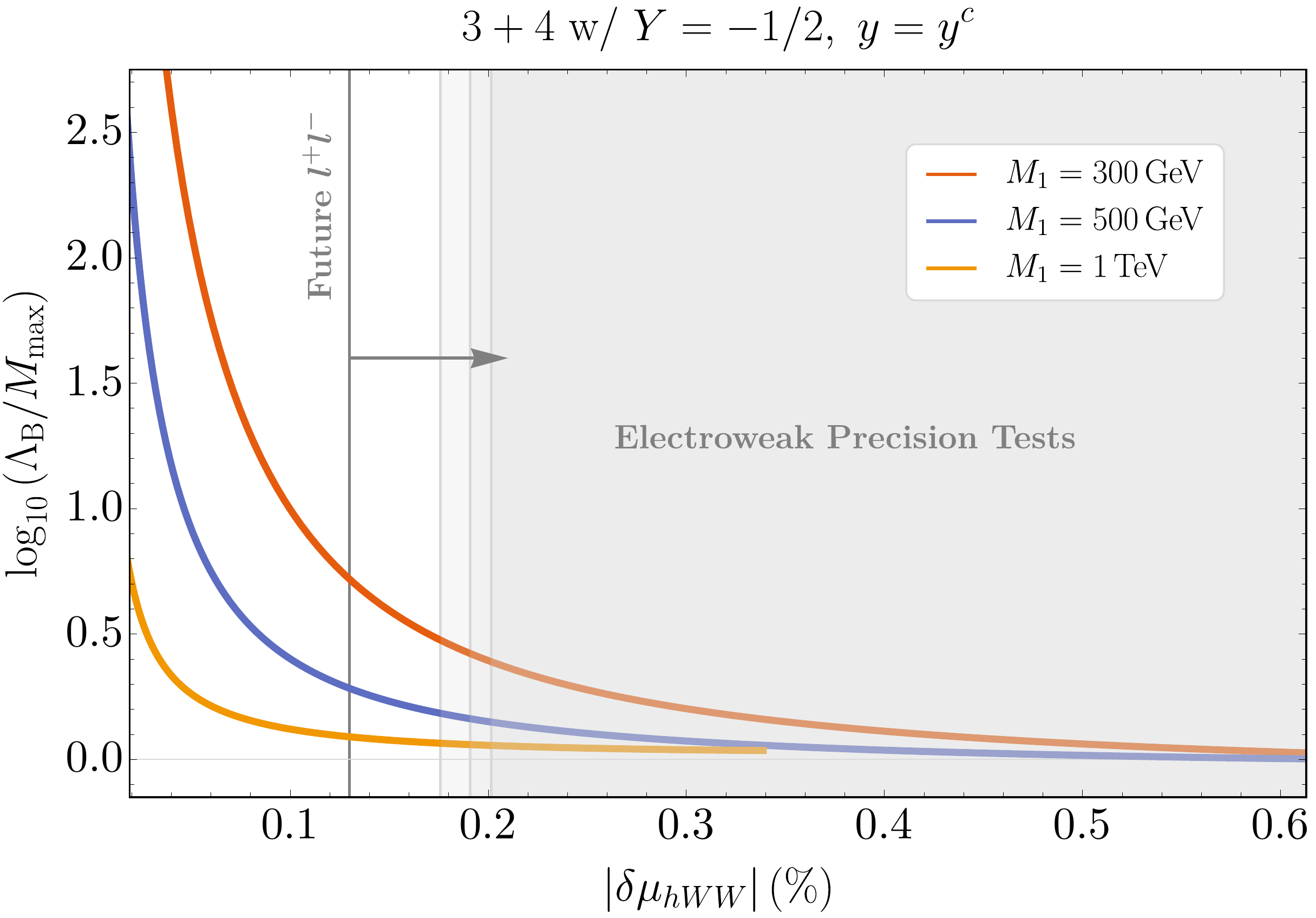}
\includegraphics[width=0.49\textwidth]{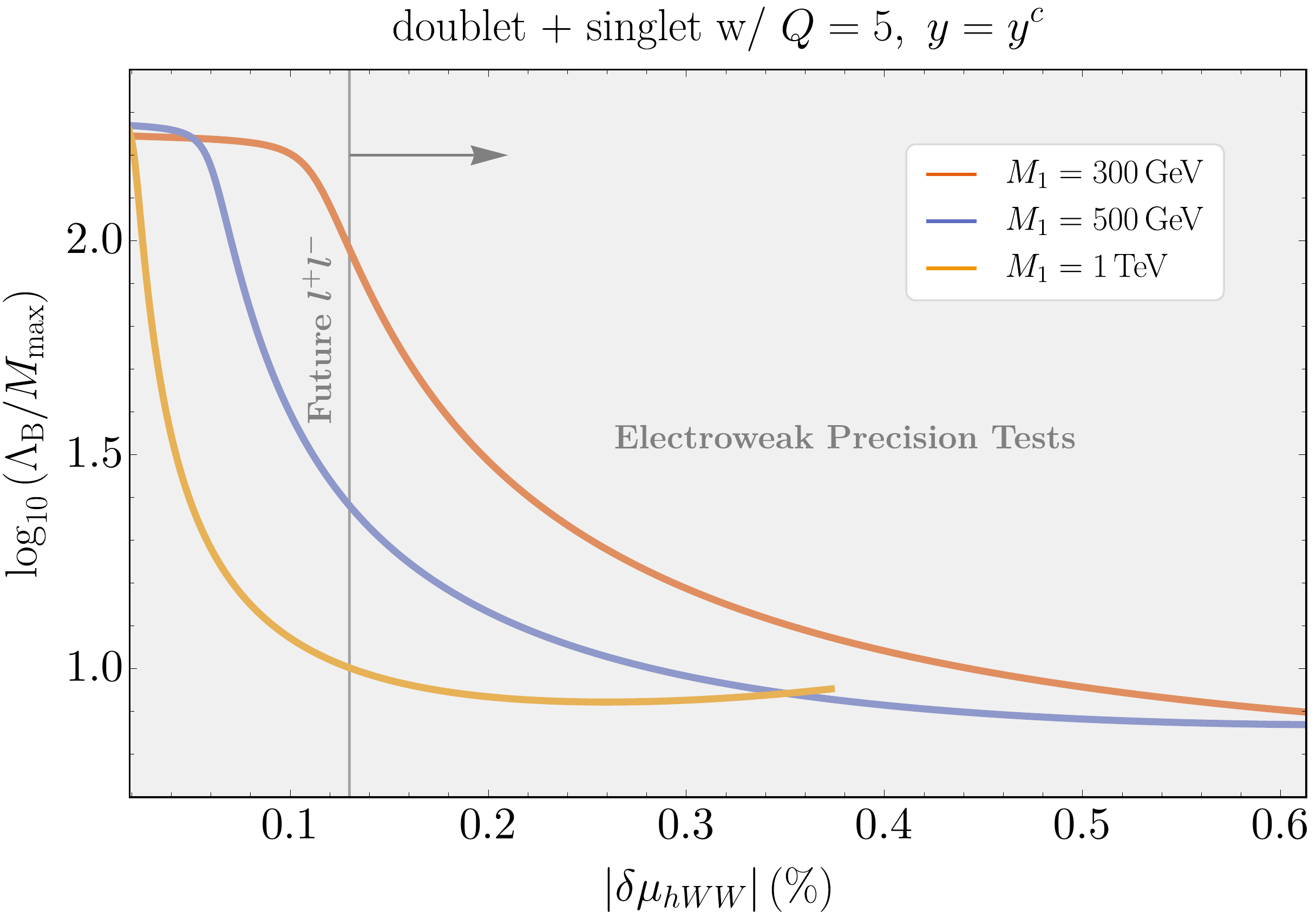}
\caption{Instability scale of the RGEs $\Lambda_B$ (i.e. upper bound on the scale of new bosons) as a function of relative $hWW$ coupling deviation in the vector-like lepton models defined in Section~\ref{sec:4f}. $M_{\rm max}$ on the $y$-axis is the largest of the new fermions masses, while $M_1$ in the legend is the smallest one. The smallest values of $M_1$ in the plot reflect a conservative estimate of collider constraints on the models. The gray shaded areas represent the constraint from EWPTs. The EWPT constraint at lowest $\delta \mu$ is on the line with lowest $M_1$. The constraint gets monotonically weaker at higher masses. The choice $y=\pm y^c$ indicated in the title of each plot maximizes $\Lambda_B$.}
\label{fig:WW}
\end{figure}

In Fig.s~\ref{fig:WWDS} and~\ref{fig:WW} we show the value of $\Lambda_B$ as a function of the relative $hWW$ coupling deviation. In all cases we find that $\Lambda_B$ is determined by the instability of the Higgs potential, which occurs before any Landau pole. The only exceptions are models with large hypercharge ($Q=5$ for the singlet), where we see a scale of the instability independent of $\delta \mu$ at small values of the coupling deviation. In these regions the first problem with the theory is a hypercharge Landau pole.

 We also find, regardless of the masses or representations of the new fermions, that coupling deviations measurable at HL-LHC are in tension with EWPTs (area shaded in gray). We stop the lines in the plot where we lose perturbative control of the theory (the Higgs quartic becomes rapidly large and negative after the threshold for vacuum decay). Larger values of $M_1$ correspond to larger Yukawas at fixed $\delta \mu$ and we stop the lines at lower values of $\delta \mu$. 

The general message that all Figures give is that we lose control of the theory ($\Lambda_B$ a factor of a few to ten above the heaviest fermion mass $M_{\rm max}$) long before we can have a deviation large enough for HL-LHC. The only exception is the bottom left panel of Fig.~\ref{fig:WWDS}. There we see that for $M_1 \leq 150$~GeV we can generate a $\delta \mu$ within reach of HL-LHC, however in this case the new bosons must exist at the same scale as the fermions. This plot is made for $M=\frac{y v}{\sqrt{2}}-M_1$. This solution for $M$ exists only at large enough $y$ and small enough $M_1$. In all other plots we take $M=\frac{y v}{\sqrt{2}}+M_1$, because the other solution does not exist for Yukawa couplings $\lesssim \mathcal{O}(1)$. In the $M=\frac{y v}{\sqrt{2}}-M_1$ plot, labelled ``second branch", we stop the $M_1=50$~GeV line at the value of $y$ beyond which $\delta \mu$ starts decreasing. For such small $M_1$ different terms in the coupling deviation are comparable and start to cancel, giving much smaller values of $\Lambda_B$ at fixed $\delta \mu$ compared to larger values of $M_1$.

We conclude that seeing a modification of $hWW$ at HL-LHC requires new bosons at the scale where the deviation is generated. This conclusion is  strengthened by the bound from EWPTs (gray shaded area in the Figures) that excludes coupling deviations larger than the few permille level. Note that even if this result appears quite strong, it is not guaranteed that the new bosons can be produced at HL-LHC. For example, a strongly coupled ($g_* \simeq 4\pi)$ new boson that can affect $hWW$ at tree level, can be roughly as heavy as $M_B \simeq 4 \pi v/\sqrt{\delta \mu_{hWW}} \simeq 13 $~TeV~$\times(\sqrt{3\%/\delta \mu_{hWW}})$.

Coupling deviations as small as those detectable at future lepton colliders give in most cases $\Lambda_B \lesssim 100$~TeV, implying the existence of new bosons well below this scale, potentially within reach of future hadron colliders. The main exception are light doublet+singlet fermions, where the new bosons can be as heavy as the GUT scale. However in this case we will detect vector-like fermions with masses $M_1 \lesssim 50$~GeV already at HL-LHC. 

One last general point common to all Figures is that the EWPTs bound is rather insensitive to $M_1$ (in the bottom right panel of Fig.~\ref{fig:WWDS} it is even the same for the three values of $M_1$ in the Figure). This is due to the fact that the bound is on almost the same combination of couplings and masses that enter $\delta \mu$, so at fixed $\delta \mu$ it is almost independent of $M_1$. This is illustrated schematically in Fig.~\ref{fig:ST}.

\begin{figure}[!t]
\centering
\includegraphics[width=0.49\textwidth]{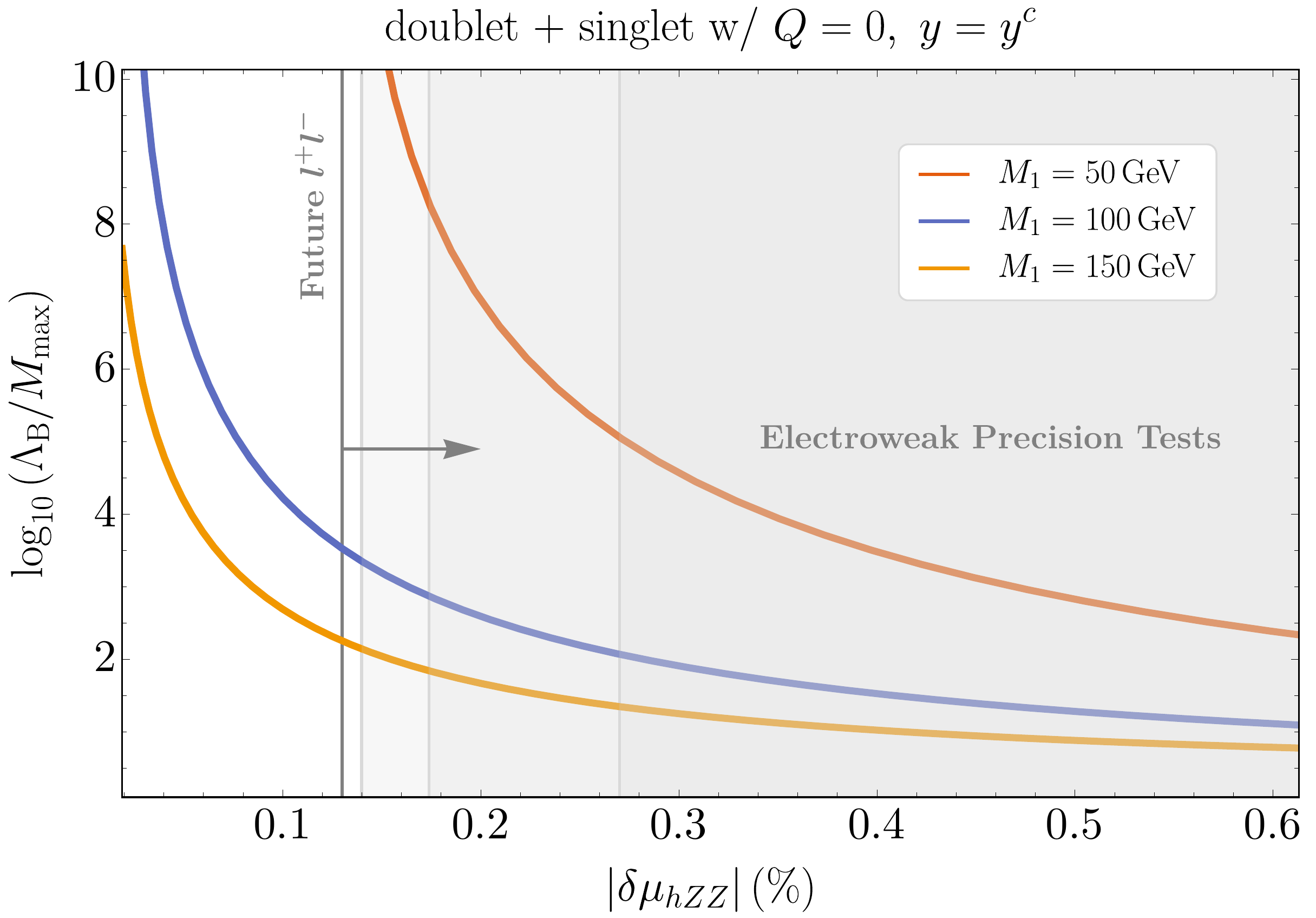}
\includegraphics[width=0.49\textwidth]{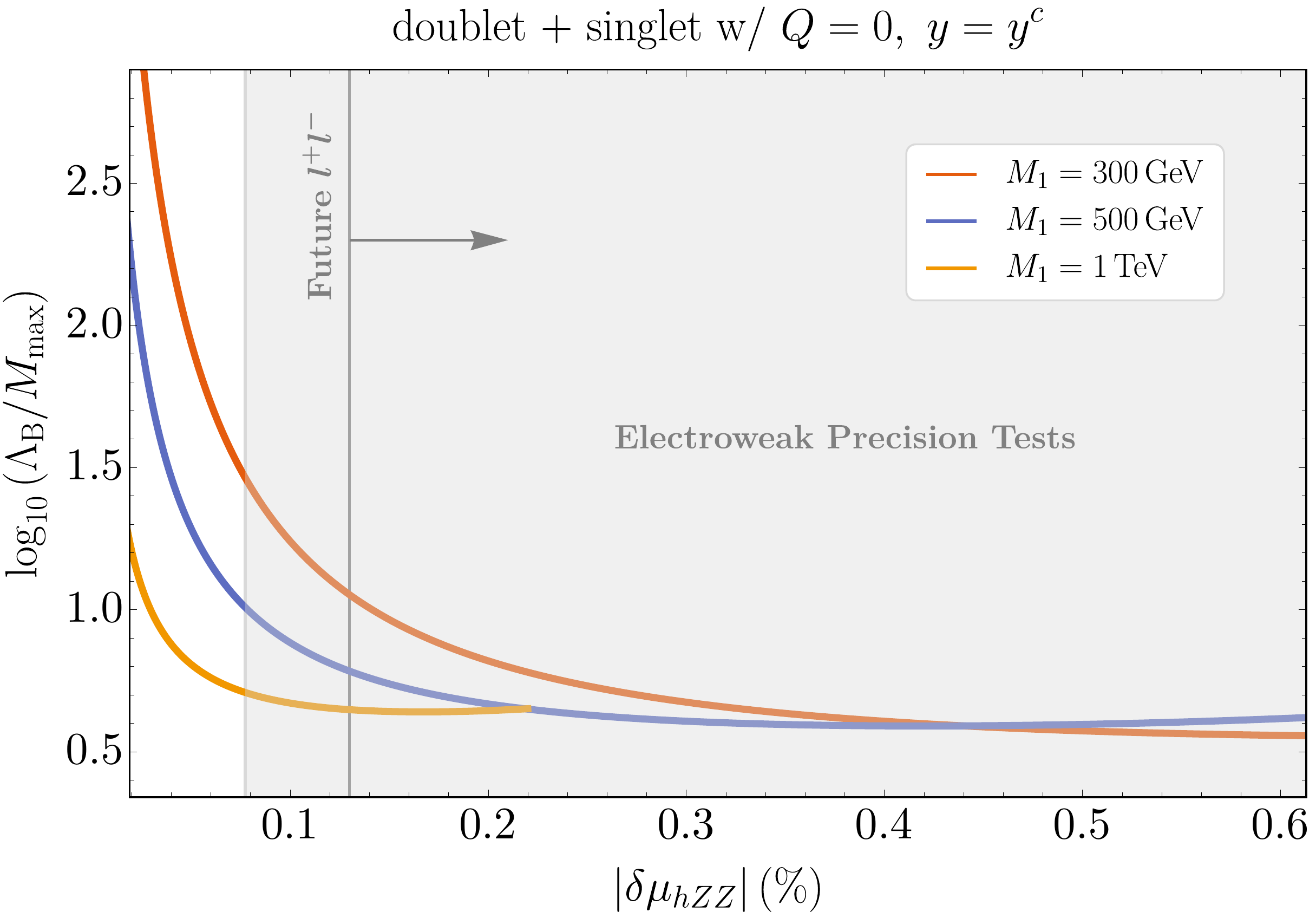}
\caption{Instability scale of the RGEs $\Lambda_B$ (i.e. upper bound on the scale of new bosons) as a function of relative $hZZ$ coupling deviation in the doublet+singlet model defined in Section~\ref{sec:4f}. $M_{\rm max}$ on the $y$-axis is the largest of the new fermions masses, while $M_1$ is the smallest. The smallest values of $M_1$ reflect a conservative estimate of collider constraints on the models.
The gray shaded areas represent the constraint from EWPTs. In the right panel the constraint is the same for all values of $M_1$. In the left panel the EWPT constraint at lowest $\delta \mu$ is on the line with lowest $M_1$. The constraint gets monotonically weaker at higher masses. The choice $y=y^c$ indicated in the title of each plot maximizes $\Lambda_B$.}
\label{fig:ZZDS}
\end{figure}

\begin{figure}[!t]
\centering
\includegraphics[width=0.49\textwidth]{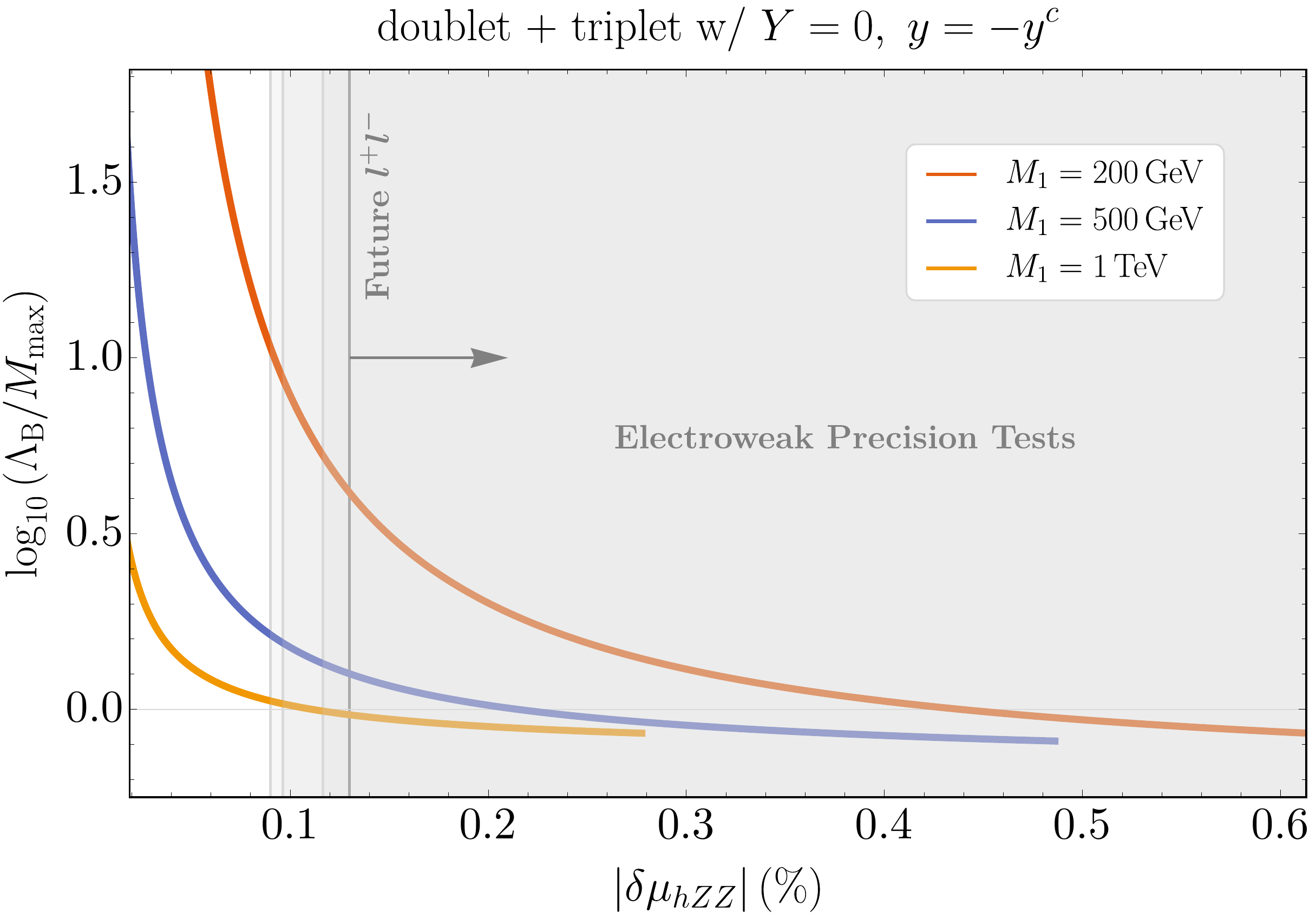}
\includegraphics[width=0.49\textwidth]{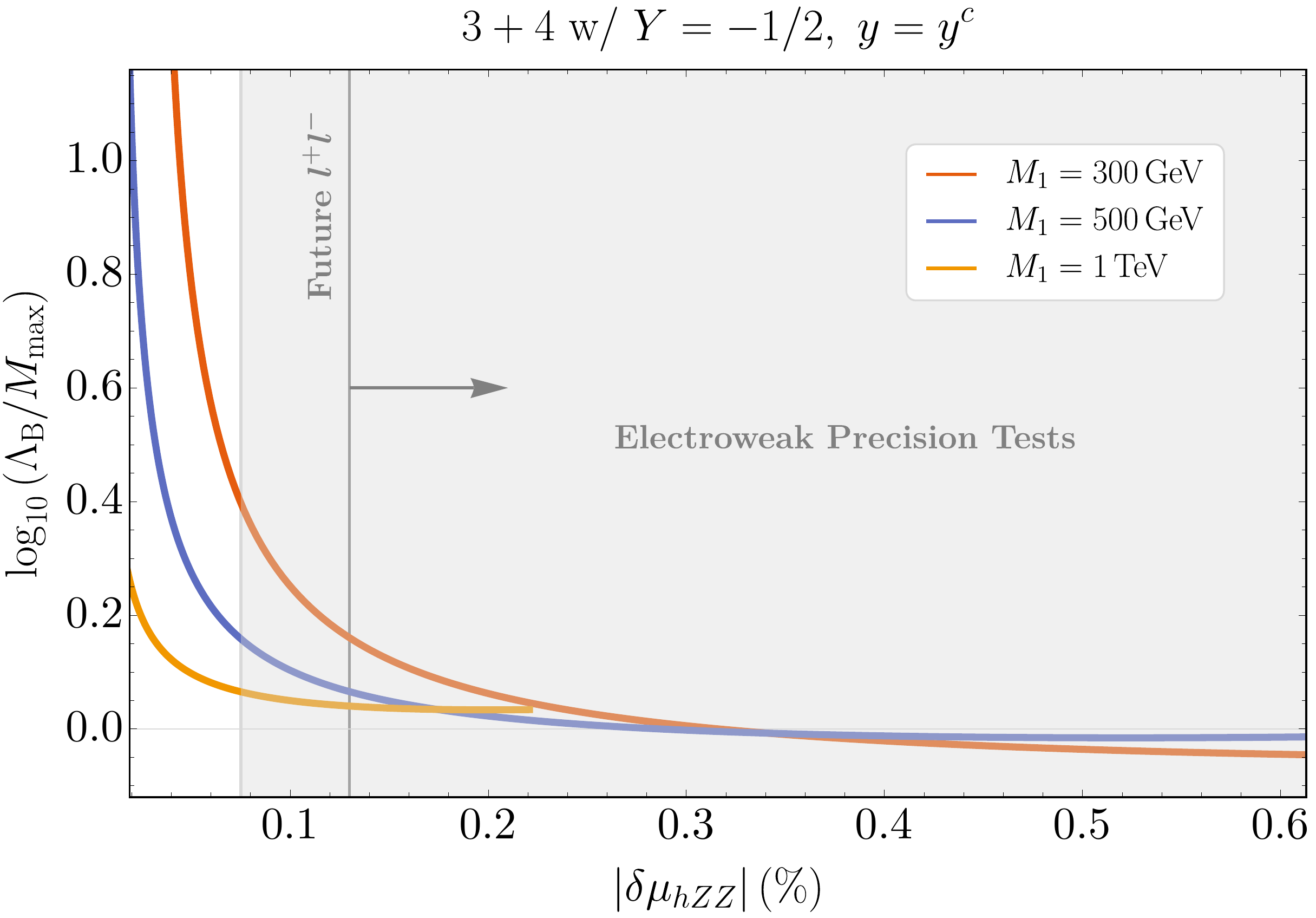}
\includegraphics[width=0.49\textwidth]{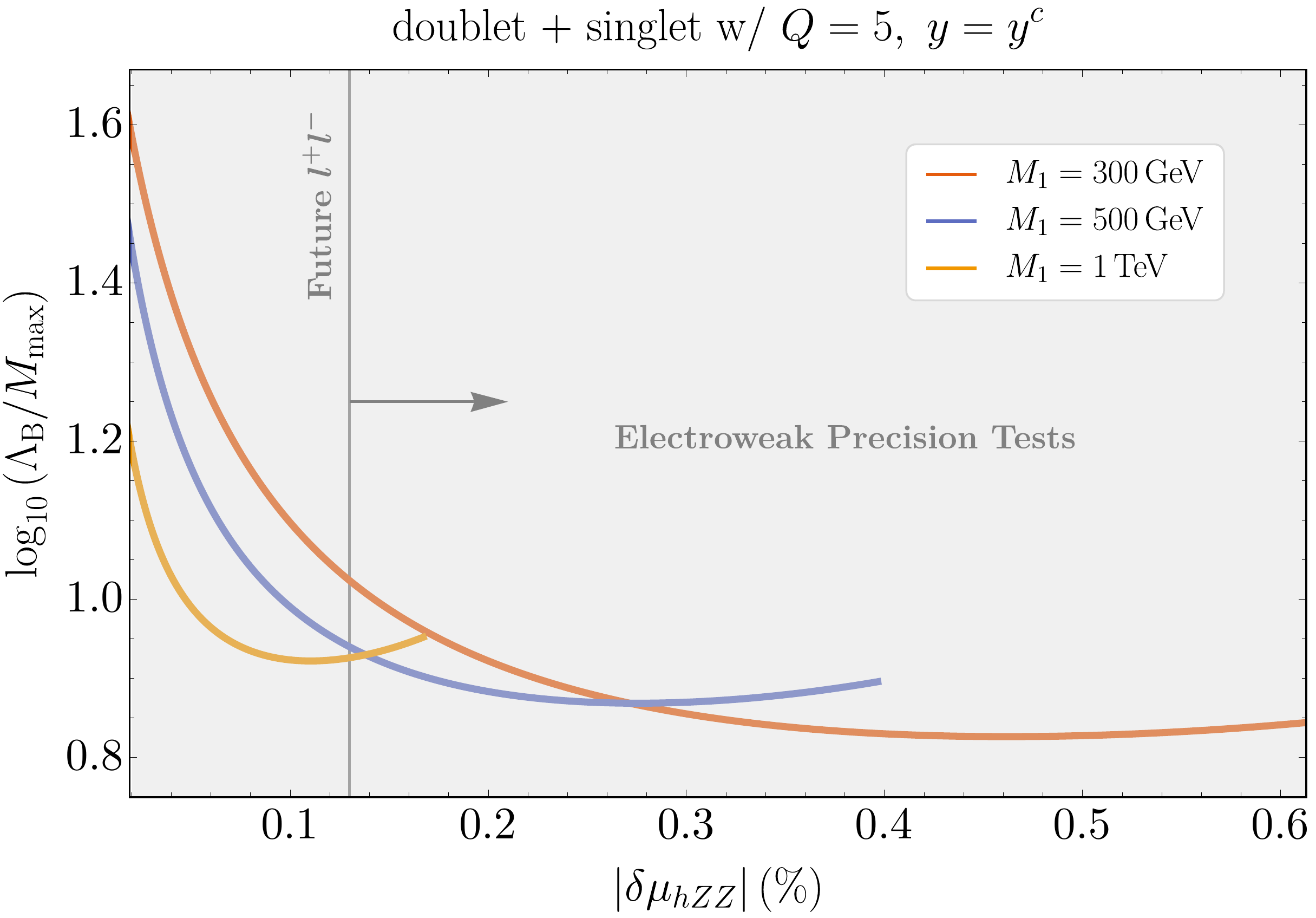}
\caption{Instability scale of the RGEs $\Lambda_B$ (i.e. upper bound on the scale of new bosons) as a function of relative $hZZ$ coupling deviation in the vector-like lepton models defined in Section~\ref{sec:4f}. $M_{\rm max}$ on the $y$-axis is the largest of the new fermions masses, while $M_1$ is the smallest. The smallest values of $M_1$ reflect a conservative estimate of collider constraints on the models. The gray shaded areas represent the constraint from EWPTs. In the right panel the constraint is the same for all values of $M_1$. In the left panel the EWPT constraint at lowest $\delta \mu$ is on the line with lowest $M_1$. The constraint gets monotonically weaker at higher masses. The choice $y=\pm y^c$ indicated in the title of each plot maximizes $\Lambda_B$.}
\label{fig:ZZ}
\end{figure}

It is interesting to compare Fig.~\ref{fig:WWDS}, where we show $\Lambda_B$ for the doublet+singlet model with a neutral singlet, and Fig.~\ref{fig:WW}, where we show $\Lambda_B$ for higher $SU(2)_L$ representations and hypercharges. Not surprisingly, scenarios where the new fermions can be lighter (namely the doublet+singlet model) have the largest value of $\Lambda_B$ for a given coupling deviation. This can be seen by comparing the first three panels of Fig.~\ref{fig:WWDS} with all other subfigures. 

What was not completely obvious from our simple arguments in Section~\ref{sec:SU2} is that at fixed $M_1$ (lightest new fermion mass) the value of $\Lambda_B$ does not vary greatly between different representations. This emerges from the comparison of the bottom right panel of Fig.~\ref{fig:WWDS} to the three panels of Fig.~\ref{fig:WW}. In these subfigures the $y^4$ term in Eq.~\eqref{eq:su2n} dominates the coupling deviation and we find the instability of the Higgs potential to determine $\Lambda_B$ in all Figures (except at $Q=5$ and small $\delta \mu$ where the hypercharge Landau pole dominates). The rescaling $y^{(c)}\to n^{1/4} y^{(c)}_n$ makes both the instability of the Higgs potential and the coupling deviation at fixed masses independent of $n$. We do not show $\Lambda_B$ for the last few fermionic theories that barely avoid a low energy Landau pole (i.e. a vector-like 5 of $SU(2)_L$ plus a vector-like 4, and above) since it is even smaller than that shown in Fig.~\ref{fig:WW}.

To conclude the discussion of $hWW$ it is worth commenting on the regions where the cutoffs slightly increase and where it appears to be independent of $\delta \mu$. The latter case corresponds to a Landau pole in the hypercharge. This shows that already at $Y=5$ this can be the dominant effect. The increase at large $\delta \mu$ is close to where we lose control of the theory, but it is still a perturbative effect. It arises from the large threshold correction to $\lambda$ at the matching scale between SM and SM+new fermions, which is proportional to $y^4$ and positive. 

The case of $hZZ$ is illustrated in Fig.s~\ref{fig:ZZDS} and~\ref{fig:ZZ} and is very similar to $hWW$. All the qualitative statements made for $hWW$ hold also in this case, with slightly lower $\Lambda_B$ for any given $\delta \mu$. New bosons are responsible for any deviation visible at HL-LHC and even permille level deviations require light new bosons. The behavior of the $\Lambda_B$ vs $\delta \mu$ curves is explained by the same arguments given for $hWW$.

\subsubsection{$\gamma\gamma$}\label{sec:photon_results}

\begin{figure}[!t]
\centering
\includegraphics[width=0.49\textwidth]{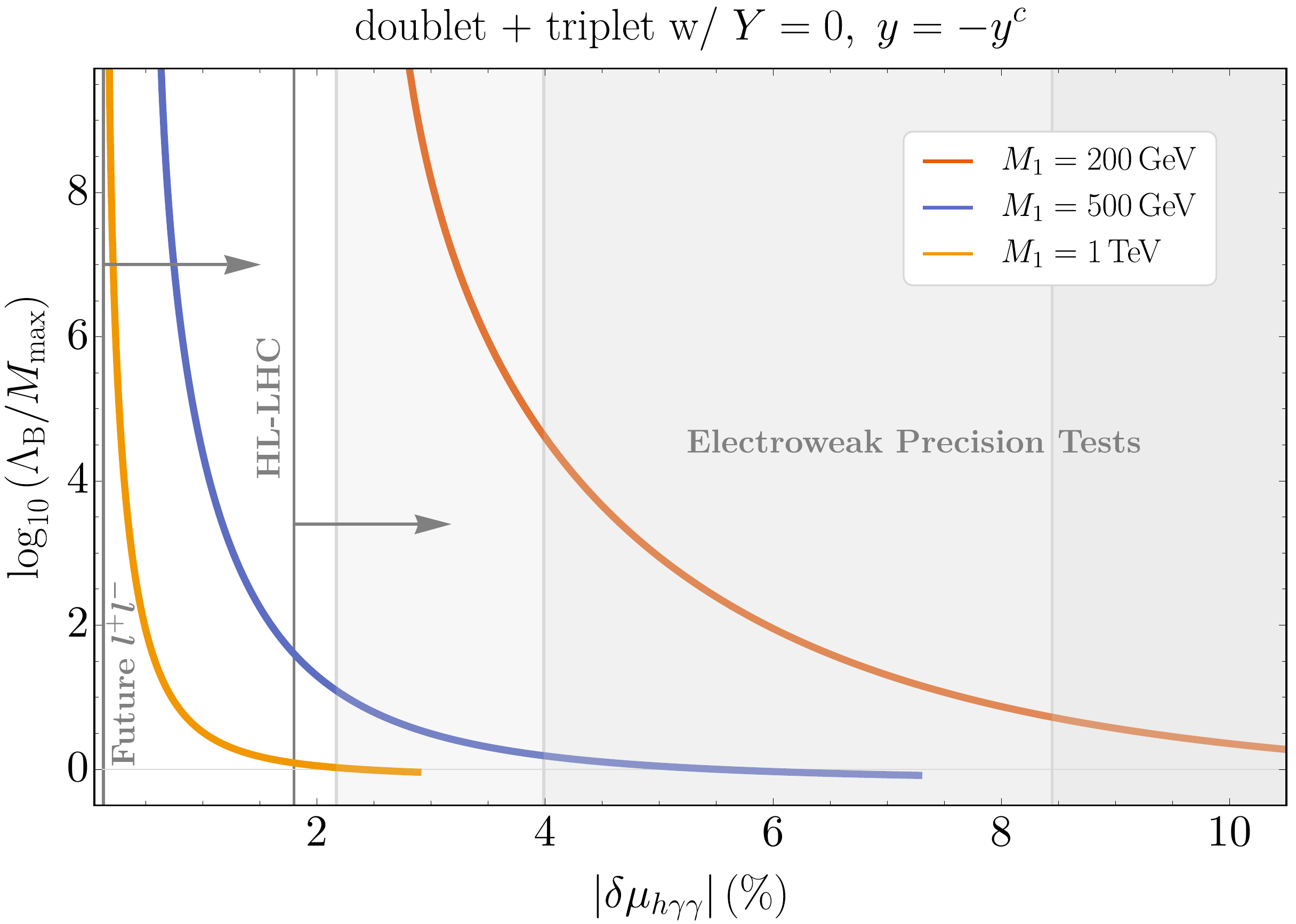}
\includegraphics[width=0.49\textwidth]{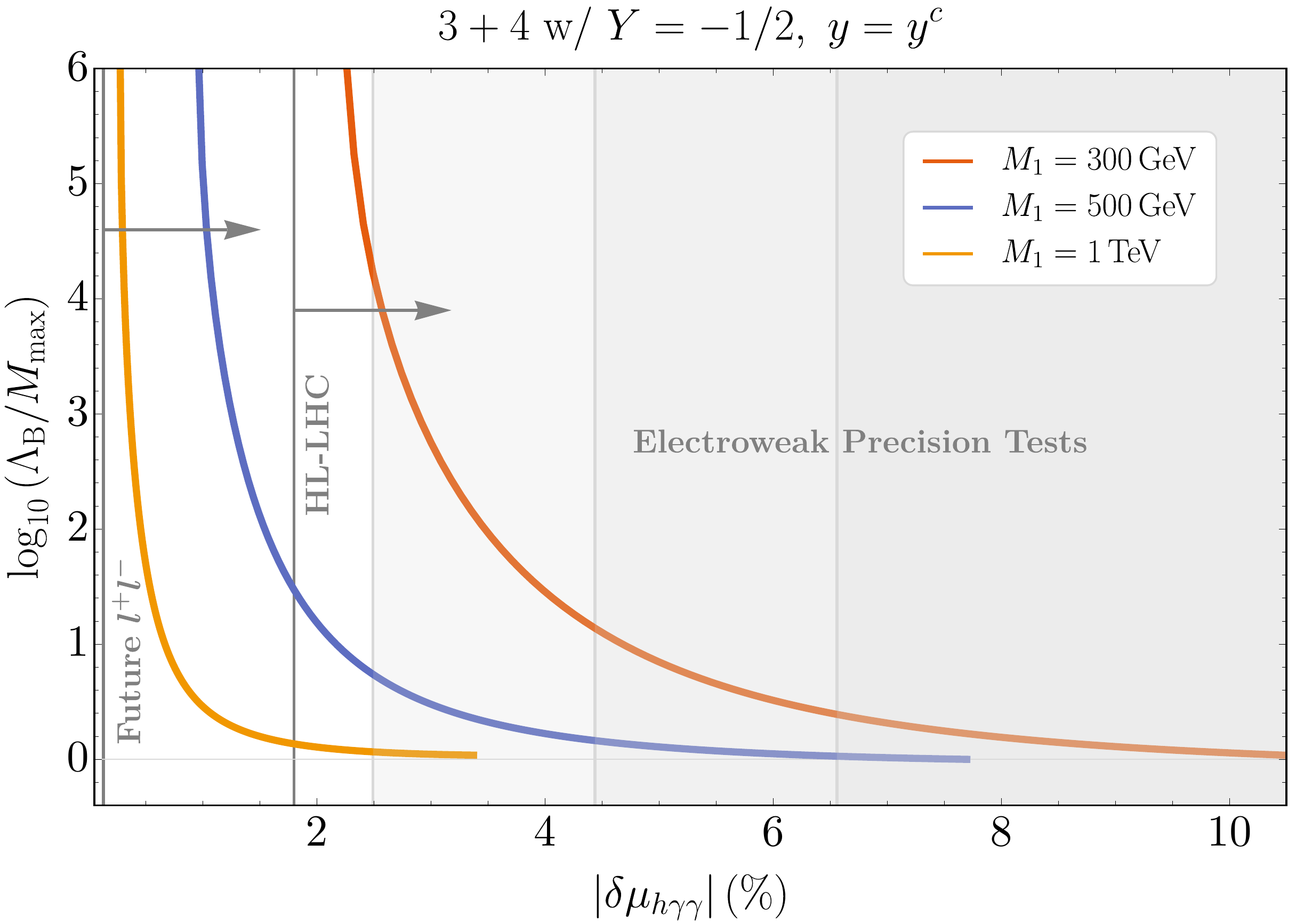}
\caption{Instability scale of the RGEs $\Lambda_B$ (i.e. upper bound on the scale of new bosons) as a function of relative $h\gamma\gamma$ coupling deviation in the vector-like lepton models defined in Section~\ref{sec:4f}. $M_{\rm max}$ on the $y$-axis is the largest of the new fermions masses, while $M_1$ is the smallest. The smallest values of $M_1$ reflect a conservative estimate of collider constraints on the models. The gray shaded areas represent the constraint from EWPTs. The EWPT constraint at lowest $\delta \mu$ is on the line with lowest $M_1$. The constraint gets monotonically weaker at higher masses. The choice $y=\pm y^c$ indicated in the title of each plot maximizes $\Lambda_B$.}
\label{fig:gammagamma}
\end{figure}

The case of $h\gamma\gamma$ is much simpler. The result is automatically finite, since the vertex does not exist in the SM at tree-level, and we can write the coupling deviation for general $n$ and $Y$ at large $M_L=M_N \gg v$ in a compact form, 
\be
\delta \mu_{h\gamma\gamma}&=&\frac{y y^c v^2}{36 M^2_L A_{\rm SM}}(-1)^{n} n\left[(n-1)(n+1)+4(n+1)Y+12Y^2 \right]\, ,
\ee
with $A_{\rm SM} \simeq 3.3$.
Any fermionic theory in the previous Subsection, different than the doublet+singlet case ($n=2$, $Y=-1/2$), gives a deviation also in $h\gamma\gamma$. Since this coupling exist in the SM only at loop level, the relative deviation is much larger than the $hWW$ or $hZZ$ case and potentially visible at HL-LHC. If one believes the fermionic theories in this paper, their first experimental manifestation might be $h\gamma\gamma$, as shown in Fig.~\ref{fig:gammagamma}. This is the reason why we mention this coupling deviation, but $h\gamma\gamma$ was already discussed in~\cite{Arkani-Hamed:2012dcq, Blum:2015rpa} and we do not have anything qualitative to add.  As discussed in the first two Sections, the fermionic theories in this paper prove by contradiction that observable deviations in $hWW$ or $hZZ$ require light new bosons. We do not aim at giving here a detailed phenomenological treatment of vector-like fermion theories besides our model-independent statement on $hWW$ and $hZZ$.

\section{Outlook}
We have discussed how observing a deviation in Higgs couplings to $WW$ and $ZZ$ gives information on the scale where new bosons must appear. We considered theories containing only new fermions and showed that a measurable deviation in $hWW$ or $hZZ$ requires large Yukawa couplings that destabilize the Higgs potential. 
In general, these theories must be completed very close to the scale of the new fermions to avoid a rapid decay of the SM vacuum. Phenomenologically, our most interesting result is that any measurable deviation at HL-LHC requires such a small scale for new bosons that it cannot be generated only by new fermions. Deviations measurable at future lepton colliders either require very light new fermions, with masses $M_\psi \lesssim 150$~GeV, or new bosons roughly below 100 TeV.

In this work we have computed a scale $\Lambda_B$ associated to the instability of the Higgs potential which gives a very conservative upper bound on where new bosons must appear. One natural way to improve this analysis would be to consider explicit new models where the instability is lifted and obtain a more precise determination of the actual masses of the new bosons. It is hard to imagine a way to do this model-independently, but we leave this line of inquiry to other researchers and our future selves, hoping that they can outsmart our present selves. 

Another possible future direction consists in refining our treatment of direct and indirect constraints on the new fermions. The scale of new bosons $\Lambda_B$ is sensitive to the lightest new fermions masses and here we did not go beyond a series of rough, but conservative estimates, thus obtaining an upper bound on $\Lambda_B$. We find that it will be interesting to further explore this point in case a credible deviation is measured in $hWW$ or $hZZ$. 

A third possibility for future work is to extend our discussion of the $h\gamma\gamma$ coupling. To a large extent it was treated in~\cite{Arkani-Hamed:2012dcq, Blum:2015rpa}, but there we did not systematically map all possible representations of the new fermions and we did not get a model-independent upper bound on $\Lambda_B$. Note, however, that the results in this paper suggest that the lowest fermions representations (already considered in~\cite{Arkani-Hamed:2012dcq, Blum:2015rpa}) will dominate the upper bound.

To conclude, $hWW$ and $hZZ$ couplings are a key observable for the future run of the LHC. Any deviation from the SM prediction signals the presence of light new bosons. If we are lucky, this could either be the first sign of the long awaited symmetry explaining $m_h^2$ or a definitive sign of unnaturalness. 

\section*{Acknowledgments}

We thank B. Bellazzini, K. Blum, L. Silvestrini, A.~M. Teixeira, L.T. Wang for useful discussions. G.~R. acknowledges funding from the European Union’s Horizon 2020 research and innovation programme under the Marie Skłodowska-Curie actions Grant Agreement no 945298-ParisRegionFP.


\appendix
\section{Fermionic Low Energy Theories}\label{app:reps}
In~\cite{Bizot:2015zaa} a thorough survey of new heavy fermions that are: 1) experimentally still allowed 2) anomaly-free and 3) can affect Higgs couplings to SM particles was conducted. In this Section we review what representations in~\cite{Bizot:2015zaa} are relevant for us (i.e. can give an observable deviation in $hZZ$ or $hWW$).

\subsection{One Fermion}\label{sec:one}
As discussed in more detail in~\cite{Bizot:2015zaa}, there are two fermion representations that taken in isolation are consistent with the three conditions stated above, a singlet of all SM gauge groups
\be
N=(1,1)_0
\ee
and a $SU(2)_L$ triplet
\be
\Sigma=(1, 3)_0\, .
\ee
They can both have a Majorana mass and couple to the Higgs via a SM lepton $\ell$. The relevant part of the Lagrangian for the singlet reads
\be
\mathcal{L}_N = - \frac{M_N}{2}N^2- y \ell H N\, .
\ee
This gives two neutral mass eigenstates. If $M_N \ll y v$ they are mixed at $\mathcal{O}(1)$ which is excluded by searches for new sterile neutrinos. The opposite limit $M_N \gg y v$ is phenomenologically viable, giving two eigenstates
\be
m_{\nu_l} \simeq \frac{y^2 v^2}{2 M_N}\, , \quad m_{\nu_h} \simeq M_N\, ,
\ee
mixed at $\mathcal{O}(y^2 v^2/M_N^2)$. Given upper bounds on neutrino masses, the SM-like neutrino must be $\nu_l$, $m_{\nu_l} \lesssim$~eV. The effects of the new state on SM couplings are suppressed by at least $m_{\nu_l}/v$ and for our purposes are unobservably small.

A similar reasoning can be followed to conclude that also $\Sigma$ does not produce observable Higgs coupling deviations, since its neutral component behaves exactly like $N$
\be
\mathcal{L}_\Sigma = - \frac{m_\Sigma}{2}\Sigma^2- y \ell H \Sigma = - \frac{m_\Sigma}{2}\Sigma^2- y \nu_l \frac{h+v}{\sqrt{2}} \Sigma^0 + ...\, .
\ee
Considering the charged components can only strengthen the bounds on $m_\Sigma$.

\subsection{Two Fermions}\label{sec:two}
Any copy of a SM fermion with a vector-like partner satisfies the three conditions stated above. Also any of the exotic leptons (plus a vector-like partner) listed here:
\be
\Lambda&=&(1,2)_{-3/2}\, , \quad \Delta=(1,3)_{-1}\, , \quad X_T=(3,2)_{7/6}\, , \quad Y_B=(3,2)_{-5/6}\, , \nn \\
X_Q&=&(3,3)_{2/3}\, , \quad Y_Q=(3,3)_{-1/3}\, ,
\ee
is a viable possibility. In practice we have listed all fermions that can couple with an existing SM fermion through a Yukawa coupling. Let us first consider the leptons, for concreteness a vector-like partner of the electron described by the two left-handed spinors $E, E^c$. The relevant part of the Lagrangian reads
\be
\mathcal{L}_E = - m_E E E^c - y \ell_\tau H E^c-\mu \tau^c E+{\rm h.c.}\, . \label{eq:massE}
\ee
We coupled $E, E^c$ just to the $\tau$ to avoid stringent constraints from flavor changing processes. Different choices are possible, but only strengthen our conclusion below (i.e. that these representations should not be considered in our work). We could immediately discard this option by noting that $H\tau\tau$ and $Z\tau\tau$ are affected at tree-level while the couplings of interest to us are modified at loop level. However let us be more precise. Current constraints on charged particles at the LHC are quite stringent, placing them firmly above the weak scale. Therefore it is sensible to integrate out $E$ and $E^c$ to understand what are the leading effects on SM coupling deviations. At tree level the equations of motion read
\be
E&=& - \frac{y l_\tau H}{m_E}+\frac{\mu^\dagger}{|m_E|^2}(D_\mu^{E^c} \tau^{c\dagger})\bar \sigma^\mu\, , \nn \\
E^c &=& -\frac{\mu \tau^c}{m_E}+\frac{y^\dagger}{|m_E|^2}(D_\mu^E l_\tau^\dagger)\bar \sigma^\mu\, .
\ee
The Lagrangian in Eq.~\eqref{eq:massE} including all terms up to $\mathcal{O}(1/m_E^2)$ only gives a shift to the $\tau$ Yukawa coupling 
\be
\mathcal{L}_E = \frac{\mu y}{m_E} \ell_\tau H \tau^c+{\rm h.c.}+\mathcal{O}(1/m_E^4)\, .
\ee
This contribution to $y_\tau$ cannot be observed as a Higgs coupling deviation because the $\tau $ mass and its coupling are affected in the same way. However, we are forced to impose
\be
\frac{\mu y}{m_E} \lesssim y_\tau^{\rm exp.} \simeq 0.01\, .
\ee
The kinetic terms of the two heavy leptons give a more interesting result
\be
\mathcal{L}_E^{\rm kin} &=& \frac{i |y|^2}{4 |m_E|^2}\left[(H^\dagger D_\mu H)(l_\tau^\dagger \bar \sigma^\mu l_\tau)+(H^\dagger D_\mu \vec \sigma H)\cdot (l_\tau^\dagger \bar \sigma^\mu \vec \sigma l_\tau)\right]- y_\tau \frac{|y|^2 |H|^2}{2 |m_E|^2}l_\tau H \tau^c \nn \\
&+& i\frac{|\mu|^2}{|m_E|^2} \tau^{c\dagger} \bar \sigma^\mu D_\mu^{E^c} \tau^c\, .
\ee
We can read in appendix B of~\cite{Blum:2015rpa} (Eq.~B.4) how these operators affect Higgs and $Z$ couplings (we call $g_A$ and $g_V$ the axial and vector coupling of the $\tau$ to the $Z$). The result is
\be
\frac{\delta g_{h\tau\tau}}{g_{h\tau\tau}}=\frac{|y|^2 v^2}{2 |m_E|^2}\, , \quad \frac{\delta g_A}{g_A}=-\frac{|y|^2 v^2}{4 |m_E|^2}-\frac{|\mu|^2}{|m_E|^2}\, ,\quad \frac{\delta g_V}{g_V}=-\frac{|y|^2 v^2}{4 |m_E|^2}+\frac{|\mu|^2}{|m_E|^2}\, .
\ee
The $hZZ$ and $hWW$ coupling deviations arise at loop level and at leading order we have
\be
\frac{\delta g_{hVV}}{g_{hVV}} \sim \frac{|y|^2 v^2}{16 \pi^2 |m_E|^2}\max[1,y^2/g^2]\, .
\ee
There are also terms proportional to $y \mu$ or $\mu^2$, but are suppressed by the insertion of the SM $\tau$ Yukawa coupling. We see immediately that while we can tune $\delta g_V$ to be small without affecting $\delta g_{hVV}$, we can't have a large $\delta g_{hVV}$ without an even larger $Z$-coupling deviation $\delta g_A$. The maximal allowed value for $\delta g_{hVV}$ given a bound on $\delta g_A$ is obtained at $\mu=0$
\be
\left.\frac{\delta g_{hVV}}{g_{hVV}}\right|_{\rm max} \sim \frac{1}{\pi^2}\left(\frac{\delta g_A}{g_A}\right)_{\rm exp}\, ,
\ee
where we took $y\sim 2$, the largest value compatible with not having a Landau pole or an instability right above the new fermions' masses. $Z$ couplings are constrained at the permille level by LEP~\cite{ParticleDataGroup:2020ssz, ParticleDataGroup:2022pth}, so this deviation is much smaller than what we can have with the same Yukawa if we introduce 3 or 4 vector-like fermions (see for example Section~\ref{sec:reps} in the main body of the paper), so for a given coupling deviation we get a smaller cutoff and we do not need to consider this case. 

\section{Large $SU(2)_L$ Representations}\label{app:SU2}
We can treat the $SU(2)_L$ representations in the main body of the paper in two equivalent ways. The most common choice in particle physics is to think about a representation of dimension $n$ as a vector with $n$ components 
\be
L = \left(\begin{array}{c}L_{(n-1)/2} \\ L_{(n-2)/2} \\ ... \\ L_m \\ ... \\ L_{-(n-1)/2}\end{array}\right)\, , \label{eq:vector}
\ee
labelled by an index $m$. We can identify each component with the state $|j,m\rangle$ where the equivalent of the total spin is $j=(n-1)/2$. It is then straightforward to write the Lagrangian for the components of $L$ and $L^c$, 
\be
\mathcal{L}_{\rm free} &=& \sum_{m=-(n-1)/2}^{(n-1)/2}\left[L_m^\dagger \overline{\sigma}^{\mu} \partial_\mu L_m+L_m^{c\dagger} \overline{\sigma}^{\mu} \partial_\mu L_m^c- M_L (-1)^{\frac{n-1}{2}-m} L_{-m}^c L_m\right]\, , \nn \\
\mathcal{L}_{\rm gauge} &=& -\frac{g}{\sqrt{2}}\sum_{m=-(n-1)/2}^{(n-1)/2}\left[\sqrt{\frac{n}{2}\left(\frac{n}{2}-1\right)-m(m+1)}L_{m+1}^\dagger\overline{\sigma}^{\mu}L_m W_\mu^+ + {\rm h.c.}\right] \nn \\
&-& \sum_{m=-(n-1)/2}^{(n-1)/2}\left[e (Y+m) L_{m}^\dagger\overline{\sigma}^{\mu}L_m A_\mu  - \frac{g}{c_W} (m-(Y+m)s_W^2) L_{m}^\dagger\overline{\sigma}^{\mu}L_m Z_\mu\right]\, ,\nn \\
\mathcal{L}_{\rm int} &=& -\sum_{m=-(n-2)/2}^{(n-2)/2}\frac{(-1)^{\frac{n-2}{2}-m}}{\sqrt{2}}\left[\sqrt{1-\frac{2m -1}{n-1}}\left(yL_{1/2-m} H^0 N^c_m+y^c L_{1/2-m}^c H^{+*} N_m\right)\right. \nn \\
&+&\left. \sqrt{1+\frac{2m +1}{n-1}}\left(y^c L_{-1/2-m}^c H^{0*} N_m+y L_{-1/2-m} H^{+} N_m^c\right)\right]\, .
\ee
To write $\mathcal{L}_{\text{free}}, \mathcal{L}_{\text{gauge}}$ and $\mathcal{L}_{\text{int}}$ we used the usual Clebsh-Gordan decomposition of the direct product of $SU(2)$ representations and the well-known form of the generators of the group for a representation of dimension $n$~\cite{Georgi:2000vve}. We have omitted the gauge interactions of $L^c$ because they can easily be deduced from $\mathcal{L}_{\rm gauge}$. 

A second option is to treat the representation of dimension $n$ as a symmetric tensor with $n-1$ indexes\footnote{The dimension $d$ of a symmetric tensor with $k$ indexes running from 1 to $N$ is $d=\left(\begin{array}{c}N+k-1 \\ k\end{array}\right)$.}, each transforming as a fundamental of $SU(2)$
\be
L=L^{i_1 i_2 ... i_{n-1}}\, .
\ee
This is the most natural way to treat $L$ if one builds up higher dimensional representations from the direct product of lower dimensional ones. Under a $SU(2)$ transformation the tensor $L$ is rotated to $L^\prime$
\be
(L^\prime)^{i_1 i_2 ... i_{n-1}}=U^{i_1 j_1}U^{i_2 j_2}...U^{i_{n-1} j_{n-1}}L^{j_1 j_2 ... j_{n-1}}\, , \quad U=e^{i \vec \alpha \cdot \frac{\vec \sigma}{2}}\, .
\ee
If we adopt the convention of lowering the indexes of the fundamental representation using the totally antisymmetric tensor $\epsilon_{ij}$
\be
L_i = \epsilon_{ij}L^j\, ,\quad \epsilon=\left(\begin{array}{cc}0 & 1 \\ -1 & 0\end{array}\right)\, ,
\ee
contractions of upper and lower indexes are $SU(2)$-invariant and we can write $\mathcal{L}_{\rm free}$ and $\mathcal{L}_{\rm int}$ in a compact form
\be
\mathcal{L}_{\rm free} &=& \sum_{(i_1 i_2 ... i_{n-1})}\left[(L^\dagger)^{i_1 i_2 ... i_{n-1}}\overline{\sigma}^{\mu} \partial_\mu L^{i_1 i_2 ... i_{n-1}}+(L^{c\dagger})^{i_1 i_2 ... i_{n-1}}\overline{\sigma}^{\mu} \partial_\mu (L^c)^{i_1 i_2 ... i_{n-1}}- M_L L^c_{i_1 i_2 ... i_{n-1}} L^{i_1 i_2 ... i_{n-1}}\right]\, , \nn \\
\mathcal{L}_{\rm int} &=& -\sum_{(i_1 i_2 ... i_{n-1})}\left[y L_{i_1 i_2 ... i_{n-1}} H^{i_1} (N^c)^{i_2 ... i_{n-1}}+y^c L_{i_1 i_2 ... i_{n-1}}^c H_{i_1}^\dagger N^{i_2 ... i_{n-1}}\right]\, .
\ee
Note that the sums extend only over the independent components of the tensors. All tensors are fully symmetric in the exchange of any pair of indexes, so for example we are not including in the sum both $L^{12i_3 i_4 ... i_{n-1}}$ and $L^{21i_3 i_4 ... i_{n-1}}$ otherwise we would be double-counting.

It is possible to write down also gauge interactions in this notation, but the resulting expressions are cumbersome and not particularly illuminating. It is much better to notice that there is a simple relation between $L^{j_1 j_2 ... j_{n-1}}$ and our original vector. Let us call $L^{\tilde k}$ the component of  $L^{j_1 j_2 ... j_{n-1}}$ with $k$ indexes equal to 1 and $n-k-1$ equal to $2$. Its electric charge is
\be
Q L^{\tilde k}=\left(Y+\frac{k}{2}-\frac{n-1-k}{2}\right)L^{\tilde k}\, ,
\ee
so we can identify $L^{\tilde k}=L_{k-(n-1)/2}$, where $L_{k-(n-1)/2}$ is one of the components of the vector in Eq.~\eqref{eq:vector}. We can then write the action of the generators on $L^{i_1i_2 ... i_{n-1}}$ starting from the well-known expressions of their action on vector components
\be
T^{\pm} L_m &=& \sqrt{\frac{n}{2}\left(\frac{n-1}{2}\right)- m(m\pm1)} L_{m\pm1}\, , \nn \\
T^3 L_m &=& m L_m\, .
\ee
The results in this work were obtained both with the tensor and with the vector notation and verified to be the same.

\section{Auxiliary Functions}\label{app:fs}
The three-body decay widths in the main body of the text and the corresponding coupling deviations have been expressed in terms of the functions listed in this Appendix.

\subsection{$WW$}
In the case of $hWW$ the functions introduced by the new leptons at $\mathcal{O}(1/M^2)$ in the vector-like mass can be written explicitly:
\be
R_W(x)&=&-180-\frac{6}{x^2}+141 x+\frac{45}{x}+\left(-36 x-\frac{9}{x}+54\right) \log (x) \nn \\
&+&\frac{\left(-720 x-\frac{36}{x}+288\right) \left[ \tan ^{-1}\left(\frac{2 \sqrt{x}-1}{\sqrt{4
   x-1}}\right) - \cot ^{-1}\left(\sqrt{4 x-1}\right) \right]}{\sqrt{4 x-1}}\, , \\
P_W^{(1)}(x)&=&-320 x^{5/2}+600 x^{3/2}-\frac{6}{x^{3/2}}+\left(72 x^{5/2}-252 x^{3/2}+90 \sqrt{x}-\frac{12}{\sqrt{x}}\right) \log (x)\nn\\
&+&\frac{\left(2016 x^{5/2}-1632 x^{3/2}+456
   \sqrt{x}-\frac{48}{\sqrt{x}}\right) \left[\tan ^{-1}\left(\frac{2 \sqrt{x}-1}{\sqrt{4 x-1}}\right)-\cot ^{-1}\left(\sqrt{4 x-1}\right)\right]}{\sqrt{4 x-1}} \nn \\
   &-&342 \sqrt{x}+\frac{68}{\sqrt{x}}\, , \\
P_W^{(2)}(x)&=&-162 \sqrt{x}+\frac{68}{\sqrt{x}}+4 x^{5/2}+96 x^{3/2}-\frac{6}{x^{3/2}}+\left(-36 x^{3/2}+54 \sqrt{x}-\frac{12}{\sqrt{x}}\right) \log (x) \nn \\
&+&\frac{\left(-480 x^{3/2}+312 \sqrt{x}-\frac{48}{\sqrt{x}}\right) \left[ \tan
   ^{-1}\left(\frac{2 \sqrt{x}-1}{\sqrt{4 x-1}}\right) - \cot ^{-1}\left(\sqrt{4 x-1}\right) \right]}{\sqrt{4x-1}}\, ,\nn \\ \mbox{} \\
P_W^{(3)}(x)&=&640 x^{5/2}-918 x^{3/2}+\left(-144 x^{5/2}+432 x^{3/2}-72 \sqrt{x}+\frac{6}{\sqrt{x}}\right) \log (x)\nn\\
&+&\frac{\left(-4032 x^{5/2}+1824 x^{3/2}-336
   \sqrt{x}+\frac{24}{\sqrt{x}}\right) \left[\tan ^{-1}\left(\frac{2 \sqrt{x}-1}{\sqrt{4 x-1}}\right)-  \cot ^{-1}\left(\sqrt{4 x-1}\right) \right]}{\sqrt{4 x-1}} \nn\\
   &+&324 \sqrt{x}-\frac{46}{\sqrt{x}}\, .
\ee
\subsection{$ZZ$}
In the case of $hZZ$ the dimensionless functions in Section~\ref{sec:coupling} are best expressed in terms of the following integrals:
\be
f_p=\int_{4\frac{m^2}{m_h^2}}^{(1-z^{1/2})^2}dy\frac{y^p}{(z-y)^2}\sqrt{1-2(z+y)+(z-y)^2}\, , \quad z=\frac{m_Z^2}{m_h^2}\, . 
\ee
For the $h \to Z^* Z \to e^+ e^- Z$ width we have the following auxiliary functions: 
\be
R_T(z)&=&\frac{1}{6}\left[f_2+2 (5z-1)f_1 +(1-z)^2f_0+2\frac{m^2}{m_h^2}(1-z)^2f_{-1}\right], \\ 
P_{Z}^{(1)}(z)&=&\frac{1}{6}\left[-f_3+(3-11z)f_2+(1-z)(11z-3)f_1+(1-z)^3f_0-2\frac{m^2}{m_h^2}(1-z)^3f_{-1}\right], \nn \\
\mbox{} \\ 
P_{Z}^{(2)}(z)&=&\frac{1}{6}\left[-f_3+(3+z)f_2+(z-1)(3+z)f_1+(1-z)^3f_0+2\frac{m^2}{m_h^2}(1-z)^3f_{-1} \right], \\ 
P_{Z}^{(3)}(z)&=&\frac{1}{3}\left[f_3+(11z-2)f_2+\left(1-4z+11z^2\right)f_1+ z(1-z)^2f_0+2\frac{m^2}{m_h^2}z(1-z)^2f_{-1}\right], \nn \\
P^{(4)}_T(z)&=&z\left[-2f_2+2f_1+ 2z(z-1)f_0+4\frac{m^2}{m_h^2}z(z-1)f_{-1}\right]. 
\ee

\bibliography{refs}

\providecommand{\href}[2]{#2}\begingroup\raggedright\begin{thebibliography}{100}

\bibitem{Arkani-Hamed:2012dcq}
N.~Arkani-Hamed, K.~Blum, R.~T. D'Agnolo, and J.~Fan, ``{2:1 for Naturalness at
  the LHC?},'' \href{http://dx.doi.org/10.1007/JHEP01(2013)149}{{\em JHEP} {\bf
  01} (2013)  149}, \href{http://arxiv.org/abs/1207.4482}{{\tt arXiv:1207.4482
  [hep-ph]}}.

\bibitem{Blum:2015rpa}
K.~Blum, R.~T. D'Agnolo, and J.~Fan, ``{Vacuum stability bounds on Higgs
  coupling deviations in the absence of new bosons},''
  \href{http://dx.doi.org/10.1007/JHEP03(2015)166}{{\em JHEP} {\bf 03} (2015)
  166}, \href{http://arxiv.org/abs/1502.01045}{{\tt arXiv:1502.01045
  [hep-ph]}}.

\bibitem{Gogoladze:2008ak}
I.~Gogoladze, N.~Okada, and Q.~Shafi, ``{Higgs Boson Mass Bounds in the
  Standard Model with Type III and Type I Seesaw},''
  \href{http://dx.doi.org/10.1016/j.physletb.2008.08.023}{{\em Phys. Lett. B}
  {\bf 668} (2008)  121--125}, \href{http://arxiv.org/abs/0805.2129}{{\tt
  arXiv:0805.2129 [hep-ph]}}.

\bibitem{Reece:2012gi}
M.~Reece, ``{Vacuum instabilities with a wrong-sign Higgs–gluon–gluon
  amplitude},'' \href{http://dx.doi.org/10.1088/1367-2630/15/4/043003}{{\em New
  J. Phys.} {\bf 15} (2013)  043003},
  \href{http://arxiv.org/abs/1208.1765}{{\tt arXiv:1208.1765 [hep-ph]}}.

\bibitem{Chen:2012faa}
C.-S. Chen and Y.~Tang, ``{Vacuum stability, neutrinos, and dark matter},''
  \href{http://dx.doi.org/10.1007/JHEP04(2012)019}{{\em JHEP} {\bf 04} (2012)
  019}, \href{http://arxiv.org/abs/1202.5717}{{\tt arXiv:1202.5717 [hep-ph]}}.

\bibitem{Joglekar:2012vc}
A.~Joglekar, P.~Schwaller, and C.~E.~M. Wagner, ``{Dark Matter and enhanced $h
  \rightarrow \gamma \gamma$ rate from vector-like Leptons},''
  \href{http://dx.doi.org/10.1007/JHEP12(2012)064}{{\em JHEP} {\bf 12} (2012)
  064}, \href{http://arxiv.org/abs/1207.4235}{{\tt arXiv:1207.4235 [hep-ph]}}.

\bibitem{Kearney:2012zi}
J.~Kearney, A.~Pierce, and N.~Weiner, ``{Vectorlike fermions and Higgs
  couplings},'' \href{http://dx.doi.org/10.1103/PhysRevD.86.113005}{{\em Phys.
  Rev. D} {\bf 86} (2012)  113005}, \href{http://arxiv.org/abs/1207.7062}{{\tt
  arXiv:1207.7062 [hep-ph]}}.

\bibitem{Batell:2012ca}
B.~Batell, S.~Gori, and L.-T. Wang, ``{Higgs couplings and precision
  electroweak data},'' \href{http://dx.doi.org/10.1007/JHEP01(2013)139}{{\em
  JHEP} {\bf 01} (2013)  139}, \href{http://arxiv.org/abs/1209.6382}{{\tt
  arXiv:1209.6382 [hep-ph]}}.

\bibitem{Altmannshofer:2013zba}
W.~Altmannshofer, M.~Bauer, and M.~Carena, ``{Exotic leptons: Higgs, flavor and
  collider phenomenology},''
  \href{http://dx.doi.org/10.1007/JHEP01(2014)060}{{\em JHEP} {\bf 01} (2014)
  060}, \href{http://arxiv.org/abs/1308.1987}{{\tt arXiv:1308.1987 [hep-ph]}}.

\bibitem{Fairbairn:2013xaa}
M.~Fairbairn and P.~Grothaus, ``{Baryogenesis and Dark Matter with Vector-like
  Fermions},'' \href{http://dx.doi.org/10.1007/JHEP10(2013)176}{{\em JHEP} {\bf
  10} (2013)  176}, \href{http://arxiv.org/abs/1307.8011}{{\tt arXiv:1307.8011
  [hep-ph]}}.

\bibitem{Xiao:2014kba}
M.-L. Xiao and J.-H. Yu, ``{Stabilizing electroweak vacuum in a vectorlike
  fermion model},'' \href{http://dx.doi.org/10.1103/PhysRevD.90.014007}{{\em
  Phys. Rev. D} {\bf 90} (2014) no.~1, 014007},
  \href{http://arxiv.org/abs/1404.0681}{{\tt arXiv:1404.0681 [hep-ph]}}.
  [Addendum: Phys.Rev.D 90, 019901 (2014)].

\bibitem{Ellis:2014dza}
S.~A.~R. Ellis, R.~M. Godbole, S.~Gopalakrishna, and J.~D. Wells, ``{Survey of
  vector-like fermion extensions of the Standard Model and their
  phenomenological implications},''
  \href{http://dx.doi.org/10.1007/JHEP09(2014)130}{{\em JHEP} {\bf 09} (2014)
  130}, \href{http://arxiv.org/abs/1404.4398}{{\tt arXiv:1404.4398 [hep-ph]}}.

\bibitem{Angelescu:2016mhl}
A.~Angelescu and G.~Arcadi, ``{Dark matter phenomenology of SM and enlarged
  Higgs sectors extended with vector-like leptons},''
  \href{http://dx.doi.org/10.1140/epjc/s10052-017-5015-2}{{\em Eur. Phys. J. C}
  {\bf 77} (2017) no.~7, 456}, \href{http://arxiv.org/abs/1611.06186}{{\tt
  arXiv:1611.06186 [hep-ph]}}.

\bibitem{Goswami:2018jar}
S.~Goswami, K.~N. Vishnudath, and N.~Khan, ``{Constraining the minimal type-III
  seesaw model with naturalness, lepton flavor violation, and electroweak
  vacuum stability},'' \href{http://dx.doi.org/10.1103/PhysRevD.99.075012}{{\em
  Phys. Rev. D} {\bf 99} (2019) no.~7, 075012},
  \href{http://arxiv.org/abs/1810.11687}{{\tt arXiv:1810.11687 [hep-ph]}}.

\bibitem{Gopalakrishna:2018uxn}
S.~Gopalakrishna and A.~Velusamy, ``{Higgs vacuum stability with vectorlike
  fermions},'' \href{http://dx.doi.org/10.1103/PhysRevD.99.115020}{{\em Phys.
  Rev. D} {\bf 99} (2019) no.~11, 115020},
  \href{http://arxiv.org/abs/1812.11303}{{\tt arXiv:1812.11303 [hep-ph]}}.

\bibitem{Borah:2020nsz}
D.~Borah, R.~Roshan, and A.~Sil, ``{Sub-TeV singlet scalar dark matter and
  electroweak vacuum stability with vectorlike fermions},''
  \href{http://dx.doi.org/10.1103/PhysRevD.102.075034}{{\em Phys. Rev. D} {\bf
  102} (2020) no.~7, 075034}, \href{http://arxiv.org/abs/2007.14904}{{\tt
  arXiv:2007.14904 [hep-ph]}}.

\bibitem{Bandyopadhyay:2020djh}
P.~Bandyopadhyay, S.~Jangid, and M.~Mitra, ``{Scrutinizing Vacuum Stability in
  IDM with Type-III Inverse seesaw},''
  \href{http://dx.doi.org/10.1007/JHEP02(2021)075}{{\em JHEP} {\bf 02} (2021)
  075}, \href{http://arxiv.org/abs/2008.11956}{{\tt arXiv:2008.11956
  [hep-ph]}}.

\bibitem{Arsenault:2022xty}
A.~Arsenault, K.~Y. Cingiloglu, and M.~Frank, ``{Vacuum stability in the
  Standard Model with vectorlike fermions},''
  \href{http://dx.doi.org/10.1103/PhysRevD.107.036018}{{\em Phys. Rev. D} {\bf
  107} (2023) no.~3, 036018}, \href{http://arxiv.org/abs/2207.10332}{{\tt
  arXiv:2207.10332 [hep-ph]}}.

\bibitem{Hiller:2022rla}
G.~Hiller, T.~H\"ohne, D.~F. Litim, and T.~Steudtner, ``{Portals into Higgs
  vacuum stability},''
  \href{http://dx.doi.org/10.1103/PhysRevD.106.115004}{{\em Phys. Rev. D} {\bf
  106} (2022) no.~11, 115004}, \href{http://arxiv.org/abs/2207.07737}{{\tt
  arXiv:2207.07737 [hep-ph]}}.

\bibitem{Arvanitaki:2011ck}
A.~Arvanitaki and G.~Villadoro, ``{A Non Standard Model Higgs at the LHC as a
  Sign of Naturalness},'' \href{http://dx.doi.org/10.1007/JHEP02(2012)144}{{\em
  JHEP} {\bf 02} (2012)  144}, \href{http://arxiv.org/abs/1112.4835}{{\tt
  arXiv:1112.4835 [hep-ph]}}.

\bibitem{Cepeda:2019klc}
M.~Cepeda {\em et al.}, ``{Report from Working Group 2}: {Higgs Physics at the
  HL-LHC and HE-LHC},''
  \href{http://dx.doi.org/10.23731/CYRM-2019-007.221}{{\em CERN Yellow Rep.
  Monogr.} {\bf 7} (2019)  221--584},
  \href{http://arxiv.org/abs/1902.00134}{{\tt arXiv:1902.00134 [hep-ph]}}.

\bibitem{ILCInternationalDevelopmentTeam:2022izu}
{\bf ILC International Development Team} Collaboration, A.~Aryshev {\em et
  al.}, ``{The International Linear Collider: Report to Snowmass 2021},''
  \href{http://arxiv.org/abs/2203.07622}{{\tt arXiv:2203.07622
  [physics.acc-ph]}}.

\bibitem{Robson:2018zje}
A.~Robson and P.~Roloff, ``{Updated CLIC luminosity staging baseline and Higgs
  coupling prospects},'' \href{http://arxiv.org/abs/1812.01644}{{\tt
  arXiv:1812.01644 [hep-ex]}}.

\bibitem{Bernardi:2022hny}
G.~Bernardi {\em et al.}, ``{The Future Circular Collider: a Summary for the US
  2021 Snowmass Process},'' \href{http://arxiv.org/abs/2203.06520}{{\tt
  arXiv:2203.06520 [hep-ex]}}.

\bibitem{CEPCPhysicsStudyGroup:2022uwl}
{\bf CEPC Physics Study Group} Collaboration, H.~Cheng {\em et al.}, ``{The
  Physics potential of the CEPC. Prepared for the US Snowmass Community
  Planning Exercise (Snowmass 2021)},'' in {\em {2022 Snowmass Summer Study}}.
\newblock 5, 2022.
\newblock \href{http://arxiv.org/abs/2205.08553}{{\tt arXiv:2205.08553
  [hep-ph]}}.

\bibitem{Forslund:2022xjq}
M.~Forslund and P.~Meade, ``{High precision higgs from high energy muon
  colliders},'' \href{http://dx.doi.org/10.1007/JHEP08(2022)185}{{\em JHEP}
  {\bf 08} (2022)  185}, \href{http://arxiv.org/abs/2203.09425}{{\tt
  arXiv:2203.09425 [hep-ph]}}.

\bibitem{deBlas:2022aow}
J.~de~Blas, J.~Gu, and Z.~Liu, ``{Higgs boson precision measurements at a
  125~GeV muon collider},''
  \href{http://dx.doi.org/10.1103/PhysRevD.106.073007}{{\em Phys. Rev. D} {\bf
  106} (2022) no.~7, 073007}, \href{http://arxiv.org/abs/2203.04324}{{\tt
  arXiv:2203.04324 [hep-ph]}}.

\bibitem{deBlas:2022ofj}
J.~de~Blas, Y.~Du, C.~Grojean, J.~Gu, V.~Miralles, M.~E. Peskin, J.~Tian,
  M.~Vos, and E.~Vryonidou, ``{Global SMEFT Fits at Future Colliders},'' in
  {\em {2022 Snowmass Summer Study}}.
\newblock 6, 2022.
\newblock \href{http://arxiv.org/abs/2206.08326}{{\tt arXiv:2206.08326
  [hep-ph]}}.

\bibitem{Staub:2008uz}
F.~Staub, ``{SARAH},'' \href{http://arxiv.org/abs/0806.0538}{{\tt
  arXiv:0806.0538 [hep-ph]}}.

\bibitem{Staub:2015kfa}
F.~Staub, ``{Exploring new models in all detail with SARAH},''
  \href{http://dx.doi.org/10.1155/2015/840780}{{\em Adv. High Energy Phys.}
  {\bf 2015} (2015)  840780}, \href{http://arxiv.org/abs/1503.04200}{{\tt
  arXiv:1503.04200 [hep-ph]}}.

\bibitem{Staub:2013tta}
F.~Staub, ``{SARAH 4 : A tool for (not only SUSY) model builders},''
  \href{http://dx.doi.org/10.1016/j.cpc.2014.02.018}{{\em Comput. Phys.
  Commun.} {\bf 185} (2014)  1773--1790},
  \href{http://arxiv.org/abs/1309.7223}{{\tt arXiv:1309.7223 [hep-ph]}}.

\bibitem{Staub:2012pb}
F.~Staub, ``{SARAH 3.2: Dirac Gauginos, UFO output, and more},''
  \href{http://dx.doi.org/10.1016/j.cpc.2013.02.019}{{\em Comput. Phys.
  Commun.} {\bf 184} (2013)  1792--1809},
  \href{http://arxiv.org/abs/1207.0906}{{\tt arXiv:1207.0906 [hep-ph]}}.

\bibitem{Staub:2010jh}
F.~Staub, ``{Automatic Calculation of supersymmetric Renormalization Group
  Equations and Self Energies},''
  \href{http://dx.doi.org/10.1016/j.cpc.2010.11.030}{{\em Comput. Phys.
  Commun.} {\bf 182} (2011)  808--833},
  \href{http://arxiv.org/abs/1002.0840}{{\tt arXiv:1002.0840 [hep-ph]}}.

\bibitem{Staub:2009bi}
F.~Staub, ``{From Superpotential to Model Files for FeynArts and
  CalcHep/CompHep},'' \href{http://dx.doi.org/10.1016/j.cpc.2010.01.011}{{\em
  Comput. Phys. Commun.} {\bf 181} (2010)  1077--1086},
  \href{http://arxiv.org/abs/0909.2863}{{\tt arXiv:0909.2863 [hep-ph]}}.

\bibitem{ALEPH:2005ab}
{\bf ALEPH, DELPHI, L3, OPAL, SLD, LEP Electroweak Working Group, SLD
  Electroweak Group, SLD Heavy Flavour Group} Collaboration, S.~Schael {\em et
  al.}, ``{Precision electroweak measurements on the $Z$ resonance},''
  \href{http://dx.doi.org/10.1016/j.physrep.2005.12.006}{{\em Phys. Rept.} {\bf
  427} (2006)  257--454}, \href{http://arxiv.org/abs/hep-ex/0509008}{{\tt
  arXiv:hep-ex/0509008}}.

\bibitem{Bizot:2015zaa}
N.~Bizot and M.~Frigerio, ``{Fermionic extensions of the Standard Model in
  light of the Higgs couplings},''
  \href{http://dx.doi.org/10.1007/JHEP01(2016)036}{{\em JHEP} {\bf 01} (2016)
  036}, \href{http://arxiv.org/abs/1508.01645}{{\tt arXiv:1508.01645
  [hep-ph]}}.

\bibitem{Giudice:2007fh}
G.~F. Giudice, C.~Grojean, A.~Pomarol, and R.~Rattazzi, ``{The
  Strongly-Interacting Light Higgs},''
  \href{http://dx.doi.org/10.1088/1126-6708/2007/06/045}{{\em JHEP} {\bf 06}
  (2007)  045}, \href{http://arxiv.org/abs/hep-ph/0703164}{{\tt
  arXiv:hep-ph/0703164}}.

\bibitem{Schwartz:2014sze}
M.~D. Schwartz, {\em {Quantum Field Theory and the Standard Model}}.
\newblock Cambridge University Press, 3, 2014.

\bibitem{Peskin:1995ev}
M.~E. Peskin and D.~V. Schroeder, {\em {An Introduction to quantum field
  theory}}.
\newblock Addison-Wesley, Reading, USA, 1995.

\bibitem{Bardin:1999ak}
D.~Y. Bardin and G.~Passarino, {\em {The standard model in the making:
  Precision study of the electroweak interactions}}.
\newblock 1999.

\bibitem{ParticleDataGroup:2022pth}
{\bf Particle Data Group} Collaboration, R.~L. Workman {\em et al.}, ``{Review
  of Particle Physics},'' \href{http://dx.doi.org/10.1093/ptep/ptac097}{{\em
  PTEP} {\bf 2022} (2022)  083C01}.

\bibitem{Ellis:1975ap}
J.~R. Ellis, M.~K. Gaillard, and D.~V. Nanopoulos, ``{A Phenomenological
  Profile of the Higgs Boson},''
  \href{http://dx.doi.org/10.1016/0550-3213(76)90382-5}{{\em Nucl. Phys. B}
  {\bf 106} (1976)  292}.

\bibitem{Shifman:1979eb}
M.~A. Shifman, A.~I. Vainshtein, M.~B. Voloshin, and V.~I. Zakharov,
  ``{Low-Energy Theorems for Higgs Boson Couplings to Photons},'' {\em Sov. J.
  Nucl. Phys.} {\bf 30} (1979)  711--716.

\bibitem{Kniehl:1995tn}
B.~A. Kniehl and M.~Spira, ``{Low-energy theorems in Higgs physics},''
  \href{http://dx.doi.org/10.1007/s002880050007}{{\em Z. Phys. C} {\bf 69}
  (1995)  77--88}, \href{http://arxiv.org/abs/hep-ph/9505225}{{\tt
  arXiv:hep-ph/9505225}}.

\bibitem{Gillioz:2012se}
M.~Gillioz, R.~Grober, C.~Grojean, M.~Muhlleitner, and E.~Salvioni, ``{Higgs
  Low-Energy Theorem (and its corrections) in Composite Models},''
  \href{http://dx.doi.org/10.1007/JHEP10(2012)004}{{\em JHEP} {\bf 10} (2012)
  004}, \href{http://arxiv.org/abs/1206.7120}{{\tt arXiv:1206.7120 [hep-ph]}}.

\bibitem{Patel:2015tea}
H.~H. Patel, ``{Package-X: A Mathematica package for the analytic calculation
  of one-loop integrals},''
  \href{http://dx.doi.org/10.1016/j.cpc.2015.08.017}{{\em Comput. Phys.
  Commun.} {\bf 197} (2015)  276--290},
  \href{http://arxiv.org/abs/1503.01469}{{\tt arXiv:1503.01469 [hep-ph]}}.

\bibitem{Patel:2016fam}
H.~H. Patel, ``{Package-X 2.0: A Mathematica package for the analytic
  calculation of one-loop integrals},''
  \href{http://dx.doi.org/10.1016/j.cpc.2017.04.015}{{\em Comput. Phys.
  Commun.} {\bf 218} (2017)  66--70},
  \href{http://arxiv.org/abs/1612.00009}{{\tt arXiv:1612.00009 [hep-ph]}}.

\bibitem{tHooft:1973alw}
G.~'t~Hooft, ``{A Planar Diagram Theory for Strong Interactions},''
  \href{http://dx.doi.org/10.1016/0550-3213(74)90154-0}{{\em Nucl. Phys. B}
  {\bf 72} (1974)  461}.

\bibitem{DeRujula:1972pc}
A.~De~Rujula and B.~E. Lautrup, ``{Excited leptons and the anomalous magnetic
  moments of muon and electron},''
  \href{http://dx.doi.org/10.1007/BF02770499}{{\em Lett. Nuovo Cim.} {\bf 3S2}
  (1972)  49--57}.

\bibitem{Fritzsch:1975sr}
H.~Fritzsch, M.~Gell-Mann, and P.~Minkowski, ``{Vector - Like Weak Currents and
  New Elementary Fermions},''
  \href{http://dx.doi.org/10.1016/0370-2693(75)90040-4}{{\em Phys. Lett. B}
  {\bf 59} (1975)  256--260}.

\bibitem{Fujikawa:1976qy}
K.~Fujikawa, ``{Test of Vector-Like Weak Lepton Currents in e+ e-
  Annihilation},'' \href{http://dx.doi.org/10.1016/0370-2693(76)90498-6}{{\em
  Phys. Lett. B} {\bf 62} (1976)  176--178}.

\bibitem{Cheng:1976ii}
T.~P. Cheng, ``{Hierarchy of Lepton Masses in a Vector-Like Theory with
  Majorana Particles},'' \href{http://dx.doi.org/10.1103/PhysRevD.14.1367}{{\em
  Phys. Rev. D} {\bf 14} (1976)  1367}.

\bibitem{Slavnov:1976dp}
A.~A. Slavnov, ``{Vector-Like Lepton Model with Supersymmetric Interaction},''
  \href{http://dx.doi.org/10.1007/BF01108500}{{\em Teor. Mat. Fiz.} {\bf 29}
  (1976)  154--160}.

\bibitem{Azimov:1976kc}
Y.~I. Azimov, L.~L. Frankfurt, and V.~A. Khoze, ``{Vector-Like Models and the
  Possible Nature of Direct Leptons},'' {\em Pisma Zh. Eksp. Teor. Fiz.} {\bf
  24} (1976)  373--376.

\bibitem{Mikaelian:1976ut}
K.~O. Mikaelian, ``{Supernovae Explosions, the New Leptons, and Righthanded
  Neutrinos},'' \href{http://dx.doi.org/10.1103/PhysRevLett.36.1089}{{\em Phys.
  Rev. Lett.} {\bf 36} (1976)  1089--1092}.

\bibitem{Ma:1977da}
E.~Ma, ``{Eigenvalue Condition for the Weinberg Angle and Possible New Leptons
  and Quarks},'' \href{http://dx.doi.org/10.1143/PTP.58.1896}{{\em Prog. Theor.
  Phys.} {\bf 58} (1977)  1896}.

\bibitem{McKay:1978wn}
D.~W. McKay and H.~Munczek, ``{SU(2) X U(1) Vector - Like Model: New Leptons,
  Quarks and the Axion Question},''
  \href{http://dx.doi.org/10.1103/PhysRevD.19.985}{{\em Phys. Rev. D} {\bf 19}
  (1979)  985}.

\bibitem{Goldberg:1981gc}
H.~Goldberg, ``{Bounds on $e^+ e^- \to \ell^* \bar{\ell}$epton and $\ell P \to
  \ell^*$ X ($\ell^*$ = Excited Lepton) and Prospects for Visible $\ell^*$
  Tracks in Cosmic Ray Emulsion Events},''
  \href{http://dx.doi.org/10.1103/PhysRevD.24.1991}{{\em Phys. Rev. D} {\bf 24}
  (1981)  1991}.

\bibitem{delAguila:1982fs}
F.~del Aguila and M.~J. Bowick, ``{The Possibility of New Fermions With
  $\Delta$ I = 0 Mass},''
  \href{http://dx.doi.org/10.1016/0550-3213(83)90316-4}{{\em Nucl. Phys. B}
  {\bf 224} (1983)  107}.

\bibitem{Enqvist:1983xh}
K.~Enqvist and J.~Maalampi, ``{Signatures of Excited Leptons in $p \bar{p}$
  Collisions},'' \href{http://dx.doi.org/10.1016/0370-2693(84)90401-5}{{\em
  Phys. Lett. B} {\bf 135} (1984)  329}.

\bibitem{Cabibbo:1983bk}
N.~Cabibbo, L.~Maiani, and Y.~Srivastava, ``{Anomalous Z Decays: Excited
  Leptons?},'' \href{http://dx.doi.org/10.1016/0370-2693(84)91850-1}{{\em Phys.
  Lett. B} {\bf 139} (1984)  459--463}.

\bibitem{Ohba:1984uyo}
I.~Ohba, ``{A MODEL OF LOCAL U(1) GAUGE INVARIANT COUPLING BETWEEN LEPTON AND
  EXCITED LEPTON},'' \href{http://dx.doi.org/10.1143/PTP.72.1291}{{\em Prog.
  Theor. Phys.} {\bf 72} (1984)  1291--1293}.

\bibitem{Kalyniak:1984zz}
P.~Kalyniak and M.~K. Sundaresan, ``{PHENOMENOLOGICAL COUPLING OF EXCITED
  LEPTON TO Z0 AND Z0 DECAY TO HARD gamma},''
  \href{http://dx.doi.org/10.1103/PhysRevD.31.2813}{{\em Phys. Rev. D} {\bf 31}
  (1985)  2813}.

\bibitem{Nakazato:1984dy}
H.~Nakazato, M.~Namiki, Y.~Yamanaka, and K.-i. Yokoyama, ``{PHENOMENOLOGICAL
  ANALYSIS OF Z0 ---\ensuremath{>} anti-lepton lepton gamma DECAY ON EXCITED
  LEPTON HYPOTHESIS},'' \href{http://dx.doi.org/10.1143/PTP.72.865}{{\em Prog.
  Theor. Phys.} {\bf 72} (1984)  865}.

\bibitem{Choudhury:1984sh}
S.~R. Choudhury, R.~G. Ellis, and G.~C. Joshi, ``{A NOTE ON EXCITED LEPTONS},''
  {\em Hadronic J.} {\bf 8} (1985)  60.

\bibitem{Ellis:1985nu}
R.~G. Ellis, M.~Matsuda, and B.~H.~J. McKellar, ``{EXCITED FERMIONS AND
  LOW-ENERGY PARITY VIOLATING PHENOMENA},''
  \href{http://dx.doi.org/10.1103/PhysRevD.32.1657}{{\em Phys. Rev. D} {\bf 32}
  (1985)  1657}.

\bibitem{Hagiwara:1985wt}
K.~Hagiwara, D.~Zeppenfeld, and S.~Komamiya, ``{Excited Lepton Production at
  LEP and HERA},'' \href{http://dx.doi.org/10.1007/BF01571391}{{\em Z. Phys. C}
  {\bf 29} (1985)  115}.

\bibitem{Krucken:1985cx}
T.~Krucken, D.~Rein, and R.~Rodenberg, ``{SIGNATURES OF EXCITED LEPTONS
  SLIGHTLY ABOVE Z MASS},''
  \href{http://dx.doi.org/10.1016/0370-2693(85)90887-1}{{\em Phys. Lett. B}
  {\bf 159} (1985)  201--204}.

\bibitem{Tanimoto:1985vj}
M.~Tanimoto, ``{The Effect of the Excited Leptons and the Composite Scalar
  Bosons on the Processes $e^+ e^- \to W^+ W^-$ or $Z^0 Z^0$},''
  \href{http://dx.doi.org/10.1016/0370-2693(85)91334-6}{{\em Phys. Lett. B}
  {\bf 160} (1985)  312--316}.

\bibitem{Rizzo:1986wf}
T.~G. Rizzo, ``{Production of Exotic $E(6)$ Leptons in $e p$ Collisions},''
  \href{http://dx.doi.org/10.1016/0370-2693(87)90712-X}{{\em Phys. Lett. B}
  {\bf 188} (1987)  95--98}.

\bibitem{Rizzo:1987ye}
T.~G. Rizzo, ``{LIMITS ON NEW FERMIONS FROM p anti-p COLLIDER DATA},''
  \href{http://dx.doi.org/10.1142/S0217732387000628}{{\em Mod. Phys. Lett. A}
  {\bf 2} (1987)  505}.

\bibitem{Kovalchuk:1987aq}
V.~A. Kovalchuk and I.~V. Stoletnyi, ``{SIGNALS OF EXCITED LEPTONS IN PHOTON
  AND Z BOSON PRODUCTION REACTIONS WITH COLLIDING E+ E- BEAMS. (IN RUSSIAN)},''
  {\em Ukr. Fiz. Zh. (Russ. Ed. )} {\bf 32} (1987)  1285--1289.

\bibitem{Stoker:1988kf}
D.~P. Stoker {\em et al.}, ``{Limits on New Lepton Pairs (L-, L0) With
  Arbitrary Neutrino Mass},''
  \href{http://dx.doi.org/10.1103/PhysRevD.39.1811}{{\em Phys. Rev. D} {\bf 39}
  (1989)  1811}.

\bibitem{Pitkanen:1988ke}
M.~Pitkanen, ``{Are Bound States of Color Excited Leptons Responsible for the
  Anomalous $e^+ e^-$ Production in Heavy Ion Collisions?},''
  \href{http://dx.doi.org/10.1007/BF00673631}{{\em Int. J. Theor. Phys.} {\bf
  29} (1990)  275}.

\bibitem{Geng:1988ar}
C.~Q. Geng, ``{Uniqueness of Quarks, Leptons and Exotic Fermions in the Chiral
  Color Models},'' \href{http://dx.doi.org/10.1103/PhysRevD.39.2402}{{\em Phys.
  Rev. D} {\bf 39} (1989)  2402}.

\bibitem{Bhattacharyya:1989am}
G.~Bhattacharyya and A.~Raychaudhuri, ``{The Width and Flavor Changing Decays
  of the $Z$ as a Test of Exotic Quarks and Leptons},''
  \href{http://dx.doi.org/10.1103/PhysRevD.42.268}{{\em Phys. Rev. D} {\bf 42}
  (1990)  268--272}.

\bibitem{Singh:1990qd}
S.~Singh, R.~K. Parashari, and N.~K. Sharma, ``{Possible deviations in tau
  polarization in e- e+ ---\ensuremath{>} tau- tau+ due to exotic lepton
  mixing},'' \href{http://dx.doi.org/10.1103/PhysRevD.42.3881}{{\em Phys. Rev.
  D} {\bf 42} (1990)  3881--3884}.

\bibitem{Nagawat:1990ui}
A.~K. Nagawat, S.~Singh, and N.~K. Sharma, ``{Possible deviations due to exotic
  lepton mixings in the process $e^{+} e^{-} \to W^{+} W^{-}$},''
  \href{http://dx.doi.org/10.1103/PhysRevD.42.2984}{{\em Phys. Rev. D} {\bf 42}
  (1990)  2984--2997}.

\bibitem{Nagawat:1990fa}
A.~K. Nagawat, S.~Singh, and N.~K. Sharma, ``{Possible deviations in Z0 pair
  production at e+ e- colliders due to exotic lepton mixings},''
  \href{http://dx.doi.org/10.1142/S0217732390000676}{{\em Mod. Phys. Lett. A}
  {\bf 5} (1990)  605--612}.

\bibitem{Singh:1990ha}
S.~Singh, A.~K. Nagawat, and N.~K. Sharma, ``{Effects of mixing between
  ordinary and exotic charged leptons in the process e+ e- ---\ensuremath{>} Z0
  Z0},'' \href{http://dx.doi.org/10.1103/PhysRevD.41.1438}{{\em Phys. Rev. D}
  {\bf 41} (1990)  1438--1448}.

\bibitem{Ginzburg:1990kv}
I.~F. Ginzburg and D.~Y. Ivanov, ``{The Excited leptons and quarks at gamma
  gamma / gamma e colliders},''
  \href{http://dx.doi.org/10.1016/0370-2693(92)90566-M}{{\em Phys. Lett. B}
  {\bf 276} (1992)  214--218}.

\bibitem{Hung:1992tz}
P.~Q. Hung, ``{Can the rho parameter allow for the existence of a nondegenerate
  fourth family?},'' \href{http://dx.doi.org/10.1103/PhysRevLett.69.3143}{{\em
  Phys. Rev. Lett.} {\bf 69} (1992)  3143--3146}.

\bibitem{Hayashi:1992cu}
M.~Hayashi, N.~V. Samsonenko, and M.~A. Rozas~Teglio, ``{Creation of excited
  leptons in colliding electron positron beams},'' {\em Russ. Phys. J.} {\bf
  35} (1992)  513--518.

\bibitem{Bhattacharyya:1992np}
G.~Bhattacharyya and A.~Raychaudhuri, ``{Probing exotic leptons through oblique
  parameters},'' \href{http://dx.doi.org/10.1016/0370-2693(92)91348-D}{{\em
  Phys. Lett. B} {\bf 296} (1992)  448--451}.

\bibitem{Boudjema:1992em}
F.~Boudjema, A.~Djouadi, and J.~L. Kneur, ``{Excited fermions at e+ e- and e P
  colliders},'' \href{http://dx.doi.org/10.1007/BF01474339}{{\em Z. Phys. C}
  {\bf 57} (1993)  425--450}.

\bibitem{Nardi:1992am}
E.~Nardi, E.~Roulet, and D.~Tommasini, ``{A Simultaneous analysis of $Z^\prime$
  and new fermions effects. Global constraints in E(6) and SO(10) models},''
  \href{http://dx.doi.org/10.1103/PhysRevD.46.3040}{{\em Phys. Rev. D} {\bf 46}
  (1992)  3040--3061}.

\bibitem{Nardi:1992nq}
E.~Nardi, ``{$Z^\prime$, new fermions and flavor changing processes.
  Constraints on E(6) models from $\mu \to e e e$},''
  \href{http://dx.doi.org/10.1103/PhysRevD.48.1240}{{\em Phys. Rev. D} {\bf 48}
  (1993)  1240--1247}, \href{http://arxiv.org/abs/hep-ph/9209223}{{\tt
  arXiv:hep-ph/9209223}}.

\bibitem{Djouadi:1993pe}
A.~Djouadi, ``{New fermions at e+ e- colliders. 1. Production and decay},''
  \href{http://dx.doi.org/10.1007/BF01411024}{{\em Z. Phys. C} {\bf 63} (1994)
  317--326}, \href{http://arxiv.org/abs/hep-ph/9308339}{{\tt
  arXiv:hep-ph/9308339}}.

\bibitem{Azuelos:1993qu}
G.~Azuelos and A.~Djouadi, ``{New fermions at e+ e- colliders. 2. Signals and
  backgrounds},'' \href{http://dx.doi.org/10.1007/BF01411025}{{\em Z. Phys. C}
  {\bf 63} (1994)  327--338}, \href{http://arxiv.org/abs/hep-ph/9308340}{{\tt
  arXiv:hep-ph/9308340}}.

\bibitem{Pleitez:1994pu}
V.~Pleitez, ``{New fermions and a vector - like third generation in SU(3) (C) x
  SU(3) (L) x U(1) ($N$) models},''
  \href{http://dx.doi.org/10.1103/PhysRevD.53.514}{{\em Phys. Rev. D} {\bf 53}
  (1996)  514--526}, \href{http://arxiv.org/abs/hep-ph/9412304}{{\tt
  arXiv:hep-ph/9412304}}.

\bibitem{Gonzalez-Garcia:1996oxz}
M.~C. Gonzalez-Garcia and S.~F. Novaes, ``{Excited fermion contribution to Z0
  physics at one loop},''
  \href{http://dx.doi.org/10.1016/S0550-3213(96)00651-7}{{\em Nucl. Phys. B}
  {\bf 486} (1997)  3--22}, \href{http://arxiv.org/abs/hep-ph/9608309}{{\tt
  arXiv:hep-ph/9608309}}.

\bibitem{doAmaralCoutinho:1996grs}
Y.~do~Amaral~Coutinho, A.~J. Ramalho, and S.~Wulck, ``{Production of exotic
  neutral leptons in radiative e p scattering},''
  \href{http://dx.doi.org/10.1103/PhysRevD.54.1215}{{\em Phys. Rev. D} {\bf 54}
  (1996)  1215--1217}.

\bibitem{Montvay:1996iu}
I.~Montvay, ``{Violations of universality in a vector - like extension of the
  standard model},'' \href{http://dx.doi.org/10.1016/0370-2693(96)00605-3}{{\em
  Phys. Lett. B} {\bf 382} (1996)  104--110},
  \href{http://arxiv.org/abs/hep-ph/9604214}{{\tt arXiv:hep-ph/9604214}}.

\bibitem{Diaz-Cruz:1997mbr}
J.~L. Diaz-Cruz, J.~J. Toscano, O.~A. Sampayo, and J.~I. Aranda, ``{Bounds on
  the mass of excited fermions},''
  \href{http://dx.doi.org/10.1142/S0217732397000042}{{\em Mod. Phys. Lett. A}
  {\bf 12} (1997)  37--45}.

\bibitem{Diener:1997nx}
K.~P.~O. Diener, B.~A. Kniehl, and A.~Pilaftsis, ``{Loop effects of exotic
  leptons on vector boson pair production at e+ e- colliders},''
  \href{http://dx.doi.org/10.1103/PhysRevD.57.2771}{{\em Phys. Rev. D} {\bf 57}
  (1998)  2771--2784}, \href{http://arxiv.org/abs/hep-ph/9709361}{{\tt
  arXiv:hep-ph/9709361}}.

\bibitem{Thomas:1998wy}
S.~D. Thomas and J.~D. Wells, ``{Phenomenology of Massive Vectorlike Doublet
  Leptons},'' \href{http://dx.doi.org/10.1103/PhysRevLett.81.34}{{\em Phys.
  Rev. Lett.} {\bf 81} (1998)  34--37},
  \href{http://arxiv.org/abs/hep-ph/9804359}{{\tt arXiv:hep-ph/9804359}}.

\bibitem{CiezaMontalvo:1998mi}
J.~E. Cieza~Montalvo and P.~P. de~Queiroz~Filho, ``{Production of a single and
  a pair of heavy exotic leptons through the p p collisions},''
  \href{http://arxiv.org/abs/hep-ph/9811521}{{\tt arXiv:hep-ph/9811521}}.

\bibitem{Rosner:1999ub}
J.~L. Rosner, ``{Mixing of charge -1/3 quarks and charged leptons with exotic
  fermions in E(6)},'' \href{http://arxiv.org/abs/hep-ph/9907438}{{\tt
  arXiv:hep-ph/9907438}}.

\bibitem{Tafirout:2000gy}
R.~Tafirout and G.~Azuelos, ``{EXOTIC: A heavy fermion and excited fermion
  Monte Carlo generator for e+ e- physics},''
  \href{http://dx.doi.org/10.1016/S0010-4655(99)00464-6}{{\em Comput. Phys.
  Commun.} {\bf 126} (2000)  244--260}. [Erratum: Comput.Phys.Commun. 133, 136
  (2000)].

\bibitem{Allanach:2001sd}
B.~C. Allanach, C.~M. Harris, M.~A. Parker, P.~Richardson, and B.~R. Webber,
  ``{Detecting exotic heavy leptons at the large hadron collider},''
  \href{http://dx.doi.org/10.1088/1126-6708/2001/08/051}{{\em JHEP} {\bf 08}
  (2001)  051}, \href{http://arxiv.org/abs/hep-ph/0108097}{{\tt
  arXiv:hep-ph/0108097}}.

\bibitem{Rakshit:2001xs}
S.~Rakshit, ``{Muon anomalous magnetic moment constrains models with excited
  leptons},'' \href{http://arxiv.org/abs/hep-ph/0111083}{{\tt
  arXiv:hep-ph/0111083}}.

\bibitem{Boos:2001vc}
E.~Boos, A.~Vologdin, D.~A. Toback, and J.~Gaspard, ``{Prospects for searching
  for excited leptons during run II of the Fermilab Tevatron},''
  \href{http://dx.doi.org/10.1103/PhysRevD.66.013011}{{\em Phys. Rev. D} {\bf
  66} (2002)  013011}, \href{http://arxiv.org/abs/hep-ph/0111034}{{\tt
  arXiv:hep-ph/0111034}}.

\bibitem{Eboli:2001hi}
O.~J.~P. Eboli, S.~M. Lietti, and P.~Mathews, ``{Excited leptons at the CERN
  large hadron collider},''
  \href{http://dx.doi.org/10.1103/PhysRevD.65.075003}{{\em Phys. Rev. D} {\bf
  65} (2002)  075003}, \href{http://arxiv.org/abs/hep-ph/0111001}{{\tt
  arXiv:hep-ph/0111001}}.

\bibitem{CiezaMontalvo:2001cy}
J.~E. Cieza~Montalvo and P.~P. de~Querioz~Filho, ``{Heavy exotic-leptons at
  LHC},'' \href{http://arxiv.org/abs/hep-ph/0112263}{{\tt
  arXiv:hep-ph/0112263}}.

\bibitem{CiezaMontalvo:2002gk}
J.~E. Cieza~Montalvo and P.~P. de~Queiroz~Filho, ``{Exotic heavy leptons
  predicted by extended models at the cern LHC},''
  \href{http://dx.doi.org/10.1103/PhysRevD.66.055003}{{\em Phys. Rev. D} {\bf
  66} (2002)  055003}.

\bibitem{Bailin:2002fd}
D.~Bailin, G.~V. Kraniotis, and A.~Love, ``{Intersecting D5-brane models with
  massive vector - like leptons},''
  \href{http://dx.doi.org/10.1088/1126-6708/2003/02/052}{{\em JHEP} {\bf 02}
  (2003)  052}, \href{http://arxiv.org/abs/hep-th/0212112}{{\tt
  arXiv:hep-th/0212112}}.

\bibitem{Singhal:2004wx}
J.~K. Singhal, ``{Bounds on charged lepton mixing with exotic charged
  leptons},'' \href{http://dx.doi.org/10.1007/BF02705250}{{\em Pramana} {\bf
  62} (2004)  1029--1040}.

\bibitem{EspiritoSanto:2004tf}
M.~C. Espirito~Santo, M.~Paulos, M.~Pimenta, J.~C. Romao, and B.~Tome, ``{The
  Sensitivity of cosmic ray air shower experiments for excited lepton
  detection},'' \href{http://arxiv.org/abs/hep-ph/0412345}{{\tt
  arXiv:hep-ph/0412345}}.

\bibitem{Diaz:2005th}
R.~A. Diaz, R.~Martinez, and F.~Ochoa, ``{331 vector-like models with mirror
  fermions as a possible solution for the discrepancy in the b-quark
  asymmetries, and for the neutrino mass and mixing pattern},''
  \href{http://arxiv.org/abs/hep-ph/0508051}{{\tt arXiv:hep-ph/0508051}}.

\bibitem{Holdom:2006mr}
B.~Holdom, ``{The Discovery of the fourth family at the LHC: What if?},''
  \href{http://dx.doi.org/10.1088/1126-6708/2006/08/076}{{\em JHEP} {\bf 08}
  (2006)  076}, \href{http://arxiv.org/abs/hep-ph/0606146}{{\tt
  arXiv:hep-ph/0606146}}.

\bibitem{Dorsner:2006fx}
I.~Dorsner and P.~Fileviez~Perez, ``{Upper Bound on the Mass of the Type III
  Seesaw Triplet in an SU(5) Model},''
  \href{http://dx.doi.org/10.1088/1126-6708/2007/06/029}{{\em JHEP} {\bf 06}
  (2007)  029}, \href{http://arxiv.org/abs/hep-ph/0612216}{{\tt
  arXiv:hep-ph/0612216}}.

\bibitem{Bajc:2006ia}
B.~Bajc and G.~Senjanovic, ``{Seesaw at LHC},''
  \href{http://dx.doi.org/10.1088/1126-6708/2007/08/014}{{\em JHEP} {\bf 08}
  (2007)  014}, \href{http://arxiv.org/abs/hep-ph/0612029}{{\tt
  arXiv:hep-ph/0612029}}.

\bibitem{Bajc:2007zf}
B.~Bajc, M.~Nemevsek, and G.~Senjanovic, ``{Probing seesaw at LHC},''
  \href{http://dx.doi.org/10.1103/PhysRevD.76.055011}{{\em Phys. Rev. D} {\bf
  76} (2007)  055011}, \href{http://arxiv.org/abs/hep-ph/0703080}{{\tt
  arXiv:hep-ph/0703080}}.

\bibitem{Abada:2007ux}
A.~Abada, C.~Biggio, F.~Bonnet, M.~B. Gavela, and T.~Hambye, ``{Low energy
  effects of neutrino masses},''
  \href{http://dx.doi.org/10.1088/1126-6708/2007/12/061}{{\em JHEP} {\bf 12}
  (2007)  061}, \href{http://arxiv.org/abs/0707.4058}{{\tt arXiv:0707.4058
  [hep-ph]}}.

\bibitem{delAguila:2008pw}
F.~del Aguila, J.~de~Blas, and M.~Perez-Victoria, ``{Effects of new leptons in
  Electroweak Precision Data},''
  \href{http://dx.doi.org/10.1103/PhysRevD.78.013010}{{\em Phys. Rev. D} {\bf
  78} (2008)  013010}, \href{http://arxiv.org/abs/0803.4008}{{\tt
  arXiv:0803.4008 [hep-ph]}}.

\bibitem{Cynolter:2008ea}
G.~Cynolter and E.~Lendvai, ``{Electroweak Precision Constraints on Vector-like
  Fermions},'' \href{http://dx.doi.org/10.1140/epjc/s10052-008-0771-7}{{\em
  Eur. Phys. J. C} {\bf 58} (2008)  463--469},
  \href{http://arxiv.org/abs/0804.4080}{{\tt arXiv:0804.4080 [hep-ph]}}.

\bibitem{Abada:2008ea}
A.~Abada, C.~Biggio, F.~Bonnet, M.~B. Gavela, and T.~Hambye, ``{mu
  ---\ensuremath{>} e gamma and tau ---\ensuremath{>} l gamma decays in the
  fermion triplet seesaw model},''
  \href{http://dx.doi.org/10.1103/PhysRevD.78.033007}{{\em Phys. Rev. D} {\bf
  78} (2008)  033007}, \href{http://arxiv.org/abs/0803.0481}{{\tt
  arXiv:0803.0481 [hep-ph]}}.

\bibitem{Biggio:2008in}
C.~Biggio, ``{The Contribution of fermionic seesaws to the anomalous magnetic
  moment of leptons},''
  \href{http://dx.doi.org/10.1016/j.physletb.2008.09.004}{{\em Phys. Lett. B}
  {\bf 668} (2008)  378--384}, \href{http://arxiv.org/abs/0806.2558}{{\tt
  arXiv:0806.2558 [hep-ph]}}.

\bibitem{Strumia:2008cf}
A.~Strumia, ``{Sommerfeld corrections to type-II and III leptogenesis},''
  \href{http://dx.doi.org/10.1016/j.nuclphysb.2008.10.007}{{\em Nucl. Phys. B}
  {\bf 809} (2009)  308--317}, \href{http://arxiv.org/abs/0806.1630}{{\tt
  arXiv:0806.1630 [hep-ph]}}.

\bibitem{delAguila:2008hw}
F.~del Aguila and J.~A. Aguilar-Saavedra, ``{Electroweak scale seesaw and heavy
  Dirac neutrino signals at LHC},''
  \href{http://dx.doi.org/10.1016/j.physletb.2009.01.010}{{\em Phys. Lett. B}
  {\bf 672} (2009)  158--165}, \href{http://arxiv.org/abs/0809.2096}{{\tt
  arXiv:0809.2096 [hep-ph]}}.

\bibitem{Kikukawa:2009mu}
Y.~Kikukawa, M.~Kohda, and J.~Yasuda, ``{The Strongly coupled fourth family and
  a first-order electroweak phase transition. I. Quark sector},''
  \href{http://dx.doi.org/10.1143/PTP.122.401}{{\em Prog. Theor. Phys.} {\bf
  122} (2009)  401--426}, \href{http://arxiv.org/abs/0901.1962}{{\tt
  arXiv:0901.1962 [hep-ph]}}.

\bibitem{He:2009tf}
X.-G. He and S.~Oh, ``{Lepton FCNC in Type III Seesaw Model},''
  \href{http://dx.doi.org/10.1088/1126-6708/2009/09/027}{{\em JHEP} {\bf 09}
  (2009)  027}, \href{http://arxiv.org/abs/0902.4082}{{\tt arXiv:0902.4082
  [hep-ph]}}.

\bibitem{Arhrib:2009xf}
A.~Arhrib, R.~Benbrik, and C.-H. Chen, ``{Lepton flavor violating tau decays in
  type-III seesaw mechanism},''
  \href{http://dx.doi.org/10.1103/PhysRevD.81.113003}{{\em Phys. Rev. D} {\bf
  81} (2010)  113003}, \href{http://arxiv.org/abs/0903.1553}{{\tt
  arXiv:0903.1553 [hep-ph]}}.

\bibitem{Arhrib:2009mz}
A.~Arhrib, B.~Bajc, D.~K. Ghosh, T.~Han, G.-Y. Huang, I.~Puljak, and
  G.~Senjanovic, ``{Collider Signatures for Heavy Lepton Triplet in Type I+III
  Seesaw},'' \href{http://dx.doi.org/10.1103/PhysRevD.82.053004}{{\em Phys.
  Rev. D} {\bf 82} (2010)  053004}, \href{http://arxiv.org/abs/0904.2390}{{\tt
  arXiv:0904.2390 [hep-ph]}}.

\bibitem{Aguilar-Saavedra:2009fxa}
J.~A. Aguilar-Saavedra, ``{Heavy lepton pair production at LHC: Model
  discrimination with multi-lepton signals},''
  \href{http://dx.doi.org/10.1016/j.nuclphysb.2009.11.021}{{\em Nucl. Phys. B}
  {\bf 828} (2010)  289--316}, \href{http://arxiv.org/abs/0905.2221}{{\tt
  arXiv:0905.2221 [hep-ph]}}.

\bibitem{Bandyopadhyay:2009xa}
P.~Bandyopadhyay, S.~Choubey, and M.~Mitra, ``{Two Higgs Doublet Type III
  Seesaw with mu-tau symmetry at LHC},''
  \href{http://dx.doi.org/10.1088/1126-6708/2009/10/012}{{\em JHEP} {\bf 10}
  (2009)  012}, \href{http://arxiv.org/abs/0906.5330}{{\tt arXiv:0906.5330
  [hep-ph]}}.

\bibitem{Patra:2009bu}
S.~Patra, ``{Testable Leptogenesis in extended Standard Model},''
  \href{http://arxiv.org/abs/0911.4577}{{\tt arXiv:0911.4577 [hep-ph]}}.

\bibitem{Cacciapaglia:2009ky}
G.~Cacciapaglia, A.~Deandrea, and J.~Llodra-Perez, ``{Higgs ---\ensuremath{>}
  Gamma Gamma beyond the Standard Model},''
  \href{http://dx.doi.org/10.1088/1126-6708/2009/06/054}{{\em JHEP} {\bf 06}
  (2009)  054}, \href{http://arxiv.org/abs/0901.0927}{{\tt arXiv:0901.0927
  [hep-ph]}}.

\bibitem{Cacciapaglia:2009cv}
G.~Cacciapaglia, A.~Deandrea, S.~R. Choudhury, and N.~Gaur, ``{T-parity odd
  heavy leptons at LHC},''
  \href{http://dx.doi.org/10.1103/PhysRevD.81.075005}{{\em Phys. Rev. D} {\bf
  81} (2010)  075005}, \href{http://arxiv.org/abs/0911.4632}{{\tt
  arXiv:0911.4632 [hep-ph]}}.

\bibitem{Ma:2009kh}
E.~Ma, ``{Inverse Seesaw Neutrino Mass from Lepton Triplets in the U(1)(Sigma)
  Model},'' \href{http://dx.doi.org/10.1142/S0217732309031867}{{\em Mod. Phys.
  Lett. A} {\bf 24} (2009)  2491--2495},
  \href{http://arxiv.org/abs/0905.2972}{{\tt arXiv:0905.2972 [hep-ph]}}.

\bibitem{Ibanez:2009du}
D.~Ibanez, S.~Morisi, and J.~W.~F. Valle, ``{Inverse tri-bimaximal type-III
  seesaw and lepton flavor violation},''
  \href{http://dx.doi.org/10.1103/PhysRevD.80.053015}{{\em Phys. Rev. D} {\bf
  80} (2009)  053015}, \href{http://arxiv.org/abs/0907.3109}{{\tt
  arXiv:0907.3109 [hep-ph]}}.

\bibitem{Picek:2009is}
I.~Picek and B.~Radovcic, ``{Novel TeV-scale seesaw mechanism with Dirac
  mediators},'' \href{http://dx.doi.org/10.1016/j.physletb.2010.03.062}{{\em
  Phys. Lett. B} {\bf 687} (2010)  338--341},
  \href{http://arxiv.org/abs/0911.1374}{{\tt arXiv:0911.1374 [hep-ph]}}.

\bibitem{Gavela:2009cd}
M.~B. Gavela, T.~Hambye, D.~Hernandez, and P.~Hernandez, ``{Minimal Flavour
  Seesaw Models},'' \href{http://dx.doi.org/10.1088/1126-6708/2009/09/038}{{\em
  JHEP} {\bf 09} (2009)  038}, \href{http://arxiv.org/abs/0906.1461}{{\tt
  arXiv:0906.1461 [hep-ph]}}.

\bibitem{Yue:2010zzb}
C.-X. Yue, H.-L. Feng, and W.~Ma, ``{Heavy charged leptons from type-III seesaw
  and pair production of the Higgs boson H at the International Linear e+ e-
  Collider},'' \href{http://dx.doi.org/10.1088/0256-307X/27/1/011202}{{\em
  Chin. Phys. Lett.} {\bf 27} (2010)  011202}.

\bibitem{Bandyopadhyay:2010wp}
P.~Bandyopadhyay and E.~J. Chun, ``{Displaced Higgs production in type III
  seesaw},'' \href{http://dx.doi.org/10.1007/JHEP11(2010)006}{{\em JHEP} {\bf
  11} (2010)  006}, \href{http://arxiv.org/abs/1007.2281}{{\tt arXiv:1007.2281
  [hep-ph]}}.

\bibitem{Inan:2010af}
S.~C. Inan, ``{Exclusive excited leptons search in two lepton final states at
  the CERN-LHC},'' \href{http://dx.doi.org/10.1103/PhysRevD.81.115002}{{\em
  Phys. Rev. D} {\bf 81} (2010)  115002},
  \href{http://arxiv.org/abs/1005.3432}{{\tt arXiv:1005.3432 [hep-ph]}}.

\bibitem{Gross:2010ce}
E.~Gross, D.~Grossman, Y.~Nir, and O.~Vitells, ``{Testing minimal lepton flavor
  violation with extra vector-like leptons at the LHC},''
  \href{http://dx.doi.org/10.1103/PhysRevD.81.055013}{{\em Phys. Rev. D} {\bf
  81} (2010)  055013}, \href{http://arxiv.org/abs/1001.2883}{{\tt
  arXiv:1001.2883 [hep-ph]}}.

\bibitem{AristizabalSierra:2010mv}
D.~Aristizabal~Sierra, J.~F. Kamenik, and M.~Nemevsek, ``{Implications of
  Flavor Dynamics for Fermion Triplet Leptogenesis},''
  \href{http://dx.doi.org/10.1007/JHEP10(2010)036}{{\em JHEP} {\bf 10} (2010)
  036}, \href{http://arxiv.org/abs/1007.1907}{{\tt arXiv:1007.1907 [hep-ph]}}.

\bibitem{Kumericki:2011hf}
K.~Kumericki, I.~Picek, and B.~Radovcic, ``{Exotic Seesaw-Motivated Heavy
  Leptons at the LHC},''
  \href{http://dx.doi.org/10.1103/PhysRevD.84.093002}{{\em Phys. Rev. D} {\bf
  84} (2011)  093002}, \href{http://arxiv.org/abs/1106.1069}{{\tt
  arXiv:1106.1069 [hep-ph]}}.

\bibitem{Bandyopadhyay:2011aa}
P.~Bandyopadhyay, S.~Choi, E.~J. Chun, and K.~Min, ``{Probing Higgs bosons via
  the type III seesaw mechanism at the LHC},''
  \href{http://dx.doi.org/10.1103/PhysRevD.85.073013}{{\em Phys. Rev. D} {\bf
  85} (2012)  073013}, \href{http://arxiv.org/abs/1112.3080}{{\tt
  arXiv:1112.3080 [hep-ph]}}.

\bibitem{Eboli:2011ia}
O.~J.~P. Eboli, J.~Gonzalez-Fraile, and M.~C. Gonzalez-Garcia, ``{Neutrino
  Masses at LHC: Minimal Lepton Flavour Violation in Type-III See-saw},''
  \href{http://dx.doi.org/10.1007/JHEP12(2011)009}{{\em JHEP} {\bf 12} (2011)
  009}, \href{http://arxiv.org/abs/1108.0661}{{\tt arXiv:1108.0661 [hep-ph]}}.

\bibitem{Ren:2011mh}
B.~Ren, K.~Tsumura, and X.-G. He, ``{A Higgs Quadruplet for Type III Seesaw and
  Implications for $\mu \to e\gamma$ and $\mu - e$ Conversion},''
  \href{http://dx.doi.org/10.1103/PhysRevD.84.073004}{{\em Phys. Rev. D} {\bf
  84} (2011)  073004}, \href{http://arxiv.org/abs/1107.5879}{{\tt
  arXiv:1107.5879 [hep-ph]}}.

\bibitem{Law:2011qe}
S.~S.~C. Law, ``{Constraints on exotic lepton doublets with minimal coupling to
  the standard model},'' \href{http://dx.doi.org/10.1007/JHEP02(2012)127}{{\em
  JHEP} {\bf 02} (2012)  127}, \href{http://arxiv.org/abs/1106.0375}{{\tt
  arXiv:1106.0375 [hep-ph]}}.

\bibitem{Ari:2011is}
V.~Ari, O.~Cakir, and V.~\c{C}etinkaya, ``{Forward-backward asymmetries of
  fourth family fermions through the Z' models at linear colliders},''
  \href{http://dx.doi.org/10.1103/PhysRevD.87.035013}{{\em Phys. Rev. D} {\bf
  87} (2013) no.~3, 035013}, \href{http://arxiv.org/abs/1108.2908}{{\tt
  arXiv:1108.2908 [hep-ph]}}.

\bibitem{Chen:2012wz}
N.~Chen and H.-J. He, ``{LHC Signatures of Two-Higgs-Doublets with Fourth
  Family},'' \href{http://dx.doi.org/10.1007/JHEP04(2012)062}{{\em JHEP} {\bf
  04} (2012)  062}, \href{http://arxiv.org/abs/1202.3072}{{\tt arXiv:1202.3072
  [hep-ph]}}.

\bibitem{Voloshin:2012tv}
M.~B. Voloshin, ``{CP Violation in Higgs Diphoton Decay in Models with
  Vectorlike Heavy Fermions},''
  \href{http://dx.doi.org/10.1103/PhysRevD.86.093016}{{\em Phys. Rev. D} {\bf
  86} (2012)  093016}, \href{http://arxiv.org/abs/1208.4303}{{\tt
  arXiv:1208.4303 [hep-ph]}}.

\bibitem{Bae:2012ir}
K.~J. Bae, T.~H. Jung, and H.~D. Kim, ``{125 GeV Higgs boson as a
  pseudo-Goldstone boson in supersymmetry with vectorlike matters},''
  \href{http://dx.doi.org/10.1103/PhysRevD.87.015014}{{\em Phys. Rev. D} {\bf
  87} (2013) no.~1, 015014}, \href{http://arxiv.org/abs/1208.3748}{{\tt
  arXiv:1208.3748 [hep-ph]}}.

\bibitem{Law:2012mj}
S.~S.~C. Law and K.~L. McDonald, ``{Inverse seesaw and dark matter in models
  with exotic lepton triplets},''
  \href{http://dx.doi.org/10.1016/j.physletb.2012.06.044}{{\em Phys. Lett. B}
  {\bf 713} (2012)  490--494}, \href{http://arxiv.org/abs/1204.2529}{{\tt
  arXiv:1204.2529 [hep-ph]}}.

\bibitem{Almeida:2012bq}
L.~G. Almeida, E.~Bertuzzo, P.~A.~N. Machado, and R.~Zukanovich~Funchal,
  ``{Does $H \to \gamma \gamma$ Taste like vanilla New Physics?},''
  \href{http://dx.doi.org/10.1007/JHEP11(2012)085}{{\em JHEP} {\bf 11} (2012)
  085}, \href{http://arxiv.org/abs/1207.5254}{{\tt arXiv:1207.5254 [hep-ph]}}.

\bibitem{Ibrahim:2012ds}
T.~Ibrahim and P.~Nath, ``{$\tau\to \mu \gamma$ decay in extensions with a
  vectorlike generation},''
  \href{http://dx.doi.org/10.1103/PhysRevD.87.015030}{{\em Phys. Rev. D} {\bf
  87} (2013) no.~1, 015030}, \href{http://arxiv.org/abs/1211.0622}{{\tt
  arXiv:1211.0622 [hep-ph]}}.

\bibitem{Endo:2012cc}
M.~Endo, K.~Hamaguchi, K.~Ishikawa, S.~Iwamoto, and N.~Yokozaki, ``{Gauge
  Mediation Models with Vectorlike Matters at the LHC},''
  \href{http://dx.doi.org/10.1007/JHEP01(2013)181}{{\em JHEP} {\bf 01} (2013)
  181}, \href{http://arxiv.org/abs/1212.3935}{{\tt arXiv:1212.3935 [hep-ph]}}.

\bibitem{Djouadi:2012ae}
A.~Djouadi and A.~Lenz, ``{Sealing the fate of a fourth generation of
  fermions},'' \href{http://dx.doi.org/10.1016/j.physletb.2012.07.060}{{\em
  Phys. Lett. B} {\bf 715} (2012)  310--314},
  \href{http://arxiv.org/abs/1204.1252}{{\tt arXiv:1204.1252 [hep-ph]}}.

\bibitem{Dermisek:2012ke}
R.~Dermisek, ``{Unification of gauge couplings in the standard model with extra
  vectorlike families},''
  \href{http://dx.doi.org/10.1103/PhysRevD.87.055008}{{\em Phys. Rev. D} {\bf
  87} (2013) no.~5, 055008}, \href{http://arxiv.org/abs/1212.3035}{{\tt
  arXiv:1212.3035 [hep-ph]}}.

\bibitem{Barr:2012ma}
S.~M. Barr and H.-Y. Chen, ``{A Simple Grand Unified Relation between Neutrino
  Mixing and Quark Mixing},''
  \href{http://dx.doi.org/10.1007/JHEP11(2012)092}{{\em JHEP} {\bf 11} (2012)
  092}, \href{http://arxiv.org/abs/1208.6546}{{\tt arXiv:1208.6546 [hep-ph]}}.

\bibitem{Majhi:2012bx}
S.~Majhi, ``{QCD corrections to excited lepton (pair) production at the LHC},''
  \href{http://dx.doi.org/10.1103/PhysRevD.88.074028}{{\em Phys. Rev. D} {\bf
  88} (2013) no.~7, 074028}, \href{http://arxiv.org/abs/1210.8307}{{\tt
  arXiv:1210.8307 [hep-ph]}}.

\bibitem{Arina:2012aj}
C.~Arina, R.~N. Mohapatra, and N.~Sahu, ``{Co-genesis of Matter and Dark Matter
  with Vector-like Fourth Generation Leptons},''
  \href{http://dx.doi.org/10.1016/j.physletb.2013.01.059}{{\em Phys. Lett. B}
  {\bf 720} (2013)  130--136}, \href{http://arxiv.org/abs/1211.0435}{{\tt
  arXiv:1211.0435 [hep-ph]}}.

\bibitem{Moreau:2012da}
G.~Moreau, ``{Constraining extra-fermion(s) from the Higgs boson data},''
  \href{http://dx.doi.org/10.1103/PhysRevD.87.015027}{{\em Phys. Rev. D} {\bf
  87} (2013) no.~1, 015027}, \href{http://arxiv.org/abs/1210.3977}{{\tt
  arXiv:1210.3977 [hep-ph]}}.

\bibitem{Feng:2013mea}
W.-Z. Feng and P.~Nath, ``{Higgs diphoton rate and mass enhancement with
  vectorlike leptons and the scale of supersymmetry},''
  \href{http://dx.doi.org/10.1103/PhysRevD.87.075018}{{\em Phys. Rev. D} {\bf
  87} (2013) no.~7, 075018}, \href{http://arxiv.org/abs/1303.0289}{{\tt
  arXiv:1303.0289 [hep-ph]}}.

\bibitem{Chang:2013eia}
X.~Chang, C.~Liu, and Y.-L. Tang, ``{Phenomenological aspects of R-parity
  violating supersymmetry with a vectorlike extra generation},''
  \href{http://dx.doi.org/10.1103/PhysRevD.87.075012}{{\em Phys. Rev. D} {\bf
  87} (2013) no.~7, 075012}, \href{http://arxiv.org/abs/1303.7055}{{\tt
  arXiv:1303.7055 [hep-ph]}}.

\bibitem{Schwaller:2013hqa}
P.~Schwaller, T.~M.~P. Tait, and R.~Vega-Morales, ``{Dark Matter and Vectorlike
  Leptons from Gauged Lepton Number},''
  \href{http://dx.doi.org/10.1103/PhysRevD.88.035001}{{\em Phys. Rev. D} {\bf
  88} (2013) no.~3, 035001}, \href{http://arxiv.org/abs/1305.1108}{{\tt
  arXiv:1305.1108 [hep-ph]}}.

\bibitem{Dermisek:2013gta}
R.~Dermisek and A.~Raval, ``{Explanation of the Muon g-2 Anomaly with
  Vectorlike Leptons and its Implications for Higgs Decays},''
  \href{http://dx.doi.org/10.1103/PhysRevD.88.013017}{{\em Phys. Rev. D} {\bf
  88} (2013)  013017}, \href{http://arxiv.org/abs/1305.3522}{{\tt
  arXiv:1305.3522 [hep-ph]}}.

\bibitem{Aguilar-Saavedra:2013twa}
J.~A. Aguilar-Saavedra, P.~M. Boavida, and F.~R. Joaquim, ``{Flavored searches
  for type-III seesaw mechanism at the LHC},''
  \href{http://dx.doi.org/10.1103/PhysRevD.88.113008}{{\em Phys. Rev. D} {\bf
  88} (2013)  113008}, \href{http://arxiv.org/abs/1308.3226}{{\tt
  arXiv:1308.3226 [hep-ph]}}.

\bibitem{McDonald:2013hsa}
K.~L. McDonald, ``{Probing Exotic Fermions from a Seesaw/Radiative Model at the
  LHC},'' \href{http://dx.doi.org/10.1007/JHEP11(2013)131}{{\em JHEP} {\bf 11}
  (2013)  131}, \href{http://arxiv.org/abs/1310.0609}{{\tt arXiv:1310.0609
  [hep-ph]}}.

\bibitem{Joglekar:2013zya}
A.~Joglekar, P.~Schwaller, and C.~E.~M. Wagner, ``{A Supersymmetric Theory of
  Vector-like Leptons},'' \href{http://dx.doi.org/10.1007/JHEP07(2013)046}{{\em
  JHEP} {\bf 07} (2013)  046}, \href{http://arxiv.org/abs/1303.2969}{{\tt
  arXiv:1303.2969 [hep-ph]}}.

\bibitem{Kyae:2013hda}
B.~Kyae and C.~S. Shin, ``{Vector-like leptons and extra gauge symmetry for the
  natural Higgs boson},'' \href{http://dx.doi.org/10.1007/JHEP06(2013)102}{{\em
  JHEP} {\bf 06} (2013)  102}, \href{http://arxiv.org/abs/1303.6703}{{\tt
  arXiv:1303.6703 [hep-ph]}}.

\bibitem{Ishiwata:2013gma}
K.~Ishiwata and M.~B. Wise, ``{Phenomenology of heavy vectorlike leptons},''
  \href{http://dx.doi.org/10.1103/PhysRevD.88.055009}{{\em Phys. Rev. D} {\bf
  88} (2013) no.~5, 055009}, \href{http://arxiv.org/abs/1307.1112}{{\tt
  arXiv:1307.1112 [hep-ph]}}.

\bibitem{Banerjee:2013kna}
P.~Banerjee and U.~A. Yajnik, ``{Production and decay rates of excited leptons
  in a left-right symmetric scenario},''
  \href{http://dx.doi.org/10.1103/PhysRevD.90.095023}{{\em Phys. Rev. D} {\bf
  90} (2014) no.~9, 095023}, \href{http://arxiv.org/abs/1310.5785}{{\tt
  arXiv:1310.5785 [hep-ph]}}.

\bibitem{Falkowski:2013jya}
A.~Falkowski, D.~M. Straub, and A.~Vicente, ``{Vector-like leptons: Higgs
  decays and collider phenomenology},''
  \href{http://dx.doi.org/10.1007/JHEP05(2014)092}{{\em JHEP} {\bf 05} (2014)
  092}, \href{http://arxiv.org/abs/1312.5329}{{\tt arXiv:1312.5329 [hep-ph]}}.

\bibitem{Batell:2013bka}
B.~Batell, S.~Jung, and C.~E.~M. Wagner, ``{Very Light Charginos and Higgs
  Decays},'' \href{http://dx.doi.org/10.1007/JHEP12(2013)075}{{\em JHEP} {\bf
  12} (2013)  075}, \href{http://arxiv.org/abs/1309.2297}{{\tt arXiv:1309.2297
  [hep-ph]}}.

\bibitem{Halverson:2014nwa}
J.~Halverson, N.~Orlofsky, and A.~Pierce, ``{Vectorlike Leptons as the Tip of
  the Dark Matter Iceberg},''
  \href{http://dx.doi.org/10.1103/PhysRevD.90.015002}{{\em Phys. Rev. D} {\bf
  90} (2014) no.~1, 015002}, \href{http://arxiv.org/abs/1403.1592}{{\tt
  arXiv:1403.1592 [hep-ph]}}.

\bibitem{Dermisek:2014cia}
R.~Dermisek, A.~Raval, and S.~Shin, ``{Effects of vectorlike leptons on $h\to
  4\ell$ and the connection to the muon g-2 anomaly},''
  \href{http://dx.doi.org/10.1103/PhysRevD.90.034023}{{\em Phys. Rev. D} {\bf
  90} (2014) no.~3, 034023}, \href{http://arxiv.org/abs/1406.7018}{{\tt
  arXiv:1406.7018 [hep-ph]}}.

\bibitem{Dermisek:2014qca}
R.~Dermisek, J.~P. Hall, E.~Lunghi, and S.~Shin, ``{Limits on Vectorlike
  Leptons from Searches for Anomalous Production of Multi-Lepton Events},''
  \href{http://dx.doi.org/10.1007/JHEP12(2014)013}{{\em JHEP} {\bf 12} (2014)
  013}, \href{http://arxiv.org/abs/1408.3123}{{\tt arXiv:1408.3123 [hep-ph]}}.

\bibitem{Biondini:2014dfa}
S.~Biondini and O.~Panella, ``{Exotic leptons at future linear colliders},''
  \href{http://dx.doi.org/10.1103/PhysRevD.92.015023}{{\em Phys. Rev. D} {\bf
  92} (2015) no.~1, 015023}, \href{http://arxiv.org/abs/1411.6556}{{\tt
  arXiv:1411.6556 [hep-ph]}}.

\bibitem{Dorsner:2014wva}
I.~Dorsner, S.~Fajfer, and I.~Mustac, ``{Light vector-like fermions in a
  minimal SU(5) setup},''
  \href{http://dx.doi.org/10.1103/PhysRevD.89.115004}{{\em Phys. Rev. D} {\bf
  89} (2014) no.~11, 115004}, \href{http://arxiv.org/abs/1401.6870}{{\tt
  arXiv:1401.6870 [hep-ph]}}.

\bibitem{Ma:2014zda}
T.~Ma, B.~Zhang, and G.~Cacciapaglia, ``{Doubly Charged Lepton from an Exotic
  Doublet at the LHC},''
  \href{http://dx.doi.org/10.1103/PhysRevD.89.093022}{{\em Phys. Rev. D} {\bf
  89} (2014) no.~9, 093022}, \href{http://arxiv.org/abs/1404.2375}{{\tt
  arXiv:1404.2375 [hep-ph]}}.

\bibitem{Gopalakrishna:2015wwa}
S.~Gopalakrishna, T.~S. Mukherjee, and S.~Sadhukhan, ``{Extra neutral scalars
  with vectorlike fermions at the LHC},''
  \href{http://dx.doi.org/10.1103/PhysRevD.93.055004}{{\em Phys. Rev. D} {\bf
  93} (2016) no.~5, 055004}, \href{http://arxiv.org/abs/1504.01074}{{\tt
  arXiv:1504.01074 [hep-ph]}}.

\bibitem{Voloshin:2015ona}
M.~B. Voloshin, ``{Condensed state of heavy vectorlike neutrinos},''
  \href{http://dx.doi.org/10.1103/PhysRevLett.115.091802}{{\em Phys. Rev.
  Lett.} {\bf 115} (2015) no.~9, 091802},
  \href{http://arxiv.org/abs/1506.04096}{{\tt arXiv:1506.04096 [hep-ph]}}.

\bibitem{Dermisek:2015oja}
R.~Dermisek, E.~Lunghi, and S.~Shin, ``{Two Higgs doublet model with vectorlike
  leptons and contributions to $pp\to WW$ and $H\to WW$},''
  \href{http://dx.doi.org/10.1007/JHEP02(2016)119}{{\em JHEP} {\bf 02} (2016)
  119}, \href{http://arxiv.org/abs/1509.04292}{{\tt arXiv:1509.04292
  [hep-ph]}}.

\bibitem{Bhattacharya:2015qpa}
S.~Bhattacharya, N.~Sahoo, and N.~Sahu, ``{Minimal vectorlike leptonic dark
  matter and signatures at the LHC},''
  \href{http://dx.doi.org/10.1103/PhysRevD.93.115040}{{\em Phys. Rev. D} {\bf
  93} (2016) no.~11, 115040}, \href{http://arxiv.org/abs/1510.02760}{{\tt
  arXiv:1510.02760 [hep-ph]}}.

\bibitem{Cogollo:2015fpa}
D.~Cogollo, ``{Exotic Leptons: Collider and Muon Magnetic Moment
  Constraints},'' \href{http://dx.doi.org/10.1142/S0217751X15501870}{{\em Int.
  J. Mod. Phys. A} {\bf 30} (2015) no.~32, 1550187},
  \href{http://arxiv.org/abs/1508.01492}{{\tt arXiv:1508.01492 [hep-ph]}}.

\bibitem{Okada:2015nca}
H.~Okada and Y.~Orikasa, ``{Two-loop Neutrino Model with Exotic Leptons},''
  \href{http://dx.doi.org/10.1103/PhysRevD.93.013008}{{\em Phys. Rev. D} {\bf
  93} (2016) no.~1, 013008}, \href{http://arxiv.org/abs/1509.04068}{{\tt
  arXiv:1509.04068 [hep-ph]}}.

\bibitem{Dermisek:2015hue}
R.~Dermisek, E.~Lunghi, and S.~Shin, ``{New decay modes of heavy Higgs bosons
  in a two Higgs doublet model with vectorlike leptons},''
  \href{http://dx.doi.org/10.1007/JHEP05(2016)148}{{\em JHEP} {\bf 05} (2016)
  148}, \href{http://arxiv.org/abs/1512.07837}{{\tt arXiv:1512.07837
  [hep-ph]}}.

\bibitem{Majhi:2015bxa}
S.~Majhi, ``{NLO QCD corrections to excited lepton production at the LHC},''
  \href{http://arxiv.org/abs/1503.02455}{{\tt arXiv:1503.02455 [hep-ph]}}.

\bibitem{Huo:2015exa}
R.~Huo, ``{Standard Model Effective Field Theory: Integrating out Vector-Like
  Fermions},'' \href{http://dx.doi.org/10.1007/JHEP09(2015)037}{{\em JHEP} {\bf
  09} (2015)  037}, \href{http://arxiv.org/abs/1506.00840}{{\tt
  arXiv:1506.00840 [hep-ph]}}.

\bibitem{Ishiwata:2015cga}
K.~Ishiwata, Z.~Ligeti, and M.~B. Wise, ``{New Vector-Like Fermions and Flavor
  Physics},'' \href{http://dx.doi.org/10.1007/JHEP10(2015)027}{{\em JHEP} {\bf
  10} (2015)  027}, \href{http://arxiv.org/abs/1506.03484}{{\tt
  arXiv:1506.03484 [hep-ph]}}.

\bibitem{Fan:2015sza}
J.~Fan, S.~M. Koushiappas, and G.~Landsberg, ``{Pseudoscalar Portal Dark Matter
  and New Signatures of Vector-like Fermions},''
  \href{http://dx.doi.org/10.1007/JHEP01(2016)111}{{\em JHEP} {\bf 01} (2016)
  111}, \href{http://arxiv.org/abs/1507.06993}{{\tt arXiv:1507.06993
  [hep-ph]}}.

\bibitem{Kumar:2015tna}
N.~Kumar and S.~P. Martin, ``{Vectorlike Leptons at the Large Hadron
  Collider},'' \href{http://dx.doi.org/10.1103/PhysRevD.92.115018}{{\em Phys.
  Rev. D} {\bf 92} (2015) no.~11, 115018},
  \href{http://arxiv.org/abs/1510.03456}{{\tt arXiv:1510.03456 [hep-ph]}}.

\bibitem{Doff:2015nru}
A.~Doff and C.~Siqueira, ``{Composite Higgs Models, Technicolor and The Muon
  Anomalous Magnetic Moment},''
  \href{http://dx.doi.org/10.1016/j.physletb.2016.01.043}{{\em Phys. Lett. B}
  {\bf 754} (2016)  294--301}, \href{http://arxiv.org/abs/1512.03256}{{\tt
  arXiv:1512.03256 [hep-ph]}}.

\bibitem{Abdullah:2015zta}
M.~Abdullah and J.~L. Feng, ``{Reviving bino dark matter with vectorlike fourth
  generation particles},''
  \href{http://dx.doi.org/10.1103/PhysRevD.93.015006}{{\em Phys. Rev. D} {\bf
  93} (2016) no.~1, 015006}, \href{http://arxiv.org/abs/1510.06089}{{\tt
  arXiv:1510.06089 [hep-ph]}}.

\bibitem{Chaudhuri:2015pna}
A.~Chaudhuri, N.~Khan, B.~Mukhopadhyaya, and S.~Rakshit, ``{Dark matter
  candidate in an extended type III seesaw scenario},''
  \href{http://dx.doi.org/10.1103/PhysRevD.91.055024}{{\em Phys. Rev. D} {\bf
  91} (2015)  055024}, \href{http://arxiv.org/abs/1501.05885}{{\tt
  arXiv:1501.05885 [hep-ph]}}.

\bibitem{Ruiz:2015zca}
R.~Ruiz, ``{QCD Corrections to Pair Production of Type III Seesaw Leptons at
  Hadron Colliders},'' \href{http://dx.doi.org/10.1007/JHEP12(2015)165}{{\em
  JHEP} {\bf 12} (2015)  165}, \href{http://arxiv.org/abs/1509.05416}{{\tt
  arXiv:1509.05416 [hep-ph]}}.

\bibitem{Angelescu:2015uiz}
A.~Angelescu, A.~Djouadi, and G.~Moreau, ``{Scenarii for interpretations of the
  LHC diphoton excess: two Higgs doublets and vector-like quarks and
  leptons},'' \href{http://dx.doi.org/10.1016/j.physletb.2016.02.064}{{\em
  Phys. Lett. B} {\bf 756} (2016)  126--132},
  \href{http://arxiv.org/abs/1512.04921}{{\tt arXiv:1512.04921 [hep-ph]}}.

\bibitem{Chen:2016lsr}
C.-H. Chen and T.~Nomura, ``{Bounds on LFV Higgs decays in a vector-like lepton
  model and searching for doubly charged leptons at the LHC},''
  \href{http://dx.doi.org/10.1140/epjc/s10052-016-4197-3}{{\em Eur. Phys. J. C}
  {\bf 76} (2016) no.~6, 353}, \href{http://arxiv.org/abs/1602.07519}{{\tt
  arXiv:1602.07519 [hep-ph]}}.

\bibitem{Tang:2016oba}
Y.-L. Tang and S.-h. Zhu, ``{Vectorlike sneutrino dark matter},''
  \href{http://dx.doi.org/10.1103/PhysRevD.93.095006}{{\em Phys. Rev. D} {\bf
  93} (2016) no.~9, 095006}, \href{http://arxiv.org/abs/1604.01903}{{\tt
  arXiv:1604.01903 [hep-ph]}}.

\bibitem{Alvarado:2016par}
C.~Alvarado, R.~M. Capdevilla, A.~Delgado, and A.~Martin, ``{Minimal models of
  loop-induced lepton flavor violation in Higgs boson decays},''
  \href{http://dx.doi.org/10.1103/PhysRevD.94.075010}{{\em Phys. Rev. D} {\bf
  94} (2016) no.~7, 075010}, \href{http://arxiv.org/abs/1602.08506}{{\tt
  arXiv:1602.08506 [hep-ph]}}.

\bibitem{Zeng:2016tmw}
Q.-G. Zeng, ``{Production of the quintuplet leptons in future high energy
  linear e + e \ensuremath{-} colliders},''
  \href{http://dx.doi.org/10.1016/j.nuclphysb.2016.02.013}{{\em Nucl. Phys. B}
  {\bf 905} (2016)  251--263}.

\bibitem{Baek:2016kud}
S.~Baek, T.~Nomura, and H.~Okada, ``{An explanation of one-loop induced h
  \textrightarrow{} \ensuremath{\mu}\ensuremath{\tau} decay},''
  \href{http://dx.doi.org/10.1016/j.physletb.2016.05.055}{{\em Phys. Lett. B}
  {\bf 759} (2016)  91--98}, \href{http://arxiv.org/abs/1604.03738}{{\tt
  arXiv:1604.03738 [hep-ph]}}.

\bibitem{Cai:2016ymq}
Y.~Cai, J.~D. Clarke, R.~R. Volkas, and T.~T. Yanagida, ``{TeV-scale
  pseudo-Dirac seesaw mechanisms in an E$6$ inspired model},''
  \href{http://dx.doi.org/10.1103/PhysRevD.94.033003}{{\em Phys. Rev. D} {\bf
  94} (2016) no.~3, 033003}, \href{http://arxiv.org/abs/1605.02743}{{\tt
  arXiv:1605.02743 [hep-ph]}}.

\bibitem{Abdullah:2016avr}
M.~Abdullah, J.~L. Feng, S.~Iwamoto, and B.~Lillard, ``{Heavy bino dark matter
  and collider signals in the MSSM with vectorlike fourth-generation
  particles},'' \href{http://dx.doi.org/10.1103/PhysRevD.94.095018}{{\em Phys.
  Rev. D} {\bf 94} (2016) no.~9, 095018},
  \href{http://arxiv.org/abs/1608.00283}{{\tt arXiv:1608.00283 [hep-ph]}}.

\bibitem{Herrero-Garcia:2016uab}
J.~Herrero-Garcia, N.~Rius, and A.~Santamaria, ``{Higgs lepton flavour
  violation: UV completions and connection to neutrino masses},''
  \href{http://dx.doi.org/10.1007/JHEP11(2016)084}{{\em JHEP} {\bf 11} (2016)
  084}, \href{http://arxiv.org/abs/1605.06091}{{\tt arXiv:1605.06091
  [hep-ph]}}.

\bibitem{Boucenna:2016qad}
S.~M. Boucenna, A.~Celis, J.~Fuentes-Martin, A.~Vicente, and J.~Virto,
  ``{Phenomenology of an $SU(2) \times SU(2) \times U(1)$ model with
  lepton-flavour non-universality},''
  \href{http://dx.doi.org/10.1007/JHEP12(2016)059}{{\em JHEP} {\bf 12} (2016)
  059}, \href{http://arxiv.org/abs/1608.01349}{{\tt arXiv:1608.01349
  [hep-ph]}}.

\bibitem{Dermisek:2016via}
R.~Dermisek, E.~Lunghi, and S.~Shin, ``{New constraints and discovery potential
  for Higgs to Higgs cascade decays through vectorlike leptons},''
  \href{http://dx.doi.org/10.1007/JHEP10(2016)081}{{\em JHEP} {\bf 10} (2016)
  081}, \href{http://arxiv.org/abs/1608.00662}{{\tt arXiv:1608.00662
  [hep-ph]}}.

\bibitem{Hebbar:2016gab}
A.~Hebbar, G.~K. Leontaris, and Q.~Shafi, ``{Masses of Third Family Vector-like
  Quarks and Leptons in Yukawa-Unified $E_6$},''
  \href{http://dx.doi.org/10.1103/PhysRevD.93.111701}{{\em Phys. Rev. D} {\bf
  93} (2016) no.~11, 111701}, \href{http://arxiv.org/abs/1604.08328}{{\tt
  arXiv:1604.08328 [hep-ph]}}.

\bibitem{vonderPahlen:2016cbw}
F.~von~der Pahlen, G.~Palacio, D.~Restrepo, and O.~Zapata, ``{Radiative Type
  III Seesaw Model and its collider phenomenology},''
  \href{http://dx.doi.org/10.1103/PhysRevD.94.033005}{{\em Phys. Rev. D} {\bf
  94} (2016) no.~3, 033005}, \href{http://arxiv.org/abs/1605.01129}{{\tt
  arXiv:1605.01129 [hep-ph]}}.

\bibitem{Gopalakrishna:2016tku}
S.~Gopalakrishna and T.~S. Mukherjee, ``{The 750 GeV diphoton excess in a two
  Higgs doublet model and a singlet scalar model, with vector-like fermions,
  unitarity constraints, and dark matter implications},''
  \href{http://arxiv.org/abs/1604.05774}{{\tt arXiv:1604.05774 [hep-ph]}}.

\bibitem{Lee:2016wiy}
C.-H. Lee and R.~N. Mohapatra, ``{Vector-Like Quarks and Leptons, SU(5)
  $\otimes$ SU(5) Grand Unification, and Proton Decay},''
  \href{http://dx.doi.org/10.1007/JHEP02(2017)080}{{\em JHEP} {\bf 02} (2017)
  080}, \href{http://arxiv.org/abs/1611.05478}{{\tt arXiv:1611.05478
  [hep-ph]}}.

\bibitem{Bahrami:2016has}
S.~Bahrami, M.~Frank, D.~K. Ghosh, N.~Ghosh, and I.~Saha, ``{Dark matter and
  collider studies in the left-right symmetric model with vectorlike
  leptons},'' \href{http://dx.doi.org/10.1103/PhysRevD.95.095024}{{\em Phys.
  Rev. D} {\bf 95} (2017) no.~9, 095024},
  \href{http://arxiv.org/abs/1612.06334}{{\tt arXiv:1612.06334 [hep-ph]}}.

\bibitem{Raby:2017igl}
S.~Raby and A.~Trautner, ``{Vectorlike chiral fourth family to explain muon
  anomalies},'' \href{http://dx.doi.org/10.1103/PhysRevD.97.095006}{{\em Phys.
  Rev. D} {\bf 97} (2018) no.~9, 095006},
  \href{http://arxiv.org/abs/1712.09360}{{\tt arXiv:1712.09360 [hep-ph]}}.

\bibitem{Megias:2017dzd}
E.~Megias, M.~Quiros, and L.~Salas, ``{$g_\mu-2$ from Vector-Like Leptons in
  Warped Space},'' \href{http://dx.doi.org/10.1007/JHEP05(2017)016}{{\em JHEP}
  {\bf 05} (2017)  016}, \href{http://arxiv.org/abs/1701.05072}{{\tt
  arXiv:1701.05072 [hep-ph]}}.

\bibitem{Poh:2017tfo}
Z.~Poh and S.~Raby, ``{Vectorlike leptons: Muon g-2 anomaly, lepton flavor
  violation, Higgs boson decays, and lepton nonuniversality},''
  \href{http://dx.doi.org/10.1103/PhysRevD.96.015032}{{\em Phys. Rev. D} {\bf
  96} (2017) no.~1, 015032}, \href{http://arxiv.org/abs/1705.07007}{{\tt
  arXiv:1705.07007 [hep-ph]}}.

\bibitem{Bertuzzo:2017wam}
E.~Bertuzzo, P.~A.~N. Machado, Y.~F. Perez-Gonzalez, and R.~Zukanovich~Funchal,
  ``{Constraints from Triple Gauge Couplings on Vectorlike Leptons},''
  \href{http://dx.doi.org/10.1103/PhysRevD.96.035035}{{\em Phys. Rev. D} {\bf
  96} (2017) no.~3, 035035}, \href{http://arxiv.org/abs/1706.03073}{{\tt
  arXiv:1706.03073 [hep-ph]}}.

\bibitem{Mann:2017wzh}
R.~Mann, J.~Meffe, F.~Sannino, T.~Steele, Z.-W. Wang, and C.~Zhang,
  ``{Asymptotically Safe Standard Model via Vectorlike Fermions},''
  \href{http://dx.doi.org/10.1103/PhysRevLett.119.261802}{{\em Phys. Rev.
  Lett.} {\bf 119} (2017) no.~26, 261802},
  \href{http://arxiv.org/abs/1707.02942}{{\tt arXiv:1707.02942 [hep-ph]}}.

\bibitem{Goswami:2017jqs}
D.~Goswami and P.~Poulose, ``{Direct searches of Type III seesaw triplet
  fermions at high energy $e^+e^-$ collider},''
  \href{http://dx.doi.org/10.1140/epjc/s10052-017-5478-1}{{\em Eur. Phys. J. C}
  {\bf 78} (2018) no.~1, 42}, \href{http://arxiv.org/abs/1702.07215}{{\tt
  arXiv:1702.07215 [hep-ph]}}.

\bibitem{Ghosh:2017vhe}
A.~Ghosh, T.~Mondal, and B.~Mukhopadhyaya, ``{Heavy stable charged tracks as
  signatures of non-thermal dark matter at the LHC : a study in some
  non-supersymmetric scenarios},''
  \href{http://dx.doi.org/10.1007/JHEP12(2017)136}{{\em JHEP} {\bf 12} (2017)
  136}, \href{http://arxiv.org/abs/1706.06815}{{\tt arXiv:1706.06815
  [hep-ph]}}.

\bibitem{Dermisek:2017ihj}
R.~Dermisek and N.~McGinnis, ``{Mass scale of vectorlike matter and
  superpartners from IR fixed point predictions of gauge and top Yukawa
  couplings},'' \href{http://dx.doi.org/10.1103/PhysRevD.97.055009}{{\em Phys.
  Rev. D} {\bf 97} (2018) no.~5, 055009},
  \href{http://arxiv.org/abs/1712.03527}{{\tt arXiv:1712.03527 [hep-ph]}}.

\bibitem{Bhattacherjee:2017cxh}
B.~Bhattacherjee, P.~Byakti, A.~Kushwaha, and S.~K. Vempati, ``{Unification
  with Vector-like fermions and signals at LHC},''
  \href{http://dx.doi.org/10.1007/JHEP05(2018)090}{{\em JHEP} {\bf 05} (2018)
  090}, \href{http://arxiv.org/abs/1702.06417}{{\tt arXiv:1702.06417
  [hep-ph]}}.

\bibitem{Barrie:2017eyd}
N.~D. Barrie, A.~Kobakhidze, S.~Liang, M.~Talia, and L.~Wu, ``{Exotic Lepton
  Searches via Bound State Production at the LHC},''
  \href{http://dx.doi.org/10.1016/j.physletb.2018.03.087}{{\em Phys. Lett. B}
  {\bf 781} (2018)  364--367}, \href{http://arxiv.org/abs/1710.11396}{{\tt
  arXiv:1710.11396 [hep-ph]}}.

\bibitem{Borah:2017dqx}
D.~Borah, S.~Sadhukhan, and S.~Sahoo, ``{Lepton Portal Limit of Inert Higgs
  Doublet Dark Matter with Radiative Neutrino Mass},''
  \href{http://dx.doi.org/10.1016/j.physletb.2017.06.006}{{\em Phys. Lett. B}
  {\bf 771} (2017)  624--632}, \href{http://arxiv.org/abs/1703.08674}{{\tt
  arXiv:1703.08674 [hep-ph]}}.

\bibitem{Ellis:2017nrp}
J.~Ellis, M.~Fairbairn, and P.~Tunney, ``{Anomaly-Free Models for Flavour
  Anomalies},'' \href{http://dx.doi.org/10.1140/epjc/s10052-018-5725-0}{{\em
  Eur. Phys. J. C} {\bf 78} (2018) no.~3, 238},
  \href{http://arxiv.org/abs/1705.03447}{{\tt arXiv:1705.03447 [hep-ph]}}.

\bibitem{Agostinho:2017biv}
N.~R. Agostinho, O.~J.~P. Eboli, and M.~C. Gonzalez-Garcia, ``{LHC Run I Bounds
  on Minimal Lepton Flavour Violation in Type-III See-saw: A Case Study},''
  \href{http://dx.doi.org/10.1007/JHEP11(2017)118}{{\em JHEP} {\bf 11} (2017)
  118}, \href{http://arxiv.org/abs/1708.08456}{{\tt arXiv:1708.08456
  [hep-ph]}}.

\bibitem{Dhargyal:2017cqo}
L.~Dhargyal, ``{Phenomenology of $U(1)_{F}$ extension of inert-doublet model
  with exotic scalars and leptons},''
  \href{http://dx.doi.org/10.1140/epjc/s10052-018-5641-3}{{\em Eur. Phys. J. C}
  {\bf 78} (2018) no.~2, 150}, \href{http://arxiv.org/abs/1709.04452}{{\tt
  arXiv:1709.04452 [hep-ph]}}.

\bibitem{Dhargyal:2018bbc}
L.~Dhargyal and S.~K. Rai, ``{Implications of a vector-like lepton doublet and
  scalar Leptoquark on $R(D^{(*)})$},''
  \href{http://arxiv.org/abs/1806.01178}{{\tt arXiv:1806.01178 [hep-ph]}}.

\bibitem{Colucci:2018qml}
S.~Colucci, F.~Giacchino, M.~H.~G. Tytgat, and J.~Vandecasteele, ``{Radiative
  corrections to vectorlike portal dark matter},''
  \href{http://dx.doi.org/10.1103/PhysRevD.98.115029}{{\em Phys. Rev. D} {\bf
  98} (2018) no.~11, 115029}, \href{http://arxiv.org/abs/1805.10173}{{\tt
  arXiv:1805.10173 [hep-ph]}}.

\bibitem{Abada:2018nio}
A.~Abada and A.~M. Teixeira, ``{Heavy neutral leptons and high-intensity
  observables},'' \href{http://dx.doi.org/10.3389/fphy.2018.00142}{{\em Front.
  in Phys.} {\bf 6} (2018)  142}, \href{http://arxiv.org/abs/1812.08062}{{\tt
  arXiv:1812.08062 [hep-ph]}}.

\bibitem{Abada:2018sfh}
A.~Abada, N.~Bernal, M.~Losada, and X.~Marcano, ``{Inclusive Displaced Vertex
  Searches for Heavy Neutral Leptons at the LHC},''
  \href{http://dx.doi.org/10.1007/JHEP01(2019)093}{{\em JHEP} {\bf 01} (2019)
  093}, \href{http://arxiv.org/abs/1807.10024}{{\tt arXiv:1807.10024
  [hep-ph]}}.

\bibitem{Carone:2018eka}
C.~D. Carone, S.~Chaurasia, and T.~V.~B. Claringbold, ``{Dark sector portal
  with vectorlike leptons and flavor sequestering},''
  \href{http://dx.doi.org/10.1103/PhysRevD.99.015009}{{\em Phys. Rev. D} {\bf
  99} (2019) no.~1, 015009}, \href{http://arxiv.org/abs/1807.05288}{{\tt
  arXiv:1807.05288 [hep-ph]}}.

\bibitem{Aboubrahim:2018hll}
A.~Aboubrahim, T.~Ibrahim, A.~Itani, and P.~Nath, ``{Observables of low-lying
  supersymmetric vectorlike leptonic generations via loop corrections},''
  \href{http://dx.doi.org/10.1103/PhysRevD.98.075009}{{\em Phys. Rev. D} {\bf
  98} (2018) no.~7, 075009}, \href{http://arxiv.org/abs/1808.00071}{{\tt
  arXiv:1808.00071 [hep-ph]}}.

\bibitem{Xu:2018pnq}
F.-Z. Xu, W.~Zhang, J.~Li, and T.~Li, ``{Search for the vectorlike leptons in
  the U(1)$_X$ model inspired by the $B$-meson decay anomalies},''
  \href{http://dx.doi.org/10.1103/PhysRevD.98.115033}{{\em Phys. Rev. D} {\bf
  98} (2018) no.~11, 115033}, \href{http://arxiv.org/abs/1809.01472}{{\tt
  arXiv:1809.01472 [hep-ph]}}.

\bibitem{Dermisek:2018hxq}
R.~Derm\'\i{}\v{s}ek and N.~McGinnis, ``{Top-bottom-tau Yukawa coupling
  unification in the MSSM plus one vectorlike family and fermion masses as IR
  fixed points},'' \href{http://dx.doi.org/10.1103/PhysRevD.99.035033}{{\em
  Phys. Rev. D} {\bf 99} (2019) no.~3, 035033},
  \href{http://arxiv.org/abs/1810.12474}{{\tt arXiv:1810.12474 [hep-ph]}}.

\bibitem{Dhargyal:2018ppo}
L.~Dhargyal, ``{Phenomenological consequences of introducing new fermions with
  exotic charges to $R(K^{(*)})$, muon (g-2), the primordial Lithium problem,
  and dark matter},'' \href{http://arxiv.org/abs/1810.10611}{{\tt
  arXiv:1810.10611 [hep-ph]}}.

\bibitem{Angelescu:2018dkk}
A.~Angelescu and P.~Huang, ``{Multistep Strongly First Order Phase Transitions
  from New Fermions at the TeV Scale},''
  \href{http://dx.doi.org/10.1103/PhysRevD.99.055023}{{\em Phys. Rev. D} {\bf
  99} (2019) no.~5, 055023}, \href{http://arxiv.org/abs/1812.08293}{{\tt
  arXiv:1812.08293 [hep-ph]}}.

\bibitem{Song:2019aav}
J.~Song and Y.~W. Yoon, ``{$W\gamma$ decay of the elusive charged Higgs boson
  in the two-Higgs-doublet model with vectorlike fermions},''
  \href{http://dx.doi.org/10.1103/PhysRevD.100.055006}{{\em Phys. Rev. D} {\bf
  100} (2019) no.~5, 055006}, \href{http://arxiv.org/abs/1904.06521}{{\tt
  arXiv:1904.06521 [hep-ph]}}.

\bibitem{Zheng:2019kqu}
S.~Zheng, ``{Minimal Vectorlike Model in Supersymmetric Unification},''
  \href{http://dx.doi.org/10.1140/epjc/s10052-020-7843-8}{{\em Eur. Phys. J. C}
  {\bf 80} (2020) no.~3, 273}, \href{http://arxiv.org/abs/1904.10145}{{\tt
  arXiv:1904.10145 [hep-ph]}}.

\bibitem{Bhattiprolu:2019vdu}
P.~N. Bhattiprolu and S.~P. Martin, ``{Prospects for vectorlike leptons at
  future proton-proton colliders},''
  \href{http://dx.doi.org/10.1103/PhysRevD.100.015033}{{\em Phys. Rev. D} {\bf
  100} (2019) no.~1, 015033}, \href{http://arxiv.org/abs/1905.00498}{{\tt
  arXiv:1905.00498 [hep-ph]}}.

\bibitem{Kawamura:2019hxp}
J.~Kawamura, S.~Raby, and A.~Trautner, ``{Complete vectorlike fourth family
  with U(1)' : A global analysis},''
  \href{http://dx.doi.org/10.1103/PhysRevD.101.035026}{{\em Phys. Rev. D} {\bf
  101} (2020) no.~3, 035026}, \href{http://arxiv.org/abs/1911.11075}{{\tt
  arXiv:1911.11075 [hep-ph]}}.

\bibitem{Bell:2019mbn}
N.~F. Bell, M.~J. Dolan, L.~S. Friedrich, M.~J. Ramsey-Musolf, and R.~R.
  Volkas, ``{Electroweak Baryogenesis with Vector-like Leptons and Scalar
  Singlets},'' \href{http://dx.doi.org/10.1007/JHEP09(2019)012}{{\em JHEP} {\bf
  09} (2019)  012}, \href{http://arxiv.org/abs/1903.11255}{{\tt
  arXiv:1903.11255 [hep-ph]}}.

\bibitem{Arnan:2019uhr}
P.~Arnan, A.~Crivellin, M.~Fedele, and F.~Mescia, ``{Generic Loop Effects of
  New Scalars and Fermions in $b\to s\ell^+\ell^-$, $(g-2)_\mu$ and a
  Vector-like $4^{\rm th}$ Generation},''
  \href{http://dx.doi.org/10.1007/JHEP06(2019)118}{{\em JHEP} {\bf 06} (2019)
  118}, \href{http://arxiv.org/abs/1904.05890}{{\tt arXiv:1904.05890
  [hep-ph]}}.

\bibitem{Kowalska:2019qxm}
K.~Kowalska and D.~Kumar, ``{Road map through the desert: unification with
  vector-like fermions},''
  \href{http://dx.doi.org/10.1007/JHEP12(2019)094}{{\em JHEP} {\bf 12} (2019)
  094}, \href{http://arxiv.org/abs/1910.00847}{{\tt arXiv:1910.00847
  [hep-ph]}}.

\bibitem{Biswas:2019ygr}
A.~Biswas, D.~Borah, and D.~Nanda, ``{Type III seesaw for neutrino masses in
  U(1)$_{B-L}$ model with multi-component dark matter},''
  \href{http://dx.doi.org/10.1007/JHEP12(2019)109}{{\em JHEP} {\bf 12} (2019)
  109}, \href{http://arxiv.org/abs/1908.04308}{{\tt arXiv:1908.04308
  [hep-ph]}}.

\bibitem{Suematsu:2019kst}
D.~Suematsu, ``{Low scale leptogenesis in a hybrid model of the scotogenic type
  I and III seesaw models},''
  \href{http://dx.doi.org/10.1103/PhysRevD.100.055008}{{\em Phys. Rev. D} {\bf
  100} (2019) no.~5, 055008}, \href{http://arxiv.org/abs/1906.12008}{{\tt
  arXiv:1906.12008 [hep-ph]}}.

\bibitem{Biggio:2019eeo}
C.~Biggio, E.~Fernandez-Martinez, M.~Filaci, J.~Hernandez-Garcia, and
  J.~Lopez-Pavon, ``{Global Bounds on the Type-III Seesaw},''
  \href{http://dx.doi.org/10.1007/JHEP05(2020)022}{{\em JHEP} {\bf 05} (2020)
  022}, \href{http://arxiv.org/abs/1911.11790}{{\tt arXiv:1911.11790
  [hep-ph]}}.

\bibitem{Fuentes-Martin:2020hvc}
J.~Fuentes-Mart\'\i{}n, G.~Isidori, M.~K\"onig, and N.~Selimovi\'c, ``{Vector
  Leptoquarks Beyond Tree Level III: Vector-like Fermions and Flavor-Changing
  Transitions},'' \href{http://dx.doi.org/10.1103/PhysRevD.102.115015}{{\em
  Phys. Rev. D} {\bf 102} (2020)  115015},
  \href{http://arxiv.org/abs/2009.11296}{{\tt arXiv:2009.11296 [hep-ph]}}.

\bibitem{DeJesus:2020yqx}
A.~S. De~Jesus, S.~Kovalenko, F.~S. Queiroz, C.~Siqueira, and K.~Sinha,
  ``{Vectorlike leptons and inert scalar triplet: Lepton flavor violation,
  $g-2$, and collider searches},''
  \href{http://dx.doi.org/10.1103/PhysRevD.102.035004}{{\em Phys. Rev. D} {\bf
  102} (2020) no.~3, 035004}, \href{http://arxiv.org/abs/2004.01200}{{\tt
  arXiv:2004.01200 [hep-ph]}}.

\bibitem{Gu:2020nic}
P.-H. Gu, ``{Weinberg dimension-5 operator by vector-like lepton doublets},''
  \href{http://arxiv.org/abs/2006.08616}{{\tt arXiv:2006.08616 [hep-ph]}}.

\bibitem{Hieu:2020hti}
T.~M. Hieu, Q.~S. Sang, and T.~Q. Trang, ``{On a standard model extension with
  vector-like fermions and Abelian symmetry},''
  \href{http://dx.doi.org/10.15625/0868-3166/30/3/15071}{{\em Commun. in Phys.}
  {\bf 30} (2020) no.~3, 231--244}.

\bibitem{Ma:2020qyz}
E.~Ma, ``{Dileptonic Scalar Dark Matter and Exotic Leptons},''
  \href{http://dx.doi.org/10.1016/j.physletb.2021.136157}{{\em Phys. Lett. B}
  {\bf 815} (2021)  136157}, \href{http://arxiv.org/abs/2010.05054}{{\tt
  arXiv:2010.05054 [hep-ph]}}.

\bibitem{Bissmann:2020lge}
S.~Bi\ss{}mann, G.~Hiller, C.~Hormigos-Feliu, and D.~F. Litim, ``{Multi-lepton
  signatures of vector-like leptons with flavor},''
  \href{http://dx.doi.org/10.1140/epjc/s10052-021-08886-3}{{\em Eur. Phys. J.
  C} {\bf 81} (2021) no.~2, 101}, \href{http://arxiv.org/abs/2011.12964}{{\tt
  arXiv:2011.12964 [hep-ph]}}.

\bibitem{Frank:2020smf}
M.~Frank and I.~Saha, ``{Muon anomalous magnetic moment in two-Higgs-doublet
  models with vectorlike leptons},''
  \href{http://dx.doi.org/10.1103/PhysRevD.102.115034}{{\em Phys. Rev. D} {\bf
  102} (2020) no.~11, 115034}, \href{http://arxiv.org/abs/2008.11909}{{\tt
  arXiv:2008.11909 [hep-ph]}}.

\bibitem{Chala:2020odv}
M.~Chala, P.~Koz\'ow, M.~Ramos, and A.~Titov, ``{Effective field theory for
  vector-like leptons and its collider signals},''
  \href{http://dx.doi.org/10.1016/j.physletb.2020.135752}{{\em Phys. Lett. B}
  {\bf 809} (2020)  135752}, \href{http://arxiv.org/abs/2005.09655}{{\tt
  arXiv:2005.09655 [hep-ph]}}.

\bibitem{Crivellin:2020ebi}
A.~Crivellin, F.~Kirk, C.~A. Manzari, and M.~Montull, ``{Global Electroweak Fit
  and Vector-Like Leptons in Light of the Cabibbo Angle Anomaly},''
  \href{http://dx.doi.org/10.1007/JHEP12(2020)166}{{\em JHEP} {\bf 12} (2020)
  166}, \href{http://arxiv.org/abs/2008.01113}{{\tt arXiv:2008.01113
  [hep-ph]}}.

\bibitem{Ashanujjaman:2020tuv}
S.~Ashanujjaman and K.~Ghosh, ``{A genuine fermionic quintuplet seesaw model:
  phenomenological introduction},''
  \href{http://dx.doi.org/10.1007/JHEP06(2021)084}{{\em JHEP} {\bf 06} (2021)
  084}, \href{http://arxiv.org/abs/2012.15609}{{\tt arXiv:2012.15609
  [hep-ph]}}.

\bibitem{Angelescu:2020yzf}
A.~Angelescu and P.~Huang, ``{Integrating Out New Fermions at One Loop},''
  \href{http://dx.doi.org/10.1007/JHEP01(2021)049}{{\em JHEP} {\bf 01} (2021)
  049}, \href{http://arxiv.org/abs/2006.16532}{{\tt arXiv:2006.16532
  [hep-ph]}}.

\bibitem{Chun:2020uzw}
E.~J. Chun and T.~Mondal, ``{Explaining $g-2$ anomalies in two Higgs doublet
  model with vector-like leptons},''
  \href{http://dx.doi.org/10.1007/JHEP11(2020)077}{{\em JHEP} {\bf 11} (2020)
  077}, \href{http://arxiv.org/abs/2009.08314}{{\tt arXiv:2009.08314
  [hep-ph]}}.

\bibitem{Chakrabarty:2020jro}
N.~Chakrabarty, ``{Doubly charged scalars and vector-like leptons confronting
  the muon g-2 anomaly and Higgs vacuum stability},''
  \href{http://dx.doi.org/10.1140/epjp/s13360-021-02168-3}{{\em Eur. Phys. J.
  Plus} {\bf 136} (2021) no.~11, 1183},
  \href{http://arxiv.org/abs/2010.05215}{{\tt arXiv:2010.05215 [hep-ph]}}.

\bibitem{Endo:2020tkb}
M.~Endo and S.~Mishima, ``{Muon $g - 2$ and CKM unitarity in extra lepton
  models},'' \href{http://dx.doi.org/10.1007/JHEP08(2020)004}{{\em JHEP} {\bf
  08} (2020) no.~08, 004}, \href{http://arxiv.org/abs/2005.03933}{{\tt
  arXiv:2005.03933 [hep-ph]}}.

\bibitem{Dermisek:2020cod}
R.~Dermisek, K.~Hermanek, N.~McGinnis, and N.~McGinnis, ``{Highly Enhanced
  Contributions of Heavy Higgs Bosons and New Leptons to Muon
  $g$\ensuremath{-}2 and Prospects at Future Colliders},''
  \href{http://dx.doi.org/10.1103/PhysRevLett.126.191801}{{\em Phys. Rev.
  Lett.} {\bf 126} (2021) no.~19, 191801},
  \href{http://arxiv.org/abs/2011.11812}{{\tt arXiv:2011.11812 [hep-ph]}}.

\bibitem{Rehman:2020ana}
M.~Rehman, M.~E. Gomez, and O.~Panella, ``{Excited lepton triplet contribution
  to electroweak observables at one loop level},''
  \href{http://dx.doi.org/10.1140/epjc/s10052-021-09153-1}{{\em Eur. Phys. J.
  C} {\bf 81} (2021) no.~5, 392}, \href{http://arxiv.org/abs/2010.01808}{{\tt
  arXiv:2010.01808 [hep-ph]}}.

\bibitem{Matsedonskyi:2020mlz}
O.~Matsedonskyi and G.~Servant, ``{High-Temperature Electroweak Symmetry
  Non-Restoration from New Fermions and Implications for Baryogenesis},''
  \href{http://dx.doi.org/10.1007/JHEP09(2020)012}{{\em JHEP} {\bf 09} (2020)
  012}, \href{http://arxiv.org/abs/2002.05174}{{\tt arXiv:2002.05174
  [hep-ph]}}.

\bibitem{Matsui:2021khj}
T.~Matsui, T.~Nomura, and K.~Yagyu, ``{Flavor dependent U(1) symmetric Zee
  model with a vector-like lepton},''
  \href{http://dx.doi.org/10.1016/j.nuclphysb.2021.115523}{{\em Nucl. Phys. B}
  {\bf 971} (2021)  115523}, \href{http://arxiv.org/abs/2102.09247}{{\tt
  arXiv:2102.09247 [hep-ph]}}.

\bibitem{Guedes:2021oqx}
G.~Guedes and J.~Santiago, ``{New leptons with exotic decays: collider limits
  and dark matter complementarity},''
  \href{http://dx.doi.org/10.1007/JHEP01(2022)111}{{\em JHEP} {\bf 01} (2022)
  111}, \href{http://arxiv.org/abs/2107.03429}{{\tt arXiv:2107.03429
  [hep-ph]}}.

\bibitem{Jana:2020qzn}
S.~Jana, N.~Okada, and D.~Raut, ``{Displaced vertex and disappearing track
  signatures in type-III seesaw},''
  \href{http://dx.doi.org/10.1140/epjc/s10052-022-10855-3}{{\em Eur. Phys. J.
  C} {\bf 82} (2022) no.~10, 927}, \href{http://arxiv.org/abs/1911.09037}{{\tt
  arXiv:1911.09037 [hep-ph]}}.

\bibitem{Das:2020gnt}
A.~Das, S.~Mandal, and T.~Modak, ``{Testing triplet fermions at the
  electron-positron and electron-proton colliders using fat jet signatures},''
  \href{http://dx.doi.org/10.1103/PhysRevD.102.033001}{{\em Phys. Rev. D} {\bf
  102} (2020) no.~3, 033001}, \href{http://arxiv.org/abs/2005.02267}{{\tt
  arXiv:2005.02267 [hep-ph]}}.

\bibitem{Das:2020uer}
A.~Das and S.~Mandal, ``{Bounds on the triplet fermions in type-III seesaw and
  implications for collider searches},''
  \href{http://dx.doi.org/10.1016/j.nuclphysb.2021.115374}{{\em Nucl. Phys. B}
  {\bf 966} (2021)  115374}, \href{http://arxiv.org/abs/2006.04123}{{\tt
  arXiv:2006.04123 [hep-ph]}}.

\bibitem{Hernandez:2021tii}
A.~E.~C. Hern\'andez, S.~F. King, and H.~Lee, ``{Fermion mass hierarchies from
  vectorlike families with an extended 2HDM and a possible explanation for the
  electron and muon anomalous magnetic moments},''
  \href{http://dx.doi.org/10.1103/PhysRevD.103.115024}{{\em Phys. Rev. D} {\bf
  103} (2021) no.~11, 115024}, \href{http://arxiv.org/abs/2101.05819}{{\tt
  arXiv:2101.05819 [hep-ph]}}.

\bibitem{Kawamura:2021ygg}
J.~Kawamura and S.~Raby, ``{Signal of four muons or more from a vector-like
  lepton decaying to a muon-philic Z' boson at the LHC},''
  \href{http://dx.doi.org/10.1103/PhysRevD.104.035007}{{\em Phys. Rev. D} {\bf
  104} (2021) no.~3, 035007}, \href{http://arxiv.org/abs/2104.04461}{{\tt
  arXiv:2104.04461 [hep-ph]}}.

\bibitem{OsmanAcar:2021plv}
A.~Osman~Acar, O.~E. Delialioglu, and S.~Sultansoy, ``{A search for the first
  generation charged vector-like leptons at future colliders},''
  \href{http://arxiv.org/abs/2103.08222}{{\tt arXiv:2103.08222 [hep-ph]}}.

\bibitem{Sen:2021fha}
C.~Sen, P.~Bandyopadhyay, S.~Dutta, and A.~KT, ``{Displaced Higgs production in
  Type-III seesaw at the LHC/FCC, MATHUSLA and muon collider},''
  \href{http://dx.doi.org/10.1140/epjc/s10052-022-10176-5}{{\em Eur. Phys. J.
  C} {\bf 82} (2022) no.~3, 230}, \href{http://arxiv.org/abs/2107.12442}{{\tt
  arXiv:2107.12442 [hep-ph]}}.

\bibitem{Morais:2021ead}
A.~P. Morais, A.~Onofre, F.~F. Freitas, J.~a. Gon\c{c}alves, R.~Pasechnik, and
  R.~Santos, ``{Deep learning searches for vector-like leptons at the LHC and
  electron/muon colliders},''
  \href{http://dx.doi.org/10.1140/epjc/s10052-023-11314-3}{{\em Eur. Phys. J.
  C} {\bf 83} (2023) no.~3, 232}, \href{http://arxiv.org/abs/2108.03926}{{\tt
  arXiv:2108.03926 [hep-ph]}}.

\bibitem{Bharadwaj:2021tgp}
H.~Bharadwaj, S.~Dutta, and A.~Goyal, ``{Leptonic g \ensuremath{-} 2 anomaly in
  an extended Higgs sector with vector-like leptons},''
  \href{http://dx.doi.org/10.1007/JHEP11(2021)056}{{\em JHEP} {\bf 11} (2021)
  056}, \href{http://arxiv.org/abs/2109.02586}{{\tt arXiv:2109.02586
  [hep-ph]}}.

\bibitem{Dermisek:2021ajd}
R.~Dermisek, K.~Hermanek, and N.~McGinnis, ``{Muon g-2 in two-Higgs-doublet
  models with vectorlike leptons},''
  \href{http://dx.doi.org/10.1103/PhysRevD.104.055033}{{\em Phys. Rev. D} {\bf
  104} (2021) no.~5, 055033}, \href{http://arxiv.org/abs/2103.05645}{{\tt
  arXiv:2103.05645 [hep-ph]}}.

\bibitem{Iguro:2021kdw}
S.~Iguro, J.~Kawamura, S.~Okawa, and Y.~Omura, ``{TeV-scale vector leptoquark
  from Pati-Salam unification with vectorlike families},''
  \href{http://dx.doi.org/10.1103/PhysRevD.104.075008}{{\em Phys. Rev. D} {\bf
  104} (2021) no.~7, 075008}, \href{http://arxiv.org/abs/2103.11889}{{\tt
  arXiv:2103.11889 [hep-ph]}}.

\bibitem{Yang:2021dtc}
B.~Yang, J.~Li, M.~Wang, and L.~Shang, ``{Search for the singlet vectorlike
  lepton in semileptonic channel at future $e^+e^-$ colliders},''
  \href{http://dx.doi.org/10.1103/PhysRevD.104.055019}{{\em Phys. Rev. D} {\bf
  104} (2021) no.~5, 055019}.

\bibitem{CarcamoHernandez:2021yev}
A.~E. C\'arcamo~Hern\'andez, S.~F. King, and H.~Lee, ``{Z mediated flavor
  changing neutral currents with a fourth vectorlike family},''
  \href{http://dx.doi.org/10.1103/PhysRevD.105.015021}{{\em Phys. Rev. D} {\bf
  105} (2022) no.~1, 015021}, \href{http://arxiv.org/abs/2110.07630}{{\tt
  arXiv:2110.07630 [hep-ph]}}.

\bibitem{Lee:2021gnw}
H.~M. Lee, J.~Song, and K.~Yamashita, ``{Seesaw lepton masses and muon $g-2$
  from heavy vector-like leptons},''
  \href{http://dx.doi.org/10.1007/s40042-021-00339-0}{{\em J. Korean Phys.
  Soc.} {\bf 79} (2021) no.~12, 1121--1134},
  \href{http://arxiv.org/abs/2110.09942}{{\tt arXiv:2110.09942 [hep-ph]}}.

\bibitem{Cherchiglia:2021syq}
A.~L. Cherchiglia, G.~De~Conto, and C.~C. Nishi, ``{Leptonic CP violation from
  a vector-like lepton},''
  \href{http://dx.doi.org/10.1007/JHEP03(2022)010}{{\em JHEP} {\bf 03} (2022)
  010}, \href{http://arxiv.org/abs/2112.03943}{{\tt arXiv:2112.03943
  [hep-ph]}}.

\bibitem{Karozas:2021dlh}
A.~Karozas, G.~K. Leontaris, and I.~Tavellaris, ``{SU(5)\texttimes{}U(1)'
  Models with a Vector-Like Fermion Family},''
  \href{http://dx.doi.org/10.3390/universe7100356}{{\em Universe} {\bf 7}
  (2021) no.~10, 356}, \href{http://arxiv.org/abs/2108.10989}{{\tt
  arXiv:2108.10989 [hep-ph]}}.

\bibitem{Sahoo:2021vug}
S.~Sahoo, S.~Singirala, and R.~Mohanta, ``{Dark matter and flavor anomalies in
  the light of vector-like fermions and scalar leptoquark},''
  \href{http://arxiv.org/abs/2112.04382}{{\tt arXiv:2112.04382 [hep-ph]}}.

\bibitem{Ashanujjaman:2021zrh}
S.~Ashanujjaman and K.~Ghosh, ``{Type-III see-saw: Search for triplet fermions
  in final states with multiple leptons and fat-jets at 13 TeV LHC},''
  \href{http://dx.doi.org/10.1016/j.physletb.2022.136889}{{\em Phys. Lett. B}
  {\bf 825} (2022)  136889}, \href{http://arxiv.org/abs/2111.07949}{{\tt
  arXiv:2111.07949 [hep-ph]}}.

\bibitem{Criado:2021qpd}
J.~C. Criado, A.~Djouadi, N.~Koivunen, K.~M\"u\"ursepp, M.~Raidal, and
  H.~Veerm\"ae, ``{Confronting spin-3/2 and other new fermions with the muon
  g-2 measurement},''
  \href{http://dx.doi.org/10.1016/j.physletb.2021.136491}{{\em Phys. Lett. B}
  {\bf 820} (2021)  136491}, \href{http://arxiv.org/abs/2104.03231}{{\tt
  arXiv:2104.03231 [hep-ph]}}.

\bibitem{Djouadi:2021wvb}
A.~Djouadi, J.~C. Criado, N.~Koivunen, K.~M\"u\"ursepp, M.~Raidal, and
  H.~Veerm\"ae, ``{New fermions in the light of the (g \ensuremath{-}
  2)\ensuremath{\mu}},'' \href{http://dx.doi.org/10.3389/fphy.2022.964131}{{\em
  Front. in Phys.} {\bf 10} (2022)  964131},
  \href{http://arxiv.org/abs/2112.12502}{{\tt arXiv:2112.12502 [hep-ph]}}.

\bibitem{Escribano:2021css}
P.~Escribano, J.~Terol-Calvo, and A.~Vicente, ``{$\boldsymbol{(g-2)_{e,\mu}}$
  in an extended inverse type-III seesaw model},''
  \href{http://dx.doi.org/10.1103/PhysRevD.103.115018}{{\em Phys. Rev. D} {\bf
  103} (2021) no.~11, 115018}, \href{http://arxiv.org/abs/2104.03705}{{\tt
  arXiv:2104.03705 [hep-ph]}}.

\bibitem{Bonilla:2021ize}
C.~Bonilla, A.~E. C\'arcamo~Hern\'andez, J.~a. Gon\c{c}alves, F.~F. Freitas,
  A.~P. Morais, and R.~Pasechnik, ``{Collider signatures of vector-like
  fermions from a flavor symmetric model},''
  \href{http://dx.doi.org/10.1007/JHEP01(2022)154}{{\em JHEP} {\bf 01} (2022)
  154}, \href{http://arxiv.org/abs/2107.14165}{{\tt arXiv:2107.14165
  [hep-ph]}}.

\bibitem{Cacciapaglia:2021gff}
G.~Cacciapaglia, C.~Cot, and F.~Sannino, ``{Naturalness of lepton
  non-universality and muon g-2},''
  \href{http://dx.doi.org/10.1016/j.physletb.2021.136864}{{\em Phys. Lett. B}
  {\bf 825} (2022)  136864}, \href{http://arxiv.org/abs/2104.08818}{{\tt
  arXiv:2104.08818 [hep-ph]}}.

\bibitem{Koren:2021ypn}
S.~Koren and U.~\"Oktem, ``{Searching for exotic production of Higgs boson + X
  to map out new physics},''
  \href{http://dx.doi.org/10.1103/PhysRevD.104.035033}{{\em Phys. Rev. D} {\bf
  104} (2021) no.~3, 035033}, \href{http://arxiv.org/abs/2102.06212}{{\tt
  arXiv:2102.06212 [hep-ph]}}.

\bibitem{Baspehlivan:2022qet}
F.~Baspehlivan, B.~Dagli, O.~E. Delialioglu, and S.~Sultansoy, ``{Why should we
  search for vector-like leptons?},''
  \href{http://arxiv.org/abs/2201.08251}{{\tt arXiv:2201.08251 [hep-ph]}}.

\bibitem{Chakraborty:2022uqo}
I.~Chakraborty, D.~K. Ghosh, N.~Ghosh, and S.~K. Rai, ``{Signals for
  vector-like leptons in an $S_3$-symmetric 2HDM at ILC},''
  \href{http://dx.doi.org/10.1140/epjc/s10052-022-10477-9}{{\em Eur. Phys. J.
  C} {\bf 82} (2022) no.~6, 538}, \href{http://arxiv.org/abs/2201.11646}{{\tt
  arXiv:2201.11646 [hep-ph]}}.

\bibitem{Drewes:2022akb}
M.~Drewes, J.~Klari\'c, and J.~L\'opez-Pav\'on, ``{New benchmark models for
  heavy neutral lepton searches},''
  \href{http://dx.doi.org/10.1140/epjc/s10052-022-11100-7}{{\em Eur. Phys. J.
  C} {\bf 82} (2022) no.~12, 1176}, \href{http://arxiv.org/abs/2207.02742}{{\tt
  arXiv:2207.02742 [hep-ph]}}.

\bibitem{Lee:2022nqz}
H.~M. Lee and K.~Yamashita, ``{A model of vector-like leptons for the muon
  $g-2$ and the W boson mass},''
  \href{http://dx.doi.org/10.1140/epjc/s10052-022-10635-z}{{\em Eur. Phys. J.
  C} {\bf 82} (2022) no.~8, 661}, \href{http://arxiv.org/abs/2204.05024}{{\tt
  arXiv:2204.05024 [hep-ph]}}.

\bibitem{Zhou:2022cql}
Q.~Zhou, X.-F. Han, and L.~Wang, ``{The CDF W-mass, muon $g-2$, and dark matter
  in a $U(1)_{L_\mu -L_\tau }$ model with vector-like leptons},''
  \href{http://dx.doi.org/10.1140/epjc/s10052-022-11051-z}{{\em Eur. Phys. J.
  C} {\bf 82} (2022) no.~12, 1135}, \href{http://arxiv.org/abs/2204.13027}{{\tt
  arXiv:2204.13027 [hep-ph]}}.

\bibitem{Li:2022vsg}
H.~Li, J.~Chao, and G.~Zhang, ``{Search for the singlet vector-like lepton in
  W\ensuremath{\nu}\ensuremath{\tau} channel at the TeV $e^+e^-$ colliders},''
  \href{http://dx.doi.org/10.1142/S0217751X22502244}{{\em Int. J. Mod. Phys. A}
  {\bf 37} (2022) no.~36, 2250224}.

\bibitem{Carvunis:2022yur}
A.~Carvunis, N.~McGinnis, and D.~E. Morrissey, ``{Relic challenges for
  vector-like fermions as connectors to a dark sector},''
  \href{http://dx.doi.org/10.1007/JHEP01(2023)014}{{\em JHEP} {\bf 01} (2023)
  014}, \href{http://arxiv.org/abs/2209.14305}{{\tt arXiv:2209.14305
  [hep-ph]}}.

\bibitem{Raju:2022zlv}
M.~Raju, A.~Mukherjee, and J.~P. Saha, ``{Investigation of $(g-2)_{\mu}$
  anomaly in the $\mu$-specific 2HDM with Vector like leptons and the
  phenomenological implications},'' \href{http://arxiv.org/abs/2207.02825}{{\tt
  arXiv:2207.02825 [hep-ph]}}.

\bibitem{Dermisek:2022xal}
R.~Dermisek, J.~Kawamura, E.~Lunghi, N.~McGinnis, and S.~Shin, ``{Leptonic
  cascade decays of a heavy Higgs boson through vectorlike leptons at the
  LHC},'' \href{http://dx.doi.org/10.1007/JHEP10(2022)138}{{\em JHEP} {\bf 10}
  (2022)  138}, \href{http://arxiv.org/abs/2204.13272}{{\tt arXiv:2204.13272
  [hep-ph]}}.

\bibitem{Brune:2022gnu}
T.~Brune, T.~W. Kephart, and H.~P\"as, ``{Muon g-2 Anomaly from Vectorlike
  Leptons in TeV scale Trinification and $E_6$ models},''
  \href{http://arxiv.org/abs/2205.05566}{{\tt arXiv:2205.05566 [hep-ph]}}.

\bibitem{Kawamura:2022fhm}
J.~Kawamura and S.~Raby, ``{W mass in a model with vectorlike leptons and
  U(1)'},'' \href{http://dx.doi.org/10.1103/PhysRevD.106.035009}{{\em Phys.
  Rev. D} {\bf 106} (2022) no.~3, 035009},
  \href{http://arxiv.org/abs/2205.10480}{{\tt arXiv:2205.10480 [hep-ph]}}.

\bibitem{Abada:2022wvh}
A.~Abada, P.~Escribano, X.~Marcano, and G.~Piazza, ``{Collider searches for
  heavy neutral leptons: beyond simplified scenarios},''
  \href{http://dx.doi.org/10.1140/epjc/s10052-022-11011-7}{{\em Eur. Phys. J.
  C} {\bf 82} (2022) no.~11, 1030}, \href{http://arxiv.org/abs/2208.13882}{{\tt
  arXiv:2208.13882 [hep-ph]}}.

\bibitem{Vatsyayan:2022rth}
D.~Vatsyayan and S.~Goswami, ``{Lowering the scale of fermion triplet
  leptogenesis with two Higgs doublets},''
  \href{http://dx.doi.org/10.1103/PhysRevD.107.035014}{{\em Phys. Rev. D} {\bf
  107} (2023) no.~3, 035014}, \href{http://arxiv.org/abs/2208.12011}{{\tt
  arXiv:2208.12011 [hep-ph]}}.

\bibitem{Li:2022kkc}
T.~Li, H.~Qin, C.-Y. Yao, and M.~Yuan, ``{Probing heavy triplet leptons of the
  type-III seesaw mechanism at future muon colliders},''
  \href{http://dx.doi.org/10.1103/PhysRevD.106.035021}{{\em Phys. Rev. D} {\bf
  106} (2022) no.~3, 035021}, \href{http://arxiv.org/abs/2205.04214}{{\tt
  arXiv:2205.04214 [hep-ph]}}.

\bibitem{Ghosh:2022vpb}
N.~Ghosh, S.~K. Rai, and T.~Samui, ``{Collider signatures of a scalar
  leptoquark and vectorlike lepton in light of muon anomaly},''
  \href{http://dx.doi.org/10.1103/PhysRevD.107.035028}{{\em Phys. Rev. D} {\bf
  107} (2023) no.~3, 035028}, \href{http://arxiv.org/abs/2206.11718}{{\tt
  arXiv:2206.11718 [hep-ph]}}.

\bibitem{Ashanujjaman:2022cso}
S.~Ashanujjaman, D.~Choudhury, and K.~Ghosh, ``{Search for exotic leptons in
  final states with two or three leptons and fat-jets at 13 TeV LHC},''
  \href{http://dx.doi.org/10.1007/JHEP04(2022)150}{{\em JHEP} {\bf 04} (2022)
  150}, \href{http://arxiv.org/abs/2201.09645}{{\tt arXiv:2201.09645
  [hep-ph]}}.

\bibitem{Li:2022xty}
D.~W. Li, C.~X. Yue, Y.~Q. Wang, and X.~C. Sun, ``{Flavorful vector-like
  leptons and the lepton-flavor-violating decays
  Z~\textrightarrow{}~l$_{i}$l$_{j}$},''
  \href{http://dx.doi.org/10.1209/0295-5075/ac5360}{{\em EPL} {\bf 137} (2022)
  no.~3, 34003}.

\bibitem{Li:2022hzl}
H.~Li, J.~Chao, and G.~Zhang, ``{Search for the singlet vector-like lepton
  through the pair production in the W\ensuremath{\nu} $_{\tau}$ channel at the
  ILC},'' \href{http://dx.doi.org/10.1209/0295-5075/ac8ecf}{{\em EPL} {\bf 139}
  (2022) no.~6, 64001}.

\bibitem{Hamaguchi:2022byw}
K.~Hamaguchi, N.~Nagata, G.~Osaki, and S.-Y. Tseng, ``{Probing new physics in
  the vector-like lepton model by lepton electric dipole moments},''
  \href{http://dx.doi.org/10.1007/JHEP01(2023)100}{{\em JHEP} {\bf 01} (2023)
  100}, \href{http://arxiv.org/abs/2211.16800}{{\tt arXiv:2211.16800
  [hep-ph]}}.

\bibitem{Arora:2022uof}
S.~Arora, M.~Kashav, S.~Verma, and B.~C. Chauhan, ``{Muon (g \ensuremath{-} 2)
  and the W-boson mass anomaly in a model based on Z4 symmetry with a
  vector-like fermion},'' \href{http://dx.doi.org/10.1093/ptep/ptac144}{{\em
  PTEP} {\bf 2022} (2022) no.~11, 113B06},
  \href{http://arxiv.org/abs/2207.08580}{{\tt arXiv:2207.08580 [hep-ph]}}.

\bibitem{Coy:2023xtw}
R.~Coy, ``{A forgotten fermion: the hypercharge \ensuremath{-}3/2 doublet, its
  phenomenology and connections to dark matter},''
  \href{http://dx.doi.org/10.1007/JHEP04(2023)133}{{\em JHEP} {\bf 04} (2023)
  133}, \href{http://arxiv.org/abs/2301.12120}{{\tt arXiv:2301.12120
  [hep-ph]}}.

\bibitem{Arcadi:2023qgf}
G.~Arcadi and A.~Djouadi, ``{A model for fermionic dark matter addressing both
  the CDF MW and the (g \ensuremath{-} 2)\ensuremath{\mu} anomalies},''
  \href{http://dx.doi.org/10.3389/fphy.2023.1143932}{{\em Front. in Phys.} {\bf
  11} (2023)  1143932}, \href{http://arxiv.org/abs/2301.06476}{{\tt
  arXiv:2301.06476 [hep-ph]}}.

\bibitem{De:2023sqa}
B.~De, ``{Dark Contributions to $h \to \mu^+\mu^-$ in the Presence of a
  $\mu$-Flavored Vector-Like Lepton},''
  \href{http://dx.doi.org/10.1088/0256-307X/40/4/049501}{{\em Chin. Phys.
  Lett.} {\bf 40} (2023) no.~4, 049501}.

\bibitem{Ajjath:2023ugn}
A.~H. Ajjath, B.~Fuks, H.-S. Shao, and Y.~Simon, ``{Precision predictions for
  exotic lepton production at the Large Hadron Collider},''
  \href{http://dx.doi.org/10.1103/PhysRevD.107.075011}{{\em Phys. Rev. D} {\bf
  107} (2023) no.~7, 075011}, \href{http://arxiv.org/abs/2301.03640}{{\tt
  arXiv:2301.03640 [hep-ph]}}.

\bibitem{Shang:2023rfv}
L.~Shang, J.~Li, X.~Jia, and B.~Yang, ``{Search for Pair-Produced vectorlike
  lepton singlet at the ILC by the XGBoost method},''
  \href{http://dx.doi.org/10.1016/j.nuclphysb.2022.116071}{{\em Nucl. Phys. B}
  {\bf 987} (2023)  116071}.

\bibitem{Bernreuther:2023uxh}
E.~Bernreuther and B.~A. Dobrescu, ``{Vectorlike leptons and long-lived bosons
  at the LHC},'' \href{http://arxiv.org/abs/2304.08509}{{\tt arXiv:2304.08509
  [hep-ph]}}.

\bibitem{Guo:2023jkz}
Q.~Guo, L.~Gao, Y.~Mao, and Q.~Li, ``{Search for vector-like leptons at a Muon
  Collider},'' \href{http://arxiv.org/abs/2304.01885}{{\tt arXiv:2304.01885
  [hep-ph]}}.

\bibitem{Antusch:2023kli}
S.~Antusch, K.~Hinze, and S.~Saad, ``{Quark-lepton Yukawa ratios and nucleon
  decay in SU(5) GUTs with type-III seesaw},''
  \href{http://dx.doi.org/10.1016/j.nuclphysb.2023.116195}{{\em Nucl. Phys. B}
  {\bf 991} (2023)  116195}, \href{http://arxiv.org/abs/2301.03601}{{\tt
  arXiv:2301.03601 [hep-ph]}}.

\bibitem{Mishra:2023cjc}
P.~Mishra, M.~K. Behera, and R.~Mohanta, ``{Neutrino phenomenology, W mass
  anomaly \& muon $(g-2)$ in minimal type-III seesaw using $T^\prime$ modular
  symmetry},'' \href{http://arxiv.org/abs/2302.00494}{{\tt arXiv:2302.00494
  [hep-ph]}}.

\bibitem{Mahapatra:2023zhi}
S.~Mahapatra, R.~N. Mohapatra, and N.~Sahu, ``{Gauged $L_e-L_{\mu}-L_{\tau}$
  symmetry, fourth generation, neutrino mass and dark matter},''
  \href{http://arxiv.org/abs/2302.01784}{{\tt arXiv:2302.01784 [hep-ph]}}.

\bibitem{Kulkarni:2023fyq}
H.~Kulkarni and S.~Raby, ``{An SU(5) \texttimes{} U(1)' SUSY GUT with a
  \textquotedblleft{}vector-like chiral\textquotedblright{} fourth family to
  fit all low energy data, including the muon g \ensuremath{-} 2},''
  \href{http://dx.doi.org/10.1007/JHEP05(2023)152}{{\em JHEP} {\bf 05} (2023)
  152}, \href{http://arxiv.org/abs/2303.07209}{{\tt arXiv:2303.07209
  [hep-ph]}}.

\bibitem{Das:2023tna}
A.~Das, S.~Mandal, and S.~Shil, ``{Testing electroweak scale seesaw models at
  $e^{-} \gamma$ and $\gamma \gamma$ colliders},''
  \href{http://arxiv.org/abs/2304.06298}{{\tt arXiv:2304.06298 [hep-ph]}}.

\bibitem{Isidori:2001bm}
G.~Isidori, G.~Ridolfi, and A.~Strumia, ``{On the metastability of the standard
  model vacuum},'' \href{http://dx.doi.org/10.1016/S0550-3213(01)00302-9}{{\em
  Nucl. Phys. B} {\bf 609} (2001)  387--409},
  \href{http://arxiv.org/abs/hep-ph/0104016}{{\tt arXiv:hep-ph/0104016}}.

\bibitem{Degrassi:2012ry}
G.~Degrassi, S.~Di~Vita, J.~Elias-Miro, J.~R. Espinosa, G.~F. Giudice,
  G.~Isidori, and A.~Strumia, ``{Higgs mass and vacuum stability in the
  Standard Model at NNLO},''
  \href{http://dx.doi.org/10.1007/JHEP08(2012)098}{{\em JHEP} {\bf 08} (2012)
  098}, \href{http://arxiv.org/abs/1205.6497}{{\tt arXiv:1205.6497 [hep-ph]}}.

\bibitem{Jackiw:1974cv}
R.~Jackiw, ``{Functional evaluation of the effective potential},''
  \href{http://dx.doi.org/10.1103/PhysRevD.9.1686}{{\em Phys. Rev. D} {\bf 9}
  (1974)  1686}.

\bibitem{Andreassen:2014gha}
A.~Andreassen, W.~Frost, and M.~D. Schwartz, ``{Consistent Use of the Standard
  Model Effective Potential},''
  \href{http://dx.doi.org/10.1103/PhysRevLett.113.241801}{{\em Phys. Rev.
  Lett.} {\bf 113} (2014) no.~24, 241801},
  \href{http://arxiv.org/abs/1408.0292}{{\tt arXiv:1408.0292 [hep-ph]}}.

\bibitem{Andreassen:2014eha}
A.~Andreassen, W.~Frost, and M.~D. Schwartz, ``{Consistent Use of Effective
  Potentials},'' \href{http://dx.doi.org/10.1103/PhysRevD.91.016009}{{\em Phys.
  Rev. D} {\bf 91} (2015) no.~1, 016009},
  \href{http://arxiv.org/abs/1408.0287}{{\tt arXiv:1408.0287 [hep-ph]}}.

\bibitem{CMS:2016kce}
{\bf CMS} Collaboration, V.~Khachatryan {\em et al.}, ``{Search for long-lived
  charged particles in proton-proton collisions at $\sqrt s=$ 13 TeV},''
  \href{http://dx.doi.org/10.1103/PhysRevD.94.112004}{{\em Phys. Rev. D} {\bf
  94} (2016) no.~11, 112004}, \href{http://arxiv.org/abs/1609.08382}{{\tt
  arXiv:1609.08382 [hep-ex]}}.

\bibitem{ATLAS:2019gqq}
{\bf ATLAS} Collaboration, M.~Aaboud {\em et al.}, ``{Search for heavy charged
  long-lived particles in the ATLAS detector in 36.1 fb$^{-1}$ of proton-proton
  collision data at $\sqrt{s} = 13$ TeV},''
  \href{http://dx.doi.org/10.1103/PhysRevD.99.092007}{{\em Phys. Rev. D} {\bf
  99} (2019) no.~9, 092007}, \href{http://arxiv.org/abs/1902.01636}{{\tt
  arXiv:1902.01636 [hep-ex]}}.

\bibitem{ATLAS:2022pib}
{\bf ATLAS} Collaboration, ``{Search for heavy, long-lived, charged particles
  with large ionisation energy loss in $pp$ collisions at $\sqrt{s} =
  13~\text{TeV}$ using the ATLAS experiment and the full Run 2 dataset},''
  \href{http://arxiv.org/abs/2205.06013}{{\tt arXiv:2205.06013 [hep-ex]}}.

\bibitem{CMS:2019hsm}
{\bf CMS} Collaboration, A.~M. Sirunyan {\em et al.}, ``{Search for vector-like
  leptons in multilepton final states in proton-proton collisions at $\sqrt{s}$
  = 13 TeV},'' \href{http://dx.doi.org/10.1103/PhysRevD.100.052003}{{\em Phys.
  Rev. D} {\bf 100} (2019) no.~5, 052003},
  \href{http://arxiv.org/abs/1905.10853}{{\tt arXiv:1905.10853 [hep-ex]}}.

\bibitem{ATLAS:2018eui}
{\bf ATLAS} Collaboration, M.~Aaboud {\em et al.}, ``{Search for
  chargino-neutralino production using recursive jigsaw reconstruction in final
  states with two or three charged leptons in proton-proton collisions at
  $\sqrt{s}=13$ TeV with the ATLAS detector},''
  \href{http://dx.doi.org/10.1103/PhysRevD.98.092012}{{\em Phys. Rev. D} {\bf
  98} (2018) no.~9, 092012}, \href{http://arxiv.org/abs/1806.02293}{{\tt
  arXiv:1806.02293 [hep-ex]}}.

\bibitem{ATLAS:2019lng}
{\bf ATLAS} Collaboration, G.~Aad {\em et al.}, ``{Searches for electroweak
  production of supersymmetric particles with compressed mass spectra in
  $\sqrt{s}=$ 13 TeV $pp$ collisions with the ATLAS detector},''
  \href{http://dx.doi.org/10.1103/PhysRevD.101.052005}{{\em Phys. Rev. D} {\bf
  101} (2020) no.~5, 052005}, \href{http://arxiv.org/abs/1911.12606}{{\tt
  arXiv:1911.12606 [hep-ex]}}.

\bibitem{ATLAS:2021moa}
{\bf ATLAS} Collaboration, G.~Aad {\em et al.}, ``{Search for
  chargino\textendash{}neutralino pair production in final states with three
  leptons and missing transverse momentum in $\sqrt{s} = 13$~TeV pp collisions
  with the ATLAS detector},''
  \href{http://dx.doi.org/10.1140/epjc/s10052-021-09749-7}{{\em Eur. Phys. J.
  C} {\bf 81} (2021) no.~12, 1118}, \href{http://arxiv.org/abs/2106.01676}{{\tt
  arXiv:2106.01676 [hep-ex]}}.

\bibitem{ATLAS:2021yqv}
{\bf ATLAS} Collaboration, G.~Aad {\em et al.}, ``{Search for charginos and
  neutralinos in final states with two boosted hadronically decaying bosons and
  missing transverse momentum in $pp$ collisions at $\sqrt {s}$ = 13\,\,TeV
  with the ATLAS detector},''
  \href{http://dx.doi.org/10.1103/PhysRevD.104.112010}{{\em Phys. Rev. D} {\bf
  104} (2021) no.~11, 112010}, \href{http://arxiv.org/abs/2108.07586}{{\tt
  arXiv:2108.07586 [hep-ex]}}.

\bibitem{ATLAS:2020pgy}
{\bf ATLAS} Collaboration, G.~Aad {\em et al.}, ``{Search for direct production
  of electroweakinos in final states with one lepton, missing transverse
  momentum and a Higgs boson decaying into two $b$-jets in $pp$ collisions at
  $\sqrt{s}=13$ TeV with the ATLAS detector},''
  \href{http://dx.doi.org/10.1140/epjc/s10052-020-8050-3}{{\em Eur. Phys. J. C}
  {\bf 80} (2020) no.~8, 691}, \href{http://arxiv.org/abs/1909.09226}{{\tt
  arXiv:1909.09226 [hep-ex]}}.

\bibitem{ATLAS:2019lff}
{\bf ATLAS} Collaboration, G.~Aad {\em et al.}, ``{Search for electroweak
  production of charginos and sleptons decaying into final states with two
  leptons and missing transverse momentum in $\sqrt{s}=13$ TeV $pp$ collisions
  using the ATLAS detector},''
  \href{http://dx.doi.org/10.1140/epjc/s10052-019-7594-6}{{\em Eur. Phys. J. C}
  {\bf 80} (2020) no.~2, 123}, \href{http://arxiv.org/abs/1908.08215}{{\tt
  arXiv:1908.08215 [hep-ex]}}.

\bibitem{ATLAS:2022zwa}
{\bf ATLAS} Collaboration, ``{Searches for new phenomena in events with two
  leptons, jets, and missing transverse momentum in $139~\text{fb}^{-1}$ of
  $\sqrt{s}=13~$TeV $pp$ collisions with the ATLAS detector},''
  \href{http://arxiv.org/abs/2204.13072}{{\tt arXiv:2204.13072 [hep-ex]}}.

\bibitem{ATLAS:2022hbt}
{\bf ATLAS} Collaboration, ``{Search for direct pair production of sleptons and
  charginos decaying to two leptons and neutralinos with mass splittings near
  the $W$-boson mass in ${\sqrt{s}=13\,}$TeV $pp$ collisions with the ATLAS
  detector},'' \href{http://arxiv.org/abs/2209.13935}{{\tt arXiv:2209.13935
  [hep-ex]}}.

\bibitem{CMS:2018szt}
{\bf CMS} Collaboration, A.~M. Sirunyan {\em et al.}, ``{Combined search for
  electroweak production of charginos and neutralinos in proton-proton
  collisions at $\sqrt{s} =$ 13 TeV},''
  \href{http://dx.doi.org/10.1007/JHEP03(2018)160}{{\em JHEP} {\bf 03} (2018)
  160}, \href{http://arxiv.org/abs/1801.03957}{{\tt arXiv:1801.03957
  [hep-ex]}}.

\bibitem{CMS:2020bfa}
{\bf CMS} Collaboration, A.~M. Sirunyan {\em et al.}, ``{Search for
  supersymmetry in final states with two oppositely charged same-flavor leptons
  and missing transverse momentum in proton-proton collisions at $\sqrt{s} =$
  13 TeV},'' \href{http://dx.doi.org/10.1007/JHEP04(2021)123}{{\em JHEP} {\bf
  04} (2021)  123}, \href{http://arxiv.org/abs/2012.08600}{{\tt
  arXiv:2012.08600 [hep-ex]}}.

\bibitem{CMS:2021cox}
{\bf CMS} Collaboration, A.~Tumasyan {\em et al.}, ``{Search for electroweak
  production of charginos and neutralinos in proton-proton collisions at $
  \sqrt{s} $ = 13 TeV},'' \href{http://dx.doi.org/10.1007/JHEP04(2022)147}{{\em
  JHEP} {\bf 04} (2022)  147}, \href{http://arxiv.org/abs/2106.14246}{{\tt
  arXiv:2106.14246 [hep-ex]}}.

\bibitem{CMS:2022sfi}
{\bf CMS} Collaboration, ``{Search for electroweak production of charginos and
  neutralinos at $\sqrt{s}$ =13 TeV in final states containing hadronic decays
  of WW, WZ, or WH and missing transverse momentum},''
  \href{http://arxiv.org/abs/2205.09597}{{\tt arXiv:2205.09597 [hep-ex]}}.

\bibitem{CMS:2021edw}
{\bf CMS} Collaboration, A.~Tumasyan {\em et al.}, ``{Search for supersymmetry
  in final states with two or three soft leptons and missing transverse
  momentum in proton-proton collisions at $ \sqrt{s} $ = 13 TeV},''
  \href{http://dx.doi.org/10.1007/JHEP04(2022)091}{{\em JHEP} {\bf 04} (2022)
  091}, \href{http://arxiv.org/abs/2111.06296}{{\tt arXiv:2111.06296
  [hep-ex]}}.

\bibitem{Peskin:1991sw}
M.~E. Peskin and T.~Takeuchi, ``{Estimation of oblique electroweak
  corrections},'' \href{http://dx.doi.org/10.1103/PhysRevD.46.381}{{\em Phys.
  Rev. D} {\bf 46} (1992)  381--409}.

\bibitem{Haller:2018nnx}
J.~Haller, A.~Hoecker, R.~Kogler, K.~M\"onig, T.~Peiffer, and J.~Stelzer,
  ``{Update of the global electroweak fit and constraints on two-Higgs-doublet
  models},'' \href{http://dx.doi.org/10.1140/epjc/s10052-018-6131-3}{{\em Eur.
  Phys. J. C} {\bf 78} (2018) no.~8, 675},
  \href{http://arxiv.org/abs/1803.01853}{{\tt arXiv:1803.01853 [hep-ph]}}.

\bibitem{Maksymyk:1993zm}
I.~Maksymyk, C.~P. Burgess, and D.~London, ``{Beyond S, T and U},''
  \href{http://dx.doi.org/10.1103/PhysRevD.50.529}{{\em Phys. Rev. D} {\bf 50}
  (1994)  529--535}, \href{http://arxiv.org/abs/hep-ph/9306267}{{\tt
  arXiv:hep-ph/9306267}}.

\bibitem{Barbieri:2004qk}
R.~Barbieri, A.~Pomarol, R.~Rattazzi, and A.~Strumia, ``{Electroweak symmetry
  breaking after LEP-1 and LEP-2},''
  \href{http://dx.doi.org/10.1016/j.nuclphysb.2004.10.014}{{\em Nucl. Phys. B}
  {\bf 703} (2004)  127--146}, \href{http://arxiv.org/abs/hep-ph/0405040}{{\tt
  arXiv:hep-ph/0405040}}.

\bibitem{Baak:2012kk}
M.~Baak, M.~Goebel, J.~Haller, A.~Hoecker, D.~Kennedy, R.~Kogler, K.~Moenig,
  M.~Schott, and J.~Stelzer, ``{The Electroweak Fit of the Standard Model after
  the Discovery of a New Boson at the LHC},''
  \href{http://dx.doi.org/10.1140/epjc/s10052-012-2205-9}{{\em Eur. Phys. J. C}
  {\bf 72} (2012)  2205}, \href{http://arxiv.org/abs/1209.2716}{{\tt
  arXiv:1209.2716 [hep-ph]}}.

\bibitem{ParticleDataGroup:2020ssz}
{\bf Particle Data Group} Collaboration, P.~A. Zyla {\em et al.}, ``{Review of
  Particle Physics},'' \href{http://dx.doi.org/10.1093/ptep/ptaa104}{{\em PTEP}
  {\bf 2020} (2020) no.~8, 083C01}.

\bibitem{Georgi:2000vve}
H.~Georgi, \href{http://dx.doi.org/10.1201/9780429499210}{{\em {Lie Algebras In
  Particle Physics : from Isospin To Unified Theories}}}.
\newblock Taylor \& Francis, Boca Raton, 2000.

\end{thebibliography}\endgroup
\bibliographystyle{utphys}


\end{document}